\documentstyle[12pt]{article}
\newcommand{\beq}{\begin{equation}}
\newcommand{\eeq}{\end{equation}}
\newcommand{\beqa}{\begin{eqnarray}}
\newcommand{\eeqa}{\end{eqnarray}}
\newcommand{\beqan}{\begin{eqnarray*}}
\newcommand{\eeqan}{\end{eqnarray*}}
\newcommand{\no}{\nonumber}

\newcommand{\ul}{\underline}
\newcommand{\ol}{\overline}
\newcommand{\ra}{\rightarrow}

\newcommand{\ben}{\begin{enumerate}}
\newcommand{\een}{\end{enumerate}}
\newcommand{\bfl}{\begin{flushleft}}
\newcommand{\efl}{\end{flushleft}}
\newcommand{\ba}{\begin{array}}
\newcommand{\ea}{\end{array}}
\newcommand{\btab}{\begin{tabular}}
\newcommand{\etab}{\end{tabular}}
\newcommand{\bit}{\begin{itemize}}
\newcommand{\eit}{\end{itemize}}

\newcommand{\cA}{{\cal A}}

\newcommand{\cO}{{\cal O}}

\newcommand{\vs}{\vspace}
\newcommand{\hs}{\hspace}
\newcommand{\ld}{\lambda}

\def \a  {\alpha}

\def \ta {\tilde{a}}
\def \tb {\tilde{b}}
\def \tc {\tilde{c}}

\def \sam {\triangle}
\newcommand{\prepr}[1] {\begin{flushright} {\bf #1} \end{flushright} \vskip
1.cm}
\newcommand{\titul}[1] {\begin{center}{\Large {\bf #1 } } \end{center}\vskip
1.cm}

\newcommand{\autor}[1] {\begin{center} \large {\bf \lineskip .3cm #1  }
                        \end{center} }

\newcommand{\lugar}[1] {\begin{center}  {\normalsize \bf \it #1   }
\end{center}}
\newcommand{\abstr}[1] {{\begin{center} \vskip .8cm {\bf \large Abstract
                        \vspace{0pt}} \end{center}}\begin{quote}
                         \normalsize #1
                        \end{quote}}
\newcounter{muni}

\begin{document}
\vspace{2mm}
\hbadness=10000
\pagenumbering{arabic}
\begin{titlepage}
{{\bf \em 20 December 1994 } }
\prepr{HEP-PH/9501360 \\ INTERNAL REPORT \\ PAR/LPTHE/94-44 }
\titul{\Large SU(2) HEAVY FLAVOUR SYMMETRY \\
 FOR B $\ra$ K ( K$^*$ ) HADRONIC FORM FACTORS}
\vs{10mm}
\autor{ M. Gourdin\footnote{\rm Postal address: LPTHE, Tour 16, $1^{er}$ Etage,
4 Place Jussieu, F-75252 Paris CEDEX 05, France.},
Y. Y. Keum\footnote{\rm Postal address: LPTHE, Tour 24, $5^{\grave{e}me}$
Etage,
 2 Place Jussieu, F-75251 Paris CEDEX 05, France. \\
\hs{5mm}
}
and X. Y. Pham$^*$ }

\vs{10mm}
\lugar{Universit\'e Pierre {\it \&} Marie Curie, Paris VI \\
Universit\'e Denis Diderot, Paris VII \\
Physique Th\'eorique et Hautes Energies, \\
Unit\'e associ\'ee au CNRS D 0280
}

\vs{10mm}
\begin{center}
{\bf  E-mail Addresses : \\ gourdin@lpthe.jussieu.fr \\ keum@lpthe.jussieu.fr
\\ pham@lpthe.jussieu.fr }
\end{center}

\vs{-14cm}
\thispagestyle{empty}
\newpage
\vs{130mm}

\vs{50mm}
\noindent
\abstr{
\vs{10mm}
We show that the factorization assumption
in colour-suppressed $B$ meson decays
is not ruled out by experimental data on
$B \ra K(K^*) + J/\Psi(\Psi^{'})$.
The problem previously pointed out might be
due to an inadequate choice of hadronic form factors.

Within the Isgur-Wise SU(2) heavy flavour symmetry framework,
we search for possible  $q^2$-dependences of
form factors that satisfy both
the large longitudinal polarization $\rho_L$ observed
in $B \ra K^* + J/\Psi$ and the relatively small ratio of rates
$R_{J/\Psi} = \Gamma(B \ra K^* + J/\Psi)/\Gamma(B \ra K + J/\Psi)$.

We find out that the puzzle could be essentially understood
if the $A_1(q^2)$ form factor is frankly decreasing
( instead of being almost constant or increasing as commonly assumed ).

Of course, the possibility of understanding experimental data is
not necessarily a proof of factorization.

\normalsize
\vs{10mm}
{\bf PACS index 13. 25. Hw, 14. 40. Nd }

}
\end{titlepage}


\newpage
\vfill \normalsize
\tableofcontents

\newpage
\setcounter{section}{0}
\renewcommand{\theequation}{\Roman{section}.\arabic{equation}}
\setcounter{equation}{0}
\section{ \hs{4mm} Introduction}

In a recent letter \cite{GKP}, Kamal and two of us (M.G. and X.Y.P.) have
shown, within the factorization approach,
the failure of commonly used $B \ra K(K^*)$ form factors
in explaining recent data on $B \ra J/\Psi + K(K^*)$ decays.
The main problem is a simultaneous fit of the large fractional
longitudinal polarization $\rho_L$ in $B \ra J/\Psi + K^*$ decay
and of the relatively small ratio of rates
$R_{J/\Psi}$ between $J/\Psi + K^*$ and $J/\Psi + K$ in the final states.
We concluded that this difficulty in understanding experimental data
might be due to a failure of the
factorization method or to a wrong choice of hadronic form factors or both.

Such an analysis has been independently performed by Aleksan, Le Yaouanc,
Oliver, P\`ene and Raynal \cite{Orsay} who also found difficulties
in explaining both $\rho_L$ and $R_{J/\Psi}$,
in spite of their large choice of heavy to light hadronic form factors
consistent with asymptotic scaling law \cite{IW}.

In our previous work \cite{GKP},
in addition to our  exploration of the usual $B \ra K(K^*)$ form factors
 available in the literature,
we also related
 the $B \ra K(K^*)$ to the $D \ra K(K^*)$ form factors using the SU(2) flavour
symmetry between the b and c quarks as first proposed
by Isgur and Wise \cite{IW}.
The input data are the hadronic form factors in the $D$ sector at $q^2 = 0$
as extracted from the semi-leptonic
$D \ra \ol{K}(\ol{K}^*) + \ell^+ + \nu_{\ell}$ decays.
In such experiments,
the $q^2$ distribution has not been measured and the analysis
of experimental data has been made assuming monopole $q^2$-dependence for all
the $D \ra \ol{K}(\ol{K}^*)$ form factors.
For that reason in \cite{GKP}, we have also used monopole forms
in the $B$ sector.
The resulting $B \ra K(K^*)$ form factors obtained in this way
are also unable to explain
simultaneously experimental data on $\rho_L$ and $R_{J/\Psi}$.

Our method, based on the Isgur-Wise relations,
has been subsequently adopted by Cheng and Tseng
\cite{Cheng} who considered various types of $q^2$ dependences
for the hadronic form factors.
However their model still encounters difficulties in reproducing
correctly experimental data.

The purpose of this paper is to make a purely phenomenological
investigation of the possible $q^2$-dependence
$-$ we shall call scenario $-$
 of the hadronic form factors in the $B$ sector
such that, assuming factorization and using
the Isgur-Wise relations \cite{IW} together with
the latest data at $q^2 = 0$ in the $D$ sector \cite{RPR},
we can obtain a good
fit for both $\rho_L$ and $R_{J/\Psi}$.

Some preliminary remarks are in order.
We are aware of the fact that the values at $q^2 = 0$ of
the $D \ra K(K^*)$ form factors
have been extracted from semi-leptonic decay experiments
assuming a monopole $q^2$-dependence
for all hadronic form factors.
This ansatz is certainly inconsistent will all theoretical
expectations coming, for instance, from QCD sum rules \cite{QCDSUM},
from lattice gauge calculations \cite{Lattice}
as well as from scaling law of heavy flavours
\cite{{Orsay},{IW},{Cheng}}.
A correct procedure would be to reanalyze
the triple angular distribution fit
in $D$ semi-leptonic decay,
with different scenarios, in order to evaluate
the sensitivity to the scenarios of
the normalization at $q^2 = 0$ of the form factors.
Such a study has not yet been done by experimentalists.
Due to the limited range of $q^2$ in the $D$ semi-leptonic decays,
it was implicitly assumed that
the values at $q^2 = 0$ of the form factors could well be insensitive
to the scenarios.
To clarify and settle the issue,
the cleanest information would come from  measurements of the $q^2$
distributions for the rates and for the various polarizations
in the semi-leptonic $D$ sector.
We are still far from such an ideal situation and for the time being,
the only pragmatic way is to use the results quoted in \cite{RPR}
\ul{with errors included }
for the values at $q^2 = 0$ of the form factors.

We propose, in this paper, four types of scenarios for each of the
$B \ra K(K^*)$ hadronic
form factors $F_1, A_1, A_2$ and $V$ in the Bauer, Stech and Wirbel
(BSW henceforth) notation
\cite{BSW}.

The $q^2$ dependences, taken as $(1 - q^2/\Lambda^2)^{-n}$,
are applied indiscriminately to all of these form factors.
The algebraic integer $n$ symbolically represents
$n_F, n_1, n_2$ and $n_V$ associated respectively to
$F_1, A_1, A_2$ and $V$.
These integers $n$ can take four values corresponding to four types of
scenarios
mentioned above :
$- 1$ for a linear dependence, $0$ for a constant,
$+ 1$ for a monopole and $+ 2$ for a dipole.
The respective pole masses $ \Lambda_F, \Lambda_1, \Lambda_2 $ and $ \Lambda_V$
for $F_1, A_1, A_2$ and $V$
are treated as phenomenological parameters.
Being related, in some way, to bound states of the $\ol{b}s$ system,
we impose to these parameters the physical constraint to be
in the range $ (5 - 6)  \hs{2mm}GeV$.
Such a requirement is satisfied by the pole masses of the BSW model
\cite{BSW}.

We now summarize the results of our finding : \\
The experimental data on $\rho_L$ and $R_{J/\Psi}$ indeed
can been fitted for three seenarios corresponding to three possibilities
$n_2 = 2, 1, 0$ for $A_2^{BK^*}$ together with :

\hs{20mm} i) $n_1$ = $- 1$ for a linear decreasing with $q^2$ of $A_1^{BK^*}$

\hs{19mm} ii) $n_V$ = $+ 2$ for a dipole increasing with $q^2$ of $V^{BK^*}$

\hs{18mm} iii) $n_F$ = $+ 1$ for a monopole increasing with $q^2$ of $F_1^{BK}$
\\
For a given selected scenario we have a non empty allowed domain
in the  $ \Lambda_F, \Lambda_1, \Lambda_2, \Lambda_V  $ parameter space.
Therefore we obtain hadronic form factors for $B \ra K (K^*)$
reproducing correctly (within experimental errors)
 $\rho_L$ and $R_{J/\Psi}$ with
the parameters $ \Lambda_F, \Lambda_1, \Lambda_2, \Lambda_V $
physically acceptable.
We now easily understand why previous attempts
\cite{{GKP}, {Orsay}, {Cheng}}  were unsuccessful,
mainly because the decrease with $q^2$ of the form factor
$A_1(q^2)$ has never been seriously considered.
Let us emphasize however that such an unusual $q^2$ behaviour has already
been obtained by Narison \cite{QCDSUM} in the QCD sum rule approach.
Of course our result is not a proof of factorization in the $B$ sector.
It only makes wrong the statement that the difficulty in
fitting simultaneously $\rho_L$ and $R_{J/\Psi}$ implies
that factorization breaks down in colour-suppressed $B$ decays.

This paper is organized as follows.
In part II we discuss in some detail the Isgur-Wise relations \cite{IW}
and, in particular,
the consistency of scenarios in the $B$ and $D$ sectors
as well as the relations
between the parameters $\Lambda_j \hs{2mm} (j = 1 , 2, V, F)$ in both sectors.
The case of the form factors $F_0$ and $A_0$,
associated to the spin zero part of the currents is equally discussed.

In part III
we give the consequences of factorization for the decay amplitudes
$ B \ra K(K^*) + ( \eta_c, J/\Psi, {\Psi}^{'} ) $
which are colour-suppressed processes.
We study the kinematics and we review the available experimental data
for these decay modes.

The part IV is devoted to the decay modes $B \ra K(K^*) + \Psi^{'} $,
in which some scenario-independent results can be obtained.
The left-right asymmetry $ {\cA}_{LR}^{'}$ between
the two transverse polarizations in
$B \ra K^* + \Psi^{'} $ is found to be large and close to its maximal value.
The fractional longitudinal polarization $\rho^{'}_L$
turns out to be a slowly varying function of
$\Lambda_2$ and the ratio of rates $R_{\Psi^{'}}$
 a function of $\Lambda_2$ and $\Lambda_F$.
The result for $R_{\Psi^{'}}$ is consistent with experiment \cite{RPR}.
Our prediction for $\rho_L^{'}$ is compared with
that of Kamal and Sanda \cite{Kamal}
who use seven different scenarios.

The part V is the central part of this paper being related to
the decay modes $B \ra K(K^*) + J/\Psi$. The study of $\rho_L$
and $R_{J/\Psi}$  allows us to select only three surviving scenarios
among the $4^3$ = 64 possible cases
and to constraint the $ \Lambda_F, \Lambda_1, \Lambda_2 , \Lambda_V$ parameter
space.

In part VI we make predictions for the decay modes $B \ra K(K^*) + \eta_c$ for
which,
unfortunately no experimental data are available.

The comparison of various charmonium states involves
different leptonic decay constants
$f_{\Psi^{'}}, f_{J/\Psi}$ and $f_{\eta_c}$.
This problem is studied in part VII in relation
with previous works \cite{Kamal} and \cite{GKP1}
and with experimental data available only
for the ratios of $\Psi^{'}$ and $J/\Psi$ production.

The $B \ra K^*$ vector and axial vector form factors studied here
can be related under reasonable assumptions to the tensor and pseudotensor
$B \ra K^*$ form factors describing
the radiative decay $B \ra K^* + \gamma$ recently observed.
This problem is briefly discussed in part VIII.

Finally in part IX, we come back to the $D$ sector
in the light of results obtained in the $B$ sector.
Of course, the hadronic form factors $F_1, A_1, A_2, V$
follow the same scenarios in the $B$
and $D$ sectors with poles masses related by
Equations (\ref{eq:19}) and (\ref{eq:30}).
We determine the normalized $q^2$ distributions
for semi-leptonic decays  $D \ra \ol{K} \ell^+ \nu_{\ell}$,
 $D \ra \ol{K}^* \ell^+ \nu_{\ell}$
and for this last mode the integrated longitudinal polarization
$\rho_L^{\it sl}$ and the left-right asymmetry
${\cA}_{LR}^{\it sl}$
between transverse polarizations
 ({\it sl} \hs{2mm}stands for semi-leptonic).
We also study the hadronic two body decay modes
$D^o \ra K^-({K^-}^*) + \pi^+(\rho^+) $
which are colour favoured
and we show that simple factorization assumption fails
 in the $D$ sector where, in addition,
final state strong interactions play an important role.

Discussions and critical remarks on the results are given in the conclusion.

\setcounter{section}{1}
\renewcommand{\theequation}{\Roman{section}.\arabic{equation}}
\setcounter{equation}{0}
\section{ \hs{3mm} The Isgur-Wise relations due to SU(2) heavy flavour symmetry
}

$1^o)$
The SU(2) flavour symmetry between the heavy b and c quarks
allows us to derive relations between
the $B \ra K(K^*)$ and $D \ra K(K^*)$
hadronic form factors at the same velocity transfer
though at different momentum transfers.
Calling $t_B$ $(t_D)$ the value of the squared momentum transfer
$q^2$ for $B(D)$ form factors,
we obtain the following kinematic relations :
\beqa
v_b \cdot v_K = v_c \cdot v_K \hs{5mm} &{\rm or}&
\hs{3mm} m_c t_B - m_b t_D - (m_b - m_c)(m_b \hs{2mm} m_c - m_K^2) = 0
\label{eq:1} \\
v_b \cdot v_{K^*} = v_c \cdot v_{K^*} \hs{3mm} &{\rm or}&
\hs{3mm} m_c t_B^* - m_b t_D^* - (m_b - m_c)(m_b \hs{2mm} m_c - m_{K^*}^2) = 0
\label{eq:2}
\eeqa

In particular, at zero squared momentum transfer in the $D$ sector,
$t_D = t_D^* = 0$, we get
\beqa
t_B \equiv t_B^o &=& ( \frac{m_b}{m_c} \ - 1)(m_b \hs{2mm} m_c - m_K^2)
\label{eq:3} \\
t_B^* \equiv t_B^{*o} &=& ( \frac{m_b}{m_c} \ - 1)(m_b \hs{2mm} m_c -
m_{K^*}^2) \label{eq:4}
\eeqa
and useful relations such as
\beq
t_D = \frac{m_c}{m_b} \ (t_B - t_B^o)  \label{eq:3a}
\eeq
\beq
t_D^{*} = \frac{m_c}{m_b} \ (t_B^{*} - t_B^{o*})  \label{eq:3b}
\eeq

The knowledge of the hadronic form factors at $q^2 = 0$ in the $D$ sector
will determine the hadronic form factors in the $B$ sector
at $q^2 = t_B^o$ for the $B \ra K$ case and
at $q^2 = t_B^{*o}$ for the $B \ra K^*$ case.

The values of $t_B^o$ and $t_B^{*o}$ depend on the quark masses $m_b$ and
$m_c$.
We choose, in this paper, $m_b = 4.7$ $GeV$, $m_c = 1.45$ $GeV$ and get :
\beqa
t_B^o &=& 14.7287 \hs{2mm} GeV^2 \label{eq:tB1} \\
t_B^{*o} &=& 13.4933 \hs{2mm} GeV^2 \label{eq:tB2}
\eeqa

\vs{2mm}
$2^o)$ We first consider the case of $B \ra K$ and $D \ra K$ form factors.
The matrix elements of the weak current involve two form factors
$f_{+}$ and $f_{-}$ defined by
\beqa
<K|J_{\mu}|D> &=& (p_D + p_K)_{\mu} \hs{2mm} f_{+}^{DK}(q^2)
+ (p_D - p_K)_{\mu} \hs{2mm} f_{-}^{DK}(q^2) \label{eq:5-a} \\
<K|J_{\mu}|B> &=& (p_B + p_K)_{\mu} \hs{2mm} f_{+}^{BK}(q^2)
+ (p_B - p_K)_{\mu} \hs{2mm} f_{-}^{BK}(q^2) \label{eq:5-b}
\eeqa
where $q = p_{B,D} - p_K$.

The Isgur-Wise relations are written as \cite{IW}
\beqa
(f_{+} + f_{-})^{BK}(t_B) &=& C_{bc} \hs{2mm} \sqrt{ \frac{m_c}{m_b} \ }
(f_{+} + f_{-})^{DK}(t_D) \label{eq:6} \\
(f_{+} - f_{-})^{BK}(t_B) &=& C_{bc} \hs{2mm} \sqrt{ \frac{m_b}{m_c} \ }
(f_{+} - f_{-})^{DK}(t_D) \label{eq:7}
\eeqa
or equivalentlly
\beq
f_{\pm}^{BK}(t_B) = C_{bc} \hs{2mm} \left[ \frac{m_b + m_c}{2 \sqrt{m_b m_c}} \
f_{\pm}^{DK}(t_D)
- \frac{m_b - m_c}{2 \sqrt{m_b m_c}} \ f_{\mp}^{DK}(t_D) \right] \label{eq:8}
\eeq
where
$C_{bc} \equiv [ \frac{\alpha_s(m_b)}{\alpha_s(m_c)} \ ]^{-6/25} $ is
 the $QCD$ correction factor.
\footnote{In this paper, we use
$C_{bc} \equiv [ \frac{\alpha_s(m_b)}{\alpha_s(m_c)} \ ]^{-6/25} \simeq 1.135$
which has been obtained by using the recent world averaged values for
$\Lambda_{\ol{MS}}$ : $\Lambda_{\ol{MS}}^5 = (225 \pm 85) \hs{2mm} MeV$,
$\Lambda_{\ol{MS}}^4 = (325 \pm 110) \hs{2mm} MeV$,
and for the quark masses : $m_b = 4.7 \hs{2mm}GeV$, $m_c = 1.45 \hs{2mm}GeV$.}

In the BSW basis \cite{BSW}, the spin one and
the spin zero parts of the weak current are separated
and two new form factors $F_1$ and $F_0$ are defined
\beqa
F_1^{PK}(q^2) &=& f_{+}^{PK}(q^2) \label{eq:9} \\
F_0^{PK}(q^2) &=& f_{+}^{PK}(q^2) + \frac{q^2}{m_P^2 - m_K^2} \ f_{-}^{PK}(q^2)
\label{eq:10}
\eeqa
where $P = B $ or $D$.

Let us define the ratio of form factors $f_{-}$ and $f_{+}$ by :
\beq
\mu^{P}(q^2) = - \hs{2mm} \frac{f_{-}^{PK}(q^2)}{f_{+}^{PK}(q^2)}
\hs{15mm} P = B,D \label{eq:11}
\eeq

Using  Eq.(\ref{eq:8}), we obtain a relation between the two quantities
$\mu^B(t_B)$ and $\mu^D(t_D)$
\beq
\mu^B(t_B) = \frac{ \sigma + \mu^D(t_D)}{1 + \sigma \hs{2mm} \mu^D(t_D)} \
\label{eq:12}
\eeq
where the constant $\sigma$ depends only on the quark masses
\beq
\sigma = \frac{m_b - m_c}{m_b + m_c} \ \label{eq:13}
\eeq

The spin one function $F_1^{BK}$ is now related to $f_{+}^{DK}$ and $\mu^D$,
and we obtain from
Eq.(\ref{eq:8})
\beq
F_1^{BK}(t_B) = C_{bc} \hs{2mm} \frac{m_b + m_c}{2 \sqrt{m_b \hs{1mm} m_c}} \
[1 + \sigma \hs{1mm} \mu^D(t_D)]
\hs{2mm} F_1^{DK}(t_D) \label{eq:14}
\eeq
and for the spin zero function $F_{0}^{BK}$, we get from  Eq.(\ref{eq:10})
\beq
F_0^{BK}(q^2) = [ 1 - \frac{q^2}{m_B^2 - m_K^2} \ \mu^B(q^2)]
\hs{3mm} F_1^{BK}(q^2) \label{eq:15}
\eeq

\vs{2mm}
$3^o)$
As explained previously, we shall use in the $D$ sector
the values of the hadronic form factors at $q^2 = 0$
as coming from semi-leptonic experimental data.
Because of the normalization constraint $F_1^{DK}(0) = F_0^{DK}(0)$
only $f_{+}^{DK}(0)$ is known and we cannot have, in that way,
any information on $f_{-}^{DK}(0)$ and $\mu^D(0)$
\footnote{
In principle, $f_{-}^{DK}(0)$ and $\mu^D(0)$ could be measured in
$D \ra \ol{K} + \mu^{+} + \nu_{\mu}$
namely by looking at polarized muon.} .

We now make an assumption which is natural in the framework of the SU(2)
heavy flavour symmetry.
If $f_{+}^{DK}(q^2)$ and $f_{-}^{DK}(q^2)$ have the same type of
$q^2$-dependence,
no matter how it is,
then the ratio $\mu^D$ is constant and, using the Isgur-Wise relations
(\ref{eq:6}) and
(\ref{eq:7}) we easily see that the same property extends to the $B$ sector and
,
in particular, the ratio $\mu^B$ is a constant related to $\mu^D$ by
Eq.(\ref{eq:12}).
Of course, $F_1^{BK}(q^2)$ and $F_1^{DK}(q^2)$ have the same type of
$q^2$ dependence.

For instance, if $F_1^{DK}(q^2)$ is written in the form
\beq
F_1^{DK}(q^2) = \frac{F_1^{DK}(0)}{ [1 - \frac{q^2}{\Lambda_{DF}^2}]^{n_F}} \
\label{eq:16}
\eeq
where $n_F$ is some algebraic integer, then, using Eq.(\ref{eq:14})
we obtain a similar expression for $F_1^{BK}(q^2)$
\beq
F_1^{BK}(q^2) = \frac{F_1^{BK}(0)}{ [1 - \frac{q^2}{\Lambda_{F}^2}]^{n_F}} \
\label{eq:17}
\eeq
with the same $n_F$.

Furthermore a relation between the pole masses $\Lambda_F$ and $\Lambda_{DF}$
has been previously given in Ref.\cite{GKP} and it comes from the identity
\beq
(1 - \frac{t_D}{\Lambda_{DF}^2}) = 1 - \frac{m_c}{m_b} \
\frac{t_B - t_B^o}{\Lambda_{DF}^2} \
= (1 + \frac{m_c}{m_b} \ \frac{t_B^o}{\Lambda_{DF}^2} \ )
(1 - \frac{t_B}{\Lambda_F^2} \ )                         \label{eq:18}
\eeq
The result is
\beq
m_c \hs{2mm} \Lambda_F^2 - m_b \hs{2mm} \Lambda_{DF}^2 = m_c \hs{2mm} t_B^o
= (m_b - m_c) (m_b \hs{2mm} m_c - m_K^2)                  \label{eq:19}
\eeq
The values at $q^2 = 0$ of the form factor $F_1$ in the $B$ and $D$ sectors are
also
related by Eq.(\ref{eq:14}).
The result depends on $n_F$ and is given by
\beq
\frac{F_1^{BK}(0)}{F_1^{DK}(0)} \ = C_{bc} \hs{2mm} \left(
\frac{m_b + m_c}{2 \sqrt{m_b \hs{2mm} m_c} } \hs{2mm}
[1 + \sigma \hs{2mm} \mu^D] \hs{2mm}
[ \frac{m_b \hs{2mm} \Lambda_{DF}^2}{m_c \hs{2mm} \Lambda_F^2} \ ]^{n_F}
\right)
\label{eq:20}
\eeq

\vs{2mm}
$4^o)$
An interesting scenario for $F_1^{DK}$, suggested by many theoretical studies
\cite{{Orsay}, {BSW}, {Neubert}}
as well as supported by experimental data \cite{{RPR}, {Witherell}}
is a monopole dependence $n_F = 1$ with a pole mass
$\Lambda_{DF}$ in the $2 \hs{2mm} GeV$ region.
While $F_1^{BK}$ and $F_1^{DK}$ increase with $q^2$ by a monopole type,
from Eq.(\ref{eq:15}) we easily see that
the form factors $F_0^{BK}$ and $F_0^{DK}$ have a different $q^2$ behaviour
due to a supplementary linearly decreasing factor in Eq.(\ref{eq:15})
$$
[1 - \frac{q^2}{m_B^2 - m_K^2} \ \hs{2mm} \mu^B]
\hs{15mm} {\rm for} \hs{10mm} F_0^{BK}
$$
$$
[1 - \frac{q^2}{m_D^2 - m_K^2} \ \hs{2mm} \mu^D]
\hs{15mm} {\rm for} \hs{10mm} F_0^{DK}
$$
In fact, for a particular value of $\mu^B$ or $\mu^D$, this factor can exactly
cancel
the pole of $F_1$ in the $B$ or in the $D$ sector, thus  making $F_0$ constant.
This is the case when
\beq
\mu^B = \frac{m_B^2 - m_K^2}{\Lambda_F^2} \hs{10mm} {\rm or} \hs{10mm}
\mu^D = \frac{m_D^2 - m_K^2}{\Lambda_{DF}^2}
\label{eq:21}
\eeq
In this way, the value of $\mu^B(\mu^D)$ is related to the pole mass
$\Lambda_F(\Lambda_{DF})$.
However such a situation, in general, does not occur
in both $B$ and $D$ sectors
because $\Lambda_F^2$ and $\Lambda_{DF}^2$ are
not independent from each other
due to Eq.(\ref{eq:19}).

If we impose the constraints (\ref{eq:21}) for $\mu^B$ and $\mu^D$,
we can determine the pole masses $\Lambda_F^2$ and $\Lambda_{DF}^2$.
The relations (\ref{eq:12}), (\ref{eq:19}) and (\ref{eq:21}) lead to
a second order equation for $\Lambda_{DF}^2$,
the roots of which are real.
However one root ( negative) is  physically unacceptable,
and the other one ( positive ) gives  $\Lambda_{DF}$
within the (2 $-$ 3) $GeV$ range, corresponding to $\Lambda_F$
within the (5 $-$ 6) $GeV$ range
and the value of $\mu^B$ is in the neighbourhood of 1.

The calculation involves the $b$ and $c$ quark masses :
with $m_b = 4.7$ $GeV$ and $m_c = 1.45$ $GeV$,
the hadronic form factor $F_0$ is constant in both $B$ and $D$ sectors
for the following pole mass values $\Lambda_{DF} = 2.32$ $GeV$,
$\Lambda_{F} = 5.67$ $GeV$ and for $\mu^B$ and $\mu^D$,
we obtain $\mu^B = 0.8585$ and $\mu^D = 0.6041$.

\vs{2mm}
$5^o)$
We do not have, to our knowledge,
a theoretical estimate of the form factors
$f_{-}^{DK}(0)$ and $f_{-}^{BK}(0)$.
The heavy quark symmetry limit can not be applied for
heavy quark to light quark, $b \ra s$, transitions.
Therefore we shall use, in the $B$ sector, a model suggested by
theoretical considerations \cite{{Cheng}, {QPXU}}
where $F_0^{BK}$ is independent of $q^2$ and $F_1^{BK}$
has a monopole $q^2$ dependence with a pole mass $\Lambda_F$.
The parameter $\mu^B$ is then a function of $\Lambda_F^2$
as given in Eq.(\ref{eq:21}) and the parameter $\mu^D$ is known
from $\mu^B$ by inverting Eq.(\ref{eq:12})
\beq
\mu^D = \frac{\mu^B - \sigma}{1 - \sigma \hs{2mm} \mu^B} \
\label{eq:22}
\eeq
where $\sigma = 0.5285$ for our choice of quark masses.

We plot in Figure 1 the ratios $\mu^B$ and $\mu^D$ as funtions
of $\Lambda_F^2$ for values of $\Lambda_F$ in the (4 $-$ 7) $GeV$ range.
They are parts of hyperbolae with asymptotia parallel to the $\Lambda_F^2$
and $\mu$ axes.
We plot, by a straight line, Eq.(\ref{eq:19})
which relates the pole masses in the $B$ and $D$ sectors.
On the same figure, we also show, with dotted points,
the quantity $\ol{\mu}^D$ which corresponds to a constant
$F_0$ form factor in the $D$ sector.

\vs{2mm}
$6^o)$
We now study the case of the $B \ra K^*$ and $D \ra K^*$ form factors.
The matrix elements of the weak current involve four form factors,
$f, g, a_{+}$ and $a_{-}$ in the Isgur-Wise basis, or $A_1, V, A_2$ and $A_0$
in the BSW basis.

The Isgur-Wise relations for $f$ and $g$ are very simple \cite{IW}
\beqa
f^{BK^*}(t_B^*) &=& C_{bc} \hs{2mm} \sqrt{ \frac{m_b}{m_c} \ }
\hs{2mm} f^{DK^*}(t_D^*) \label{eq:23} \\
g^{BK^*}(t_B^*) &=& C_{bc} \hs{2mm} \sqrt{ \frac{m_c}{m_b} \ }
\hs{2mm} g^{DK^*}(t_D^*) \label{eq:24}
\eeqa
Using the relations between the Isgur-Wise and BSW bases :
\beq
f^{PK^*} = (m_P + m_{K^*}) \hs{2mm} A_1^{PK^*} \hs{5mm},\hs{5mm}
g^{PK^*} = \frac{1}{(m_P + m_{K^*})} \ \hs{2mm} V^{PK^*}
\hs{5mm}; \hs{5mm} P = B, D  \label{eq:25}
\eeq
we get
\beqa
A_1^{BK^*}(t_B^*) &=& C_{bc} \hs{2mm}
\sqrt{ \frac{m_b}{m_c} \ } \hs{2mm}
 \frac{m_D + m_{K^*}}{m_B + m_{K^*}} \
\hs{2mm} A_1^{DK^*}(t_D^*) \label{eq:26}  \\
\cr
V^{BK^*}(t_B^*) &=& C_{bc} \hs{2mm}
\sqrt{ \frac{m_c}{m_b} \ } \hs{2mm}
 \frac{m_B + m_{K^*}}{m_D + m_{K^*}} \
\hs{2mm} V^{DK^*}(t_D^*) \label{eq:27}
\eeqa

In particular, defining the ratio
\beq
y(q^2) \equiv \frac{V(q^2)}{A_1(q^2)} \ \label{eq:28}
\eeq
the values of $y$ in the $B$ and $D$ sectors,
$y^B$ and $y^D$ are related by :
\beq
y^B(t_B^*) = \frac{m_c}{m_b} \ \hs{2mm}
\left(  \frac{m_B + m_{K^*}}{m_D + m_{K^*}} \right)^2 \hs{2mm}
y^D(t_D^*) \label{eq:29}
\eeq
The dependence with respect to $q^2$ of $A_1$ and $V$
in the $B$ and $D$ sectors is preserved by
the Isgur-Wise relations (\ref{eq:26}) and (\ref{eq:27}).
We choose the form factors $A_1^{BK^*}(q^2)$ and $V^{BK^*}(q^2)$
such as :
\beqa
A_1^{BK^*}(q^2) &=& \frac{A_1^{BK^*}(0)}{ [1 - \frac{q^2}{\Lambda_1^2}]^{n_1}}
\ \label{eq:31} \\
\cr
V^{BK^*}(q^2) &=& \frac{V^{BK^*}(0)}{ [1 - \frac{q^2}{\Lambda_V^2}]^{n_V}} \
\label{eq:32}
\eeqa
and analogous expressions in the $D$ sectors with pole masses $\Lambda_{D1}$
and $\Lambda_{DV}$ related to $\Lambda_1$ and $\Lambda_V$
by a formula  similar to Eq.(\ref{eq:19})
\beq
m_c \hs{2mm} \Lambda_B^2 - m_b \hs{2mm} \Lambda_{D}^2 = m_c \hs{2mm} t_B^{o*}
= (m_b - m_c) (m_b \hs{2mm} m_c - m_{K^*}^2)                  \label{eq:30}
\eeq
where $\Lambda_B = \Lambda_1$ or $\Lambda_V$ and
$\Lambda_D = \Lambda_{D1}$ or $\Lambda_{DV}$.
The algebraic integers $n_1$ and $n_V$ will be restricted to the values
$-1, 0, 1 $ and $2$.
The normalizations at $q^2 = 0$ of these form factors
in the $B$ and $D$ sectors
are related using Eqs.(\ref{eq:26}) and (\ref{eq:27})
\beqa
\frac{A_1^{BK^*}(0)}{A_1^{DK^*}(0)} \ &=& C_{bc} \hs{2mm}
\sqrt{ \frac{m_b}{m_c} \ } \hs{2mm}
\frac{m_D + m_{K^*}}{m_B + m_{K^*}} \  \hs{2mm}
[1 - \frac{t_B^{*o}}{\Lambda_1^2}]^{n_1}            \label{eq:33} \\
\cr
\frac{V^{BK^*}(0)}{V^{DK^*}(0)} \ &=& C_{bc} \hs{2mm}
\sqrt{ \frac{m_c}{m_b} \ } \hs{2mm}
\frac{m_B + m_{K^*}}{m_D + m_{K^*}} \  \hs{2mm}
[1 - \frac{t_B^{*o}}{\Lambda_V^2}]^{n_V}            \label{eq:34}
\eeqa

\vs{2mm}
$7^o)$
For the two other form factors $a_{+}$ and $a_{-}$ describing
the $P \ra K^*$ transition, the situation is formally similar to
$f_{+}$ and $f_{-}$ in the $P \ra K$ case previously considered.
The Isgur-Wise relations are written as \cite{IW}
\beqa
(a_{+} + a_{-})^{BK^*}(t_B^*) = C_{bc} \hs{2mm} \left( \frac{m_c}{m_b}
\right)^{3/2}
\hs{2mm} (a_{+} + a_{-})^{DK^*}(t_D^*)  \label{eq:35} \\
\cr
(a_{+} - a_{-})^{BK^*}(t_B^*) = C_{bc} \hs{2mm} \left( \frac{m_c}{m_b}
\right)^{1/2}
\hs{2mm} (a_{+} - a_{-})^{DK^*}(t_D^*)  \label{eq:36}
\eeqa
from which we deduce
\beq
a_{\pm}^{BK^*}(t_B^*) =
\frac{1}{2} \ \hs{2mm} C_{bc} \hs{2mm} \sqrt{ \frac{m_c}{m_b} \ }
\left[ (1 + \frac{m_c}{m_b} ) \hs{1mm} a_{\pm}^{DK^*}(t_D^*)
- (1 - \frac{m_c}{m_b} ) \hs{1mm} a_{\mp}^{DK^*}(t_D^*)
\right]  \label{eq:37}
\eeq

In the BSW basis \cite{BSW}, the spin one and the spin zero parts of
the weak current are separated,
with $A_1, V$ and $A_2$ for the spin one part
and  $A_0$ for the spin zero part.
The form factors $A_2$ and $A_0$ are linear combinations of the form factors
$a_{+}$, $a_{-}$
in the Isgur-Wise basis :
\beqa
A_2^{PK^*}(q^2) &=& - \hs{1mm} (m_P + m_{K^*}) \hs{2mm} a_{+}^{PK^*}(q^2)
\label{eq:38} \\
\cr
A_0^{PK^*}(q^2) &=& \frac{m_P + m_{K^*}}{2 \hs{1mm} m_{K^*}} \hs{2mm}
A_1^{PK^*}(q^2)
- \frac{m_P - m_{K^*}}{2 \hs{1mm} m_{K^*}} \hs{2mm} A_2^{PK^*}(q^2)
+ \frac{q^2}{2 \hs{1mm} m_{K^*}} \hs{2mm} a_{-}^{PK^*}(q^2) \label{eq:39}
\eeqa
where $P = B$ or $D$.

At $q^2 = 0$ the normalization of $A_0$ is determined by $A_1(0)$ and $A_2(0)$
:
\beq
A_0^{PK^*}(0) = \frac{m_P + m_{K^*}}{2 \hs{1mm} m_{K^*}} \hs{2mm} A_1^{PK^*}(0)
- \frac{m_P - m_{K^*}}{2 \hs{1mm} m_{K^*}} \hs{2mm} A_2^{PK^*}(0),
\hs{10mm} P = B, D  \label{eq:40}
\eeq
and we can not have any information on the value at $q^2 = 0$
of the form factor $a_{-}$.
\footnote{similar to the $f_{-}(q^2)$ case, the form factor $a_{-}(q^2)$
can be measured in $D \ra \ol{K}^* + \mu^{+} + \nu_{\mu}$. }
Let us now define the ratio of the form factors $a_{-}$ and $a_{+}$ by :
\beq
\lambda^P(q^2) = - \hs{1mm} \frac{a_{-}^{PK^*}(q^2)}{a_{+}^{PK^*}(q^2)} \,
\hs{10mm} P = B, D \label{eq:41}
\eeq

Using Eq.(\ref{eq:37}) we obtain a relation between the two quantities
$\lambda^B(t_B^*)$ and $\lambda^D(t_D^*)$ similar to Eq.(\ref{eq:12})
\beq
\lambda^B(t_B^*) = \frac{\sigma + \lambda^D(t_D^*)}
{1 + \sigma \hs{1mm} \lambda^D(t_D^*)}  \label{eq:42}
\eeq
with $\sigma$ defined in Eq.(\ref{eq:13}).

The function $A_2^{BK^*}(t_B^*)$ is now expressed in terms of
$A_2^{DK^*}(t_D^*)$ and $\lambda^D(t_D^*)$.

Using  Eqs.(\ref{eq:37}) and (\ref{eq:38}), we get
\beq
A_2^{BK^*}(t_B^*) = \frac{1}{2} \hs{1mm} C_{bc} \hs{2mm}
 \sqrt{ \frac{m_c}{m_b} }
\hs{2mm} \left( 1 + \frac{m_c}{m_b} \right) \hs{2mm}
\frac{m_B + m_{K^*}}{m_D + m_{K^*}} \
\hs{2mm} [ 1 + \sigma \hs{2mm} \lambda^D(t_D^*)] \hs{2mm} A_2^{DK^*}(t_D^*)
                                                               \label{eq:43}
\eeq
With Eqs.(\ref{eq:39}) and (\ref{eq:41}) we obtain the spin zero function
$A_0^{BK^*}$ :
\beq
A_0^{BK^*}(q^2) = \frac{m_B + m_{K^*}}{2 \hs{2mm} m_{K^*}} \hs{2mm}
A_1^{BK^*}(q^2) -
\frac{m_B - m_{K^*}}{2 \hs{2mm} m_{K^*}} \hs{2mm}
[1 - \frac{q^2}{m_B^2 - m_{K^*}^2} \hs{2mm} \lambda^B(q^2)]
\hs{2mm} A_2^{BK^*}(q^2)                                       \label{eq:44}
\eeq
Defining the ratio :
\beq
x(q^2) \equiv \frac{A_2(q^2)}{A_1(q^2)} \ \label{eq:45}
\eeq
and using  Eqs.(\ref{eq:26}) and (\ref{eq:43}),
the values of $x$ in the $B$ and $D$ sectors, $x^B$ and $x^D$, are related by
\beq
x^B(t_B^*) = \frac{1}{2} \hs{2mm} \frac{m_c}{m_b} \hs{2mm}
\left(1 + \frac{m_c}{m_b} \right) \hs{2mm}
\left( \frac{m_B + m_{K^*}}{m_D + m_{K^*}} \right)^2 \hs{2mm}
[1 + \sigma \hs{2mm} \lambda^D(t_D^*)] \hs{2mm} x^D(t_D^*) \label{eq:46}
\eeq

\vs{2mm}
$8^o)$
We now make an assumption for $a_{+}$ and $a_{-}$ similar to the one
made previously for $f_{+}$ and $f_{-}$.
If $a_{+}^{DK^*}(q^2)$ and $a_{-}^{DK^*}(q^2)$ have the same type of
$q^2$-dependence, no matter how it is,
then the ratio $\lambda^D$ is constant and using the Isgur-Wise
relations (\ref{eq:35}) and (\ref{eq:36}), the same property extends to
the $B$ sector with, in particular, a constant value for the ratio
$\lambda^B$ related to $\lambda^D$ by Eq.(\ref{eq:42}).
Of course, $A_2^{BK^*}(q^2)$ and $A_2^{DK^*}(q^2)$ have also the same type
of $q^2$-dependence.

The form factor $A_2^{BK^*}(q^2)$ is written in the form
\beq
A_2^{BK^*}(q^2) = \frac{A_2^{BK^*}(0)}
{ \left[ 1 - \frac{q^2}{\Lambda_2^2} \right]^{n_2} } \   \label{eq:47}
\eeq
and the normalizations at $q^2 =0$ of $A_2$ in the $B$ and $D$ sectors are
related using Eq.(\ref{eq:43})
\beq
\frac{A_2^{BK^*}(0)}{A_2^{DK^*}(0)} = \frac{1}{2} \hs{2mm}
\sqrt{ \frac{m_c}{m_b}} \hs{2mm} C_{bc} \hs{2mm}
\left( 1 + \frac{m_c}{m_b} \right) \hs{2mm}
\frac{m_B + m_{K^*}}{m_D + m_{K^*}} \hs{2mm}
[ 1 + \sigma \hs{1mm} \lambda^D] \hs{2mm}
[1 - \frac{t_B^{*o}}{\Lambda_2^2} ]^{n_2} \label{eq:48}
\eeq

The second term in Eq.(\ref{eq:44}) for $A_0$ is written as
\beq
A_0^{BK^*}(q^2) - \frac{m_B + m_{K^*}}{2 \hs{2mm} m_{K^*}} \hs{2mm}
A_1^{BK^*}(q^2) = - \hs{1mm}
\frac{m_B - m_{K^*}}{2 \hs{2mm} m_{K^*}} \hs{2mm}
[1 - \frac{q^2}{m_B^2 - m_{K^*}^2} \hs{2mm} \lambda^B]
\hs{2mm} \frac{A_2^{BK^*}(0)}
{ \left[ 1 - \frac{q^2}{\Lambda_2^2} \right]^{n_2} } \  \label{eq:49}
\eeq
and it exhibits a different $q^2$ behaviour compared to
$A_2$ due to a supplementary
linearly decreasing factor
$( 1 - \frac{q^2 \hs{2mm} \lambda^B}{m_B^2 - m_{K^*}^2} )$.
For a particular value of $\lambda^B$ or $\lambda^D$,
this factor cancels
exactly one power of $A_2^{BK^*}$ or $A_2^{DK^*}$.
This is the case when
\beq
\lambda^B = \frac{m_B^2 - m_{K^*}^2}{\Lambda_2^2} \hs{10mm} or
\hs{10mm}
\lambda^D = \frac{m_D^2 - m_{K^*}^2}{\Lambda_{D2}^2}
\label{eq:50}
\eeq

In this way the value of $\lambda^B(\lambda^D)$ is related to the pole mass
$\Lambda_2(\Lambda_{D2})$.
However such a situation does not occur, in general,
in both $B$ and $D$ sectors because
$\Lambda_2^2$ and $\Lambda_{D2}^2$ are not independent
from each other due to
a relation similar to  Eq.(\ref{eq:30}).
Conversely if we impose the two constraints (\ref{eq:50}),
we can determine the pole masses
$\Lambda_2^2$ and $\Lambda_{D2}^2$.
The relations (\ref{eq:30}) and (\ref{eq:42}) lead to
a second order equation for
$\Lambda_{D2}^2$.
However one root is negative and must be rejected.
The second one gives  $\Lambda_{D2}$ in the (2 $-$ 3) $GeV$
range corresponding to
$\Lambda_2$ in the (5 $-$ 6) $GeV$ range.
The value of $\lambda^B$ is not far from 1,
the heavy quark symmetry prediction.
Using as previously $m_b = 4.7 \hs{2mm} GeV$ and $m_c = 1.45 \hs{2mm} GeV$,
we obtain $\Lambda_{D2} = 2.59$ $GeV$,
$\Lambda_2 = 5.93$ $GeV$, $\lambda^B = 0.7698$ and $\lambda^D = 0.4068$.

\vs{2mm}
$9^o)$
In the absence of a theoretical estimate for $a_{-}^{DK^*}(0)$ or
$a_{-}^{BK^*}(0)$,
we shall use in the $B$ sector a model where $\lambda^B$ is related to
the pole mass $\Lambda_2$ by Eq.(\ref{eq:50}).
The parameter $\lambda^D$ is then computed from $\lambda^B$
by inverting Eq.(\ref{eq:42})
\beq
\lambda^D = \frac{ \lambda^B - \sigma}
{1 - \sigma \hs{2mm} \lambda^B} \ \label{eq:51}
\eeq

We plot in Figure 2 the ratios $\lambda^B$ and $\lambda^D$ as functions of
$\Lambda_2^2$, for values of $\Lambda_2$ in the range (4- 7) $GeV$.
They are parts of hyperbolae with asymptotia  parallel to
the $\Lambda_2^2$ and $\lambda$ axes.
Eq.(\ref{eq:30}) which relates the pole masses in the $B$ and $D$ sectors
is represented by a straight line in Figure 2.
On the same Figure we also show, with dotted points, the quantity
$ \ol{\lambda}^D$
which corresponds to the second relation (\ref{eq:50}).

\setcounter{section}{2}
\renewcommand{\theequation}{\Roman{section}.\arabic{equation}}
\setcounter{equation}{0}
\section{ \hs{3mm} Factorization and Kinematics, Experimental Data }

\vs{2mm}
$1^o)$
The two-body decays of the charged and neutral $B$ mesons
discussed in this paper are described,
at the tree level, by the colour-suppressed diagram of Figure 3.
Of course penguin diagrams also contribute to these decays
at the one loop level.
However the colourless charmonium states $\ol{c}c$
have to be excited from the vacuum and
for which two or three gluons are needed.
For that reason the penguin contribution will be neglected in this paper.

\vs{2mm}
$2^o)$
We consider the decay modes
\beqa
B^{+} &\ra& K^{+}({K^*}^{+}) + \eta_c \hs{1mm}
(J/\Psi, \hs{1mm} \Psi^{'}) \no \\
& &                                                    \label{eq:52} \\
B^{o} &\ra& K^{o}({K^*}^{o}) + \eta_c \hs{1mm}
(J/\Psi, \hs{1mm} \Psi^{'}) \no
\eeqa
and we compute the decay amplitudes assuming factorization.
We obtain an expression of the form
\beq
< \ol{c}c + \ol{s}q \hs{2mm} | T | \hs{2mm} \ol{b}q > \hs{5mm} \propto \hs{5mm}
< \ol{c}c | J^{\mu} | 0 > \hs{2mm} < \ol{s}q | J_{\mu} | B >  \label{eq:53}
\eeq
The first factor in the right hand side of Eq.(\ref{eq:53})
involves the leptonic decay constants
$f_{\eta_c}, f_{J/\Psi}$ and $f_{\Psi^{'}}$
for $\eta_c, J/\Psi$ and $\Psi^{'}$ respectively.
The second factor is governed by the hadronic from factors for the $B \ra K$
or $B \ra K^*$ transitions.
As a consequence, the branching ratios have the following structure
\beq
BR = BR_{0} \hs{2mm} \cdot \hs{2mm}
\left( \frac{f_{\ol{c}c}}{m_B} \right)^2
\hs{2mm} \cdot \hs{2mm} PS
\hs{2mm} \cdot \hs{2mm} FF               \label{eq:54}
\eeq
The scale $BR_{0}$ contains the Fermi coupling constant $G_{F}$,
the Cabibbo-Kobayashi-Maskawa ( CKM ) factors $V_{cb}^*V_{cs}$
as indicated on Figure 3, the $B$ meson life time $\tau_B$ and
the BSW phenomenological constant $a_2$
describing colour-suppressed processes
\beq
BR_{0} = \left[ \frac{G_F m_B^2}{\sqrt{2}} \right]^2 \hs{2mm}
|V_{cb}|^2 \hs{2mm} |V_{cs}|^2 \hs{2mm}
\frac{m_B}{8  \pi} \hs{2mm}
a_2^2 \hs{2mm} \frac{\tau_B}{\hbar}                          \label{eq:55}
\eeq
Being interested only in ratios of decay widths,
we shall not numerically compute $BR_0$.

The quantity $PS$ is a dimensionless phase space factor depending only on
 masses of the involved particles.
Because of the small $K^{+}({K^*}^{+}) - K^o({K^*}^{o})$ mass differences,
the numerical value of $PS$ is slightly different for $B^{+}$ and $B^o$ decays.
However the differences in these quantities are typically $\cO(10^{-3})$,
hence we ignore the mass difference between charged and neutral strange mesons.
We give results for $B^{+}$ decay.

The last factor $FF$ depends on the hadronic form factors and it contains
the dynamics of the weak decays.
The results are :
\beqa
&(a)&  B^+ \ra K^+ + \eta_c
\hs{7mm} : \hs{10mm} PS = 0.3265
\hs{5mm}, \hs{10mm} FF = |F_0^{BK}(m_{\eta_c}^2)|^2  \label{eq:56} \\
&(b)&  B^+ \ra K^+ + J/\Psi
\hs{3mm} : \hs{9mm} PS = 0.1296
\hs{5mm}, \hs{10mm} FF = |F_1^{BK}(m_{J/\Psi}^2)|^2  \label{eq:57} \\
&(c)&  B^+ \ra K^+ + \Psi^{'}
\hs{6mm} : \hs{9mm} PS = 0.0575
\hs{5mm}, \hs{10mm} FF = |F_1^{BK}(m_{\Psi^{'}}^2)|^2  \label{eq:58} \\
&(d)&  B^+ \ra {K^*}^+ + \eta_c
\hs{5mm} : \hs{9mm} PS = 0.1218
\hs{5mm}, \hs{10mm} FF = |A_0^{BK^*}(m_{\eta_c}^2)|^2  \label{eq:59} \\
&(e)&  B^+ \ra {K^*}^+ + J/\Psi
\hs{1mm} : \hs{9mm} PS = 0.1399
\hs{5mm}, \hs{10mm}  \no \\
& &  \hs{40mm} FF = |A_1^{BK^*}(m_{J/\Psi}^2)|^2
\hs{2mm} [(a - b x)^2 + 2 \hs{1mm}(1 + c^2 y^2) ]  \label{eq:60} \\
&(f)&  B^+ \ra {K^*}^+ + \Psi^{'}
\hs{4mm} : \hs{8mm} PS = 0.1407
\hs{5mm}, \hs{10mm}  \no \\
& & \hs{40mm} FF = |A_1^{BK^*}(m_{\Psi^{'}}^2)|^2
\hs{2mm} [(a^{'} - b^{'} x^{'})^2 + 2 \hs{1mm}(1 + {c^{'}}^2 {y^{'}}^2)]
\label{eq:61}
\eeqa
The analytic expressions for $a, b, c$ are previously given in Ref.\cite{GKP}
and $a^{'}, b^{'}, c^{'}$ are obtained respectively from $a, b, c$
by the simple substitution $m_{\Psi^{'}}$ to $m_{J/\Psi}$.
We get numerically :
\beqa
&& a = 3.1652 \hs{16mm} b = 1.3081 \hs{16mm} c = 0.4356  \label{eq:62} \\
&& a^{'} = 2.0514 \hs{15mm} b^{'} = 0.5538 \hs{15mm} c^{'} = 0.3092
\label{eq:63}
\eeqa
The ratios of form factors $x, \hs{1mm} y, \hs{1mm} x^{'}, \hs{1mm} y^{'}$
are defined by :
\beqa
x \equiv x^B(m_{J/\Psi}^2)
= \frac{A_2^{B{K^*}}(m_{J/\Psi}^2)}{A_1^{B{K^*}}(m_{J/\Psi}^2)}
\hs{8mm},
& & \hs{9mm}
y \equiv y^B(m_{J/\Psi}^2)
= \frac{V^{B{K^*}}(m_{J/\Psi}^2)}{A_1^{B{K^*} }(m_{J/\Psi}^2)}
                                                    \label{eq:64} \\
\cr
x^{'} \equiv x^B(m_{\Psi^{'}}^2)
= \frac{A_2^{B{K^*}}(m_{\Psi^{'}}^2)}{A_1^{BK^*}(m_{\Psi^{'}}^2)}
\hs{10mm},
& & \hs{10mm}
y^{'} \equiv y^B(m_{\Psi^{'}}^2)
= \frac{V^{BK^*}(m_{\Psi^{'}}^2)}{A_1^{BK^*}(m_{\Psi^{'}}^2)}
                                                    \label{eq:65}
\eeqa

In the $K^* + J/\Psi$ and $K^* + \Psi^{'}$ modes,
we have three possible polarization states,
one is longitudinal and two are transverse for both final particles.
We shall define two interesting quantities, the fractional longitudinal
polarization :
\beq
\rho_L = \frac{\Gamma(B \ra K^* + J/\Psi)_{LL}}{\Gamma(B \ra K^* + J/\Psi)}
\hs{10mm}, \hs{10mm}
\rho_L^{'} = \frac{\Gamma(B \ra K^* + \Psi^{'})_{LL}}{\Gamma(B \ra K^* +
\Psi^{'})} \label{eq:66}
\eeq
and the left-right asymmetry :
\beqa
&& \cA_{LR} = \frac{\Gamma(B \ra K^* + J/\Psi)_{--} - \Gamma(B \ra K^* +
J/\Psi)_{++}}
{\Gamma(B \ra K^* + J/\Psi)_{--} + \Gamma(B \ra K^* + J/\Psi)_{++}} \hs{20mm}
\no \\
&&
\label{eq:67} \\
&& \cA_{LR}^{'} = \frac{\Gamma(B \ra K^* + \Psi^{'})_{--} - \Gamma(B \ra K^* +
\Psi^{'})_{++}}
{\Gamma(B \ra K^* + \Psi^{'})_{--} + \Gamma(B \ra K^* + \Psi^{'})_{++}}  \no
\eeqa

We get
\beq
\rho_L = \frac{(a - b x )^2}{(a - b x)^2 + 2 [1 + c^2 \hs{1mm} y^2]}
\hs{10mm}, \hs{10mm}
\rho_L^{'} = \frac{(a^{'} - b^{'} x^{'} )^2}
{(a^{'} - b^{'} x^{'})^2 + 2 [1 + {c^{'}}^2 \hs{1mm} {y^{'}}^2]}  \label{eq:68}
\eeq
\beq
\cA_{LR} = \frac{2 c y }{1 + c^2 \hs{2mm} y^2} \hs{25mm},\hs{20mm}
\cA_{LR}^{'} = \frac{2 c^{'} y^{'}}
{1 + {c^{'}}^2 {y^{'}}^2}                                         \label{eq:69}
\eeq
We shall also define two ratios of rates
\beq
R_{J/\Psi} = \frac{\Gamma(B \ra K^* + J/\Psi)}
{\Gamma(B \ra K + J/\Psi)}   \hs{10mm}, \hs{10mm}
R_{\Psi^{'}} = \frac{\Gamma(B \ra K^* + \Psi^{'})}
{\Gamma(B \ra K + \Psi^{'})}
\label{eq:70}
\eeq

\vs{2mm}
$3^o)$
The experimental data for decay rates as averaged by PDG \cite{RPR} are given
in Table 1.

\vs{3mm}
\begin{table}[thb]
\begin{center}
\begin{tabular}{|c||c||c||}
\hline
Modes
& $B^{+}$
& $B^{o}$  \\
\hline
\hline
$K + \Psi^{'} $ &  $(0.69 \pm 0.31) \hs{1mm} \cdot \hs{1mm} 10^{-3} $
& $ < \hs{2mm} 0.8 \hs{1mm} \cdot \hs{1mm} 10^{-3} $ \\
\hline
$K^* + \Psi^{'} $  &  $ < \hs{2mm} 3.0 \hs{1mm} \cdot \hs{1mm} 10^{-3} $
& $(1.4 \pm 0.9) \hs{1mm} \cdot \hs{1mm} 10^{-3}$  \\
\hline
\hline
$K + J/\Psi$  &  $(1.02 \pm 0.14) \hs{1mm} \cdot \hs{1mm} 10^{-3}$
& $(0.75 \pm 0.21) \hs{1mm} \cdot \hs{1mm} 10^{-3}$  \\
\hline
$K^* + J/\Psi$  &  $(1.7 \pm 0.5) \hs{1mm} \cdot \hs{1mm} 10^{-3}$
& $(1.58 \pm 0.28) \hs{1mm} \cdot \hs{1mm} 10^{-3}$  \\
\hline
\hline
\end{tabular}

\vs{3mm}
Table 1. \\
\end{center}
\end{table}
%
The modes $K\eta_c$ and $K^*\eta_c$ have not yet been observed experimentally.

The ratios $R_{\Psi^{'}}$ and $R_{J/\Psi}$ can be estimated from the data of
Table 1 and
the results are shown in Table 2.

\vs{3mm}
\begin{table}[thb]
\begin{center}
\begin{tabular}{|c||c||c||c||}
\hline
Ratios
& $B^{+}$
& $B^{o}$
& $B^{+}, B^o$ combined  \\
\hline
\hline
$R_{\Psi^{'}}$ &   $ < \hs{2mm} 4.35 \pm 1.95 $
& $ > 1.75 \pm 1.12 $
& $ 2.03 \pm 1.59 $ \\
\hline
\hline
$R_{J/\Psi}$  &  $1.67 \pm 0.54$
& $2.11 \pm 0.70$
& $1.83 \pm 0.43$ \\
\hline
\hline
\end{tabular}

\vs{3mm}
Table 2. \\
\end{center}
\end{table}
%
%
A direct measurement of $R_{J/\Psi}$ by CLEO II \cite{CLEO II} is consistent
with our estimate given in the last column of Table 2,

\hs{25mm} CLEO II  \hs{30mm} $R_{J/\Psi} = 1.71 \pm 0.34$ \\
In what follows we shall use the constraint $R_{J/\Psi} \leq 2.2$.

A second type of useful experimental data which turns out to be crucial is the
fractional
longitudinal polarization for the mode $B \ra K^* + J/\Psi$ measured by 3
groups :

\hs{20mm} CLEO II \cite{CLEO II}  \hs{20mm} $\rho_L = 0.80 \pm 0.08 \pm 0.05$

\hs{23mm} CDF \cite{CDF} \hs{24mm} $\rho_L = 0.66 \pm 0.10 ^{+ \hs{2mm}
0.10}_{- \hs{2mm} 0.08} $

\hs{20mm} ARGUS \cite{ARGUS} \hs{21mm} $\rho_L = 0.97 \pm 0.16 \pm 0.15$

Averaging these results by the standard weighted least-squares procedure,
we obtain :
$$ \rho_L = 0.780 \pm 0.073 $$
In what follows we shall use the one standard deviation lower limit $\rho_L
\geq 0.7 $.

\normalsize\setcounter{section}{3}
\renewcommand{\theequation}{\Roman{section}.\arabic{equation}}
\setcounter{equation}{0}
\section{ \hs{3mm} The decay modes $B \ra K(K^*) + \Psi^{'}$ }

\vs{2mm}
$1^o)$
We first consider the decay $ B \ra K^* + \Psi^{'}$ which is described
by the three hadronic form factors $A_1^{BK^*}$, $A_2^{BK^*}$
and $V^{BK^*}$ taken
at $q^2 = m_{\Psi^{'}}^2$.
In part II we have explained how the Isgur-Wise relations
due to SU(2) heavy flavour symmetry
allow one to compute the form factors in the $B$ sector
at $q^2 = t_B^{*o}$ from the values
of the form factors in the $D$ sector at $q^2 = 0$.
It turns out that the numerical value of
$t_B^{*o} = 13.4933 \hs{2mm} GeV^2$, as given by Eq.(\ref{eq:tB2}),
is remarkably close to $m_{\Psi^{'}}^2 = 13.5866$ $GeV^2$.
Different plausible choices of $m_b, m_c$
satisfying $m_b - m_c = 3.4 \pm 0.2$  $GeV$
(as dictated by HQET scheme) yield similar results,
$t_B^{*o}$ is close to $m_{\Psi^{'}}^2$.
It is then justified to neglect the variation of the form factors
between $t_B^{*o}$ and $m_{\Psi^{'}}^2$
and we obtain from Eqs.(\ref{eq:26}) and (\ref{eq:27}) :
\beqa
&& A_1^{BK^*}(m_{\Psi^{'}}^2) = 0.8056 \hs{2mm} C_{bc} \hs{2mm} A_1^{DK^*}(0)
\label{eq:72} \\
&& V^{BK^*}(m_{\Psi^{'}}^2) = 1.2413 \hs{2mm} C_{bc} \hs{2mm} V^{DK^*}(0)
\label{eq:73}
\eeqa
With the values in the $D$ sector as given
by the Particle Data Group \cite{RPR} ( PDG henceforth ) :
\beq
A_1^{DK^*}(0) = 0.56 \pm 0.04 \hs{15mm},\hs{15mm}
V^{DK^*}(0) = 1.1 \pm 0.2   \label{eq:74}
\eeq
we get :
\beqa
A_1^{BK^*}(m_{\Psi^{'}}^2) &=& (0.4511 \pm 0.0322) \hs{2mm} C_{bc}
\label{eq:75} \\
V^{BK^*}(m_{\Psi^{'}}^2) &=& (1.3654 \pm 0.2483)  \hs{2mm} C_{bc}
\label{eq:76}
\eeqa
In an analogous way, we have :
\beq
y^B(m_{\Psi^{'}}^2) = 1.5409 \hs{2mm} y^D(0)  \label{eq:77}
\eeq
and with the PDG values \cite{RPR}, $y^D(0) = 1.89 \pm 0.25$, we get
\beq
y^{'} \equiv y^B(m_{\Psi^{'}}^2) = 2.9123 \pm 0.3852  \label{eq:78}
\eeq
The values of $A_1^{BK^*}$ and $V^{BK^*}$ at $q^2 = m_{\Psi^{'}}^2$ are
simply given by the Isgur-Wise relations from the values of $A_1^{DK^*}$
and $V^{DK^*}$ at $q^2 = 0$ and the relative error is the same in both $B$ and
$D$ sectors.

For $A_2^{BK^*}$ and $x^B$ the situation is not so simple
because of the presence of the parameter $\lambda^D$,
\beqa
& &  A_2^{BK^*}(m_{\Psi^{'}}^2) =
[ 0.8121 + 0.4291 \hs{2mm} \lambda^D] \hs{2mm}
C_{bc} \hs{2mm} A_2^{DK^*}(0) \label{eq:79} \\
& & x^{'} \equiv x^B(m_{\Psi^{'}}^2) =
[ 1.0081 + 0.5328 \hs{2mm} \lambda^D ] \hs{2mm} x^D(0)  \label{eq:80}
\eeqa

Using the model introduced in Part II,
$\lambda^B$ and $\lambda^D$ are functions of the pole
mass $\Lambda_2$ of the form factor $A_2^{BK^*}$
and represented on Figure 2.
Using the PDG values \cite{RPR} :
\beq
A_2^{DK^*}(0) = 0.40 \pm 0.08 \hs{15mm}, \hs{15mm}
x^D(0) = 0.73 \pm 0.15  \label{eq:81}
\eeq
we obtain the quantities $A_2^{BK^*}(m_{\Psi^{'}}^2)$ and $x^B(m_{\Psi^{'}}^2)$
as two functions of $\Lambda_2$.
They are represented on Figure 4 for $\Lambda_2$ in the range (5 - 6) $GeV$.

\vs{2mm}
$2^o)$
The knowledge of $y^{'} = y^B(m_{{\Psi}^{'}}^2)$ determines
the left-right asymmetry $\cA_{LR}^{'}$ previously defined.
The result
\beq
\cA_{LR}^{'} = 0.9945 \pm 0.0137   \label{eq:82}
\eeq
shows that the dominant transverse amplitude has the helicity $\lambda = -1$.
In the one standard deviation limit, $\cA_{LR}^{'} > 0.98$.

As a second consequence of the knowledge of $y^{'}$,
we can derive an upper bound for the longitudinal polarization $\rho_L^{'}$ :
\beq
\rho_L^{'}  \leq \frac{ {a^{'}}^2 }
{ {a^{'}}^2 + 2 [1 + {c^{'}}^2 \hs{1mm} {y^{'}}^2 ] } \ \label{eq:83}
\eeq
In the one standard deviation limit for $y^{'}$, we obtain
\beq
\rho_L^{'} \leq 0.5664           \label{eq:84}
\eeq
which is significantly smaller than the theoretical upper bound
$\rho_L^{'} \leq {a^{'}}^2/({a^{'}}^2 + 2) = 0.678$.
These two results (\ref{eq:82}) and (\ref{eq:84}) are clearly
scenario independent and they are direct consequences of the Isgur-Wise
SU(2) heavy flavour symmetry assuming the numerical value of $y^D(0)$
as given by experiment to be correct.

\vs{2mm}
$3^o)$
Furthermore not only  the upper bound Eq.(\ref{eq:84}) but also exact values
for
 $\rho_L^{'}$ can be computed from $x^{'} = x^B(m_{\Psi^{'}}^2)$
and $y^{'} = y^B(m_{\Psi^{'}}^2)$.
It is a function of $\Lambda_2$ represented on Figure 5 for $\Lambda_2$
in the (5 $-$ 6) $GeV$ range.
The error on $\rho_L^{'}$, $\sam \rho_L^{'}$, combines, in a quadratic way,
the error on $x^{'}$ and $y^{'}$ due to those on $x^D(0)$ and $y^D(0)$.
We observe that $\rho_L^{'}$ is a slowly increasing function of $\Lambda_2$
which takes the value $\rho_L^{'} = 0.403 \pm 0.042 $
for $\Lambda_2 = 6$ $GeV$.
In Figure 5 we also represent
our one standard deviation upper bound (\ref{eq:84}).

Estimates for $\rho_L^{'}$ have been obtained by Kamal and Santra \cite{Kamal}.
However their method and results are different from ours.
Seven scenarios are considered for relating the $J/\Psi$ and $\Psi^{'}$ modes
and the allowed domains in the $x^{'}, y^{'}$ plane for each scenario are
limited by the constraint $\rho_L \geq 0.68$.
The upper bound for $\rho_L^{'}$ found in \cite{Kamal} is the theoretical upper
bound
$\rho_{L,\hs{1mm}MAX}^{'} = {a^{'}}^2/({a^{'}}^2 + 2)  = 0.678$.
Their lower bound is slightly scenario-dependent and it varies from $0.48$ to
$0.55$.
In our case, the allowed domain for $\rho_L^{'}$,
with one standard deviation, is smaller than their results and we have :
$0.286 \leq \rho_L^{'} \leq 0.445 $.

Let us emphasize again that our prediction is scenario-independent.
It is determined by using both the Isgur-Wise relation
and the experimental values of $x^D(0)$, $y^D(0)$.
It is interesting to compare the experimentally observed $\rho_L$
with the theoretical prediction for $\rho_L^{'}$  :
$$
\rho_L = 0.78 \pm 0.073  \hs{15mm},\hs{15mm}
0.286 \leq \rho_L^{'} \leq 0.445
$$

\vs{2mm}
$4^o)$
Let us consider now the decay mode $B \ra K + \Psi^{'}$
described by the hadronic form factor $F_1^{BK}(m_{\Psi^{'}}^2)$.
The Isgur-Wise relation (\ref{eq:14}) gives $F_1^{BK}(t_B^o)$
in terms of $F_1^{DK}(0)$ and  $\Lambda_F$ via the parameter $\mu^D$
\beq
F_1^{BK}(t_B^o) = [1.1779 + 0.6225 \hs{2mm} \mu^D]
\hs{2mm} C_{bc} \hs{2mm} F_1^{DK}(0)                \label{eq:85}
\eeq
We use the PDG value \cite{RPR} $F_1^{DK}(0) = 0.75 \pm 0.03$.

In order to obtain $F_1^{BK}(m_{\Psi^{'}}^2)$,
we must extrapolate $F_1^{BK}$ from $t_B^o = 14.7288$ $GeV^2$
to $m_{\Psi^{'}}^2 = 13.5866$.
For that purpose we use,  as explained in Part II,
a monopole $q^2$-dependence with a pole mass $\Lambda_F$.
The result is shown on Figure 6 with $\Lambda_F$ in the (5 $-$ 6) $GeV$ range.
We observe that  $F_1^{BK}(m_{\Psi^{'}}^2)$ is a slowly decreasing function of
$\Lambda_F$.

\vs{2mm}
$5^o)$
We finally study the ratio of rates
$$
R_{\Psi^{'}} = \frac{\Gamma(B \ra K^* + \Psi^{'})}{\Gamma(B \ra K + \Psi^{'})}
$$
Using the phase space factors given in Part III, we obtain
\beq
R_{\Psi^{'}} =  2.4455 \hs{2mm}
\frac{ (a^{'} - b^{'} x^{'})^2 + 2 [1 + {c^{'}}^2 {y^{'}}^2 ] }{ {z^{'}}^2 } \
                                             \label{eq:86}
\eeq
where the ratio of form factors, $z^{'}$, is defined by
\beq
z^{'} = z^B(m_{\Psi^{'}}^2) =
\frac{F_1^{BK}(m_{\Psi^{'}}^2)}{A_1^{BK^*}(m_{\Psi^{'}}^2)} \label{eq:87}
\eeq
The ratio $z^{'}$, like $F_1^{BK}(m_{\Psi^{'}}^2)$,
 is a function of $\Lambda_F^2$
and the ratio $x^{'}$ is a function of $\Lambda_2^2$.
It follows that $R_{\Psi^{'}}$ depends on both parameters $\Lambda_F^2$
and $\Lambda_2^2$.
Restricting $\Lambda_F$ and $\Lambda_2$ in the range (5 $-$ 6) $GeV$,
we find the ratio $R_{\Psi^{'}}$ between
 $ 1.44 \pm 0.28 $ (for  $ \Lambda_F = \Lambda_2 = 5 \hs{2mm} GeV $)
and  $ 2.92 \pm 0.54 $ (for  $ \Lambda_F = \Lambda_2 = 6 \hs{2mm} GeV $).

Both extreme values are obviously compatible with the experimental estimate
given in Part III :
\beq
( R_{{\Psi}^{'}} )_{exp} = 2.03 \pm 1.59
\eeq
The ratio of rates $R_{{\Psi}^{'}}$ is represented on Figure 7
as a function of $\Lambda_2$ and $\Lambda_F$
both in the range $(5 - 6) \hs{2mm} GeV$.

\setcounter{section}{4}
\renewcommand{\theequation}{\Roman{section}.\arabic{equation}}
\setcounter{equation}{0}
\section{ \hs{3mm} The decay modes $B \ra K(K^*) + J/\Psi$ }

\vs{2mm}
$1^o)$
We consider the decay mode $B \ra K^* + J/\Psi$ described
 by the three form factors
$A_1^{BK^*}(m_{J/\Psi}^2)$, $A_2^{BK^*}(m_{J/\Psi}^2)$
and $V^{BK^*}(m_{J/\Psi}^2)$.
These quantities are related to the values
at $t_B^{*o}$ of the same form factors
, already obtained in Part IV,
via the scenario dependent parameters
$\alpha_1$, $\alpha_2$ and $\beta$.
\beqa
A_1^{BK^*}(m_{J/\Psi}^2) &=& \alpha_1 \hs{2mm} A_1^{BK^*}(t_B^{*o}) \no \\
A_2^{BK^*}(m_{J/\Psi}^2) &=& \alpha_2 \hs{2mm} A_2^{BK^*}(t_B^{*o})
\label{eq:88} \\
V^{BK^*}(m_{J/\Psi}^2) &=& \beta \hs{2mm} V^{BK^*}(t_B^{*o})   \no
\eeqa
and for their ratios
\beqa
x \hs{2mm} \equiv \hs{2mm} x^B(m_{J/\Psi}^2)
&=& p \hs{2mm} x^B(t_B^{*o})  \label{eq:89} \\
y \hs{2mm} \equiv \hs{2mm} y^B(m_{J/\Psi}^2)
&=& q \hs{2mm} y^B(t_B^{*o})  \label{eq:90}
\eeqa
where
\beq
p = \frac{\alpha_2}{\alpha_1} \ \hs{10mm}, \hs{10mm}
q = \frac{\beta}{\alpha_1}        \label{eq:91}
\eeq
We shall consider the scenarios qualitatively described in the Part II.
For that purpose, we introduce the function $r(\Lambda)$ defined by :
\beq
r(\Lambda) = \frac{\Lambda^2 - t_B^{*o}}{\Lambda^2 - m_{J/\Psi}^2}
                                   \label{eq:92}
\eeq
and we get :
\beq
\alpha_1 = [r(\Lambda_1)]^{n_1} \hs{6mm}, \hs{6mm}
\alpha_2 = [r(\Lambda_2)]^{n_2} \hs{6mm}, \hs{6mm}
\beta = [r(\Lambda_V)]^{n_V}     \label{eq:93}
\eeq
For the algebraic integers $n_1, n_2, n_V$,
we shall consider the four cases
$n_{i} = -1, 0, +1, +2$ with $i = 1, 2, V$.
On physical grounds, we impose to the pole masses $\Lambda_1, \Lambda_2,
\Lambda_V$
to be inside a cube $5 \hs{2mm} GeV \leq \Lambda_{i} \leq 6 \hs{2mm} GeV$.

\vs{2mm}
$2^o)$
We first study the scenario constraints due to
the longitudinal polarization fraction $\rho_L$
in $B \ra K^* + J/\Psi$.
For the $4^3 = 64$ possible triplets $[n_1, n_2, n_V]$,
we  compute $\rho_L$ using the values of
$x^B(t_B^{*o})$ and $y^B(t_B^{*o})$ obtained in Eqs.(\ref{eq:80})
and (\ref{eq:78}) and restricting the pole mass parameters
$\Lambda_1$, $\Lambda_2$ and $\Lambda_V$
inside the cube
$5 \hs{2mm} GeV \leq \Lambda_{i} \leq 6 \hs{2mm} GeV$.
We impose the experimental constraint in the form
$\rho_L + \sam \rho_L \geq 0.7$ where the error $\sam \rho_L$
is computed in quadrature [See Appendix A-1] from the errors on $x^{'}$
and $y^{'}$ due to the experimental errors of $x^D(0)$ and $y^D(0)$.
After numerically scaning over the whole $n_{i}$ and $\Lambda_{i}$ spaces,
finally our results can be summarized in the following :

\ben
\item
No solution is obtained when $n_1 = 0, 1, 2$ for the 16 values of the couple
$[n_2, n_V]$.

\item
Solutions exist only when $n_1 = -1$, i.e., when the form factor $A_1$ exhibits
a linear decrease with $q^2$.
Of course, in this case, $\Lambda_1$ is no more a pole mass but simply a slope
coefficient
and it is reasonable now to relax the constraint $\Lambda_1 \leq 6 \hs{2mm}
GeV$ and
to use only $\Lambda_1 \geq 5 \hs{2mm} GeV$ in order to exclude a too fast
variation with
$q^2$ of $A_1$.

\item
We obtain solutions for only 4 triplets $[n_1, n_2, n_V]$ :
$$
[-1, 2, 2] \hs{3mm};\hs{3mm} [-1, 1, 2]
\hs{3mm};\hs{3mm} [-1, 0, 2]
\hs{3mm};\hs{3mm} [-1, 2, 1]
$$
The allowed domains in the $\Lambda_1$, $\Lambda_2$, and $\Lambda_V$ space
are respectively represented on Figures 8, 9, 10 and 11.

\item
In the four surviving triplets mentioned above, the maximal value of $\rho_L$
occurs at $\Lambda_1 = 5 \hs{2mm} GeV$, $\Lambda_2 = 6 \hs{2mm} GeV$,
$\Lambda_V = 5 \hs{2mm} GeV$  and in the most favorable situation of two dipole
$q^2$ dependence for $A_2$ and $V$,
we  obtain $\rho_L = 0.7162 \pm 0.0236$.
Therefore $\rho_L = 0.74$ is the maximal value, within one standard deviation,
we can get in our approach, considering only the quantity $\rho_L$.

\item
It is interesting to notice that in the case of two monopole $q^2$-dependence
for $A_2$ and $V$, the maximal value of $\rho_L$ is obtained at the point
$\Lambda_1 = \Lambda_V = 5 \hs{2mm} GeV$, $\Lambda_2 = 6 \hs{2mm} GeV$
with the result $\rho_L = 0.6635 \pm 0.0339 $,
e.g., a one standard deviation value very close to $0.7$.
However we consider this case as only marginal,
since, in any way, it will be eliminated
when the second quantity $R_{J/\Psi}$ enters in the fit.
\een

\vs{2mm}
$3^o)$
The second quantity is the ratio of rates $R_{J/\Psi}$ which has the form :
\beq
R_{J/\Psi} = 1.0793 \hs{2mm}
\frac{(a - b \hs{2mm}x)^2 + 2 ( 1 + c^2 \hs{2mm} y^2)}{z^2}   \label{eq:94}
\eeq
where the ratio $z$ is defined by :
\beq
z \equiv z^B(m_{J/\Psi}^2) =
\frac{F_1^{BK}(m_{J/\Psi}^2)}{A_1^{BK^*}(m_{J/\Psi}^2)} \label{eq:95}
\eeq
We impose the experimental constraint under the form
$R_{J/\Psi} - \sam R_{J/\Psi} \leq 2.2$
where the theoretical error $\sam R_{J/\Psi}$ is computed,
in quadrature [See Appendix A-3], from the
experimental errors on $x^D(0)$, $y^D(0)$ and $z^D(0)$.

Our constraint $\rho_L + \sam \rho_L \geq 0.7$ has selected four scenarios
and the allowed domain at fixed $\Lambda_V, \Lambda_2$ is defined by :
\beq
\Lambda_1 \leq \Lambda_{1,\hs{1mm}MAX}(\Lambda_V, \Lambda_2) \label{eq:96}
\eeq
These domains have been represented on Figures 8  to 11 for
$5 \hs{2mm} GeV \leq  \Lambda_V, \Lambda_2 \leq 6 \hs{2mm} GeV$.

On the other hand, the constraint on $R_{J/\Psi}$ implies
a lower limit for $\Lambda_1$
\beq
\Lambda_1 \geq \Lambda_{1,\hs{1mm}MIN}(\Lambda_V, \Lambda_2, \Lambda_F)
\label{eq:97}
\eeq
Physical values of $\Lambda_1$ exist when and only when the lower limit
(\ref{eq:97})
is smaller than the upper limit (\ref{eq:96}).

At fixed $\Lambda_V, \Lambda_2$,
the quantity $\Lambda_{1,\hs{1mm}MIN}$ is
an increasing function of $\Lambda_F$ and we shall restrict
$\Lambda_F \geq 5 \hs{2mm} GeV$.
Therefore at fixed $\Lambda_V, \Lambda_2$, the physical domain for $\Lambda_1$
is
defined by
\beq
\Lambda_{1,\hs{1mm}MIN}(\Lambda_V, \Lambda_2, \Lambda_F = 5 \hs{2mm} GeV)
\leq \Lambda_1 \leq  \Lambda_{1,\hs{1mm}MAX}(\Lambda_V, \Lambda_2)
\label{eq:98}
\eeq

After a numerical scanning, we do not obtain solution satifying the constraint
$R_{J/\Psi} - \sam R_{J/\Psi} \leq 2.2$ with the scenario $[n_1, n_2, n_V] =
[-1, 2, 1]$.
For the three remaining scenarios
$[n_1, n_2, n_V] = [-1, 2, 2], [-1, 1, 2]$ and $[-1, 0, 2]$,
the physical regions are then represented on Figures 12, 13 and 14.

At fixed $\Lambda_V, \Lambda_2$, we have
\beq
5 \hs{2mm} GeV \leq \Lambda_F(\Lambda_V, \Lambda_2)
\leq \Lambda_{F,\hs{1mm}MAX}(\Lambda_V, \Lambda_2) \label{eq:99}
\eeq
where $\Lambda_{F,\hs{1mm}MAX}$ is determined by
\beq
\Lambda_{1,\hs{1mm}MIN} [ \Lambda_V, \Lambda_2,
\Lambda_{F,\hs{1mm}MAX}(\Lambda_V, \Lambda_2) ]
= \Lambda_{1,\hs{1mm}MAX}(\Lambda_V, \Lambda_2)       \label{eq:100}
\eeq
The quantity $\Lambda_{F,\hs{1mm}MAX}(\Lambda_V, \Lambda_2)$ is represented
on Figures 15, 16 and 17
for the three surviving scenarios.

\vs{2mm}
$4^o)$
Comment and Illustration

Starting with 64 scenarios for the  $q^2$-dependence
of the hadronic form factors
$A_1, A_2, V$, $[n_1, n_2, n_V]$ with $n_{i} = -1, 0, 1, 2$,
only survive 3 scenarios $[-1, n_2, 2]$ with $n_2 = 0, 1, 2$
for which it is possible
to find a non empty domain in the parameter space
$\Lambda_1, \Lambda_2, \Lambda_V, \Lambda_F$ such that
both experimental constraints $\rho_L + \sam \rho_L \geq 0.7$ and
$R_{J/\Psi} - \sam R_{J/\Psi} \leq 2.2$ are simultaneously satisfied.
The hadronic form factors $F_1^{BK}$ has been chosen with a monopole
$q^2$-dependence, $n_F = 1$,
consistent with data in the $D$ sector
and the condition $F_0^{BK}$ constant determines the parameter $\mu^B$
$-$ and then $\mu^D$ $-$ as a function of $\Lambda_F$.
A similar relation determines $\lambda^B$ $-$ and then $\lambda^D$ $-$
as a function of $\Lambda_2$.

\vs{3mm}
\begin{table}[thb]
\begin{center}
\begin{tabular}{||c||c|c|c|c||}
\hline
$\Lambda_1(GeV)$ &
$\rho_L$  &
$\rho_L + \sam \rho_L$ &
$ R_{J/\Psi} $  &
$ R_{J/\Psi} - \sam R_{J/\Psi} $ \\
\hline
\hline
\multicolumn{5}{||c||}{$n_2 = 2$ } \\
\hline
$\Lambda_{1, \hs{1mm}MAX} = 8.112 $ &
$ 0.665 \pm 0.035 $ &
$ 0.700 $ &
$2.089 \pm 0.508$ &
$1.581$ \\
\hline
$\Lambda_{1} = 6.810 $ &
$ 0.675 \pm 0.033 $ &
$ 0.708 $ &
$2.263 \pm 0.422$ &
$1.841$ \\
\hline
$\Lambda_{1, \hs{1mm}MIN} = 5.426 $ &
$ 0.694 \pm 0.028 $ &
$ 0.722 $ &
$2.694 \pm 0.494$ &
$2.200$ \\
\hline
\hline
\multicolumn{5}{||c||}{$n_2 = 1$ } \\
\hline
$\Lambda_{1, \hs{1mm}MAX} = 6.113 $ &
$ 0.663 \pm 0.037 $ &
$ 0.700 $ &
$2.324 \pm 0.524$ &
$1.800$ \\
\hline
$\Lambda_{1} = 5.770 $ &
$ 0.671 \pm 0.035 $ &
$ 0.705 $ &
$2.482 \pm 0.475$ &
$2.007$ \\
\hline
$\Lambda_{1, \hs{1mm}MIN} = 5.426 $ &
$ 0.681 \pm 0.033 $ &
$ 0.714 $ &
$2.714 \pm 0.514$ &
$2.200$ \\
\hline
\hline
\multicolumn{5}{||c||}{$n_2 = 0$ } \\
\hline
$\Lambda_{1, \hs{1mm}MAX} = 5.292 $ &
$ 0.660 \pm 0.040 $ &
$ 0.700 $ &
$2.625 \pm 0.542$ &
$2.083$ \\
\hline
$\Lambda_{1} = 5.237 $ &
$ 0.663 \pm 0.039 $ &
$ 0.702 $ &
$2.680 \pm 0.529$ &
$2.151$ \\
\hline
$\Lambda_{1, \hs{1mm}MIN} = 5.183 $ &
$ 0.665 \pm 0.038 $ &
$ 0.703 $ &
$2.739 \pm 0.539$ &
$2.200$ \\
\hline
\hline
\end{tabular}

\vs{3mm}
Table 3. \\
\end{center}
\end{table}

The allowed domains for $\Lambda_1, \Lambda_2, \Lambda_V$ and $\Lambda_F$
have been represented in 3 dimensional plots on Figures 12 to 17.
However it might be useful to produce some numerical values obtained
for $\rho_L$ and $R_{J/\Psi}$ in these domains and for that purpose.
we have choosen, as an illustration, $\Lambda_2 = 6 \hs{2mm} GeV$,
$\Lambda_V = \Lambda_F = 5 \hs{2mm} GeV$, and for $\Lambda_1$,
three values, $\Lambda_{1,\hs{1mm}MAX}$, $\Lambda_{1,\hs{1mm}MIN}$
and an intermediate value between the extremes.
The results are represented in Table 3.
A glance at Table 3 shows how difficult it is to fit
simultaneously the large $\rho_L$ and the relatively small
$R_{J/\Psi}$, their opposite trends making the fit so difficult
have been equally noticed \cite{Orsay}.

The relative error on $R_{J/\Psi}$ is larger than the one of $\rho_L$,
and this feature is very useful for obtaining a fit.
It is essentially due to the fact that $R_{J/\Psi}$, in addition to
the errors on $x^D(0)$ and $y^D(0)$
that also enter in $\sam \rho_L$,
has an uncertainty on $z^D(0)$ which is important.
While the relative error on $\rho_L$ is between $4 \%$ and $6 \%$,
the one of $R_{J/\Psi}$ is between $ 18 \%$ and $ 24\%$.

{}From these numerical results and those illustrated on Figures 12 to 17,
it is clear that the scenario with a dipole form factor $A_2$ is the one with
the largest phase space in $\Lambda_1, \Lambda_2, \Lambda_V, \Lambda_F$.
In this case it is relatively easy to accommodate
both $\rho_L$ and $R_{J/\Psi}$.
We notice that the largest value of $\rho_L$ we can obtain in this model is
$\rho_L = 0.694 \pm 0.028$ and for $R_{J/\Psi}$ the smallest value is
$2.089 \pm 0.508$.

For the scenario with a monopole form factor $A_2$, the situation is less
confortable
even if it is possible to fit data on $\rho_L$ and $R_{J/\Psi}$.
Now the quantity $\rho_L$  varies by less than $3 \%$ and $R_{J/\Psi}$ by about
$17 \%$.

In the third case of a constant form factor $A_2$,
the allowed domain for the parameters $\Lambda_1, \Lambda_2, \Lambda_V,
\Lambda_F$
is very small and this possiblity, even if both $\rho_L$ and $R_{J/\Psi}$
satisfy the constraints, appears to be very marginal, the values of
 $\rho_L$ and $R_{J/\Psi}$ being at the limits of the constraints.

\vs{2mm}
$5^o)$
The left-right asymmetry $\cA_{LR}$ for transverse polarization
has not been experimentally measured.
It has been defined in Eq.(\ref{eq:69})
and it depends only on the ratio $y = y^B(m_{J/\Psi}^2)$
unambiguously related in our model to $y^D(0)$.
For our three selected models, we have
\beq
y^B(m_{J/\Psi}^2) = \frac{(\Lambda_V^2 - t_B^{*o})^2 \hs{2mm}(\Lambda_1^2 -
t_B^{*o})}
{(\Lambda_V^2 - m_{J/\Psi}^2)^2 \hs{2mm}(\Lambda_1^2 - m_{J/\Psi}^2 )} \hs{2mm}
\frac{m_c}{m_b} \hs{2mm}
\left( \frac{m_B + m_{K^*}}{m_D + m_{K^*}} \right)^2
\hs{2mm} y^D(0)                                     \label{eq:101}
\eeq
The difference between the scenarios $n_2 = 2, 1, 0$ is an allowed domain
for $\Lambda_1$, $\Lambda_2$
and $ \Lambda_V $ illustrated on Figures 12, 13 and 14.
Scanning inside these domains, we make the predictions
\beqa
n_2 = 2 : & & \hs{20mm} 0.867 < \hs{2mm} \cA_{LR}  \hs{2mm} < 0.945
\label{eq:102} \\
n_2 = 1 : & & \hs{20mm} 0.837 < \hs{2mm} \cA_{LR}  \hs{2mm} < 0.910
\label{eq:103} \\
n_2 = 0 : & & \hs{20mm} 0.837 < \hs{2mm} \cA_{LR}  \hs{2mm} < 0.856
\label{eq:104}
\eeqa
The left-right asymmetry is large in the three selected cases,
not as large as in the $K^* + \Psi^{'}$ case where it is close to one.
We observe that the differences between the three scenarios are moderate.

\setcounter{section}{5}
\renewcommand{\theequation}{\Roman{section}.\arabic{equation}}
\setcounter{equation}{0}
\section{ \hs{3mm} The decay modes $B \ra K(K^*) + \eta_c$ }

\vs{2mm}
$1^o)$
We start by considering the decay mode $B \ra K + \eta_c$
governed by the form factor $F_0^{BK}(m_{\eta_c}^2)$.
The Isgur-Wise relations determine $F_0^{BK}(t_B^o)$
from $F_0^{DK}(0)$ in a $\mu^D$ dependent way :
\beq
F_0^{BK}(t_B^o) = C_{bc} \hs{2mm}
[ 0.8460 - 0.0057 \hs{2mm} \mu^D ] \hs{2mm} F_0^{DK}(0) \label{eq:105}
\eeq
We notice that the coefficient of $\mu^D$ is very small as compared with
the $\mu^D$-independent term in the bracket of Eq.(\ref{eq:105}).
It follows that $F_0^{BK}(t_B^o)$ only weakly depends on $\mu^D$.

In Part II, we have choosen a monopole form factor for $F_1^{BK}(q^2)$
and the parameter $\mu^B$ has been related to the pole mass $\Lambda_F$
in such a way to obtain a form factor  $F_0^{BK}$
independent on $q^2$.
{}From the previous considerations, the constant value of $F_0^{BK}$
will be a weakly dependent function of $\Lambda_F$.
With $\Lambda_F$ in the range (5 - 6) $GeV$, $F_0^{BK}$ increases slowly from
$0.6286 \pm 0.0251 $
(for $ \Lambda_F = 5 \hs{2mm}GeV $)
to $0.6327 \pm 0.0253$
(for $ \Lambda_F = 6 \hs{2mm}GeV $).
The quantity $F_1^{BK}(0) = F_0^{BK}(0)$ has been represented on Figure 6.

\vs{2mm}
$2^o)$
We now consider the decay mode $B \ra K^* + \eta_c$
which is described by the hadronic form factors $A_0^{BK^*}(m_{\eta_c}^2)$.
The Isgur-Wise relation (\ref{eq:44}) gives $A_0^{BK^*}$
in terms of $A_1^{BK^*}$, $A_2^{BK^*}$ and the parameter $\lambda^B$.
With the choice made in Part II for  $\lambda^B$
\beq
 \lambda^B = \frac{m_B^2 - m_{K^*}^2}{\Lambda_2^2}  \label{eq:108}
\eeq
we get :
\beq
A_0^{BK^*}(m_{\eta_c}^2) = 3.4602 \hs{2mm} A_1^{BK^*}(m_{\eta_c}^2)
- 2.4602 \hs{2mm} \left[ 1 - \frac{m_{\eta_c}^2}{\Lambda_2^2} \right]
\hs{2mm} A_2^{BK^*}(m_{\eta_c}^2)      \label{eq:109}
\eeq

In Part V, we have obtained constraints on the scenarios allowed
by the requirement to fit $\rho_L$ and $R_{J/\Psi}$.
They correspond to $n_1 = -1$ and $n_2 = 2, 1, 0$.

We introduce the function $S(\Lambda)$ in order to relate
the values of the hadronic form factors at  $q^2 = m^2_{\eta_c}$
to the values at $q^2 = t_B^{*o}$
where they are known from part II :
\beq
S(\Lambda) = \frac{\Lambda^2 - t_B^{*o}}{\Lambda^2 - m_{\eta_c}^2}
\label{eq:110}
\eeq
and we get :
\beq
A_0^{BK^*}(m_{\eta_c}^2) =
\frac{3.4602}{S(\Lambda_1)} \hs{2mm} A_1^{BK^*}(t_B^{*o})
- 2.4602 \hs{2mm} \left[ 1 - \frac{m_{\eta_c}^2}{\Lambda_2^2} \right]
\hs{2mm} [S(\Lambda_2)]^{n_2}
\hs{2mm} A_2^{BK^*}(t_B^{*o})      \label{eq:111}
\eeq

The form factor value $A_0^{BK^*}(m_{\eta_c}^2)$ is scenario-dependent,
firstly by the value of $n_2$ ; $n_2 = 2, 1, 0$,
secondly by the values of the pole masses $\Lambda_1$ and $\Lambda_2$
 in the restricted domains described in Part V.
We vary $\Lambda_1$ and $\Lambda_2$ inside these domains and obtain :
\beqa
 n_2 = 2 : \hs{15mm} & &  A_0^{BK^*}(m^2_{\eta_c})
= 1.3768 \pm 0.2928 \label{eq:112} \\
 n_2 = 1 : \hs{15mm} & &  A_0^{BK^*}(m^2_{\eta_c})
= 1.3979 \pm 0.2677 \label{eq:113} \\
 n_2 = 0 : \hs{15mm} & &  A_0^{BK^*}(m^2_{\eta_c})
= 1.3611 \pm 0.1790 \label{eq:114}
\eeqa
We observe that the results for $A_0^{BK^*}(m_{\eta_c}^2)$
is only weakly scenario-dependent due to cancellations between the two terms of
Eq.(\ref{eq:111})

\vs{2mm}
$3^o)$
The ratio $R_{\eta_c}$ of these two decay modes
\beq
R_{\eta_c} = \frac{\Gamma(B \ra K^* + \eta_c)}{\Gamma(B \ra K + \eta_c)}
\label{eq:115}
\eeq
is given by :
\beq
R_{\eta_c} = 0.3732 \hs{2mm} \left|
\frac{A_0^{BK^*}(m_{\eta_c}^2)}{F_0^{BK}(m_{\eta_c}^2)}
\right|^2  \label{eq:116}
\eeq
Using the values (\ref{eq:112}), (\ref{eq:113}), (\ref{eq:114})
for $A_0^{BK^*}(m_{\eta_c}^2)$,
and  Eq.(\ref{eq:105}) for $F_0^{BK}(m_{\eta_c}^2)$, we obtain :
\beqa
n_2 = 2 : \hs{15mm} & & R_{\eta_c} = 1.7903 \pm 0.7748 \label{eq:117} \\
n_2 = 1 : \hs{15mm} & & R_{\eta_c} = 1.8456 \pm 0.7221 \label{eq:118} \\
n_2 = 0 : \hs{15mm} & & R_{\eta_c} = 1.7497 \pm 0.4810 \label{eq:119}
\eeqa
Taking these results all together, we obtain the  one standard deviation bounds
\beq
1.02 \leq \hs{2mm} R_{\eta_c} \hs{2mm} \leq 2.56 \label{eq:120}
\eeq
We comment, in the next Part VII, on the difference between these results and
the bounds on  $R_{\eta_c}$ previously obtained in Ref.\cite{GKP1}.

\setcounter{section}{6}
\renewcommand{\theequation}{\Roman{section}.\arabic{equation}}
\setcounter{equation}{0}
\section{ \hs{3mm} Comparison of Different Charmonium States Production }

$1^o)$
Ratio of decay widths with the same strange meson, $K$ or $K^*$, and
different charmonium states are interesting quantities involving
the leptonic decay constant $f_{\ol{c}c}$.
We define four such ratios
referred to the most accurately measured $J/\Psi$ production :
\beq
S = \frac{\Gamma(B \ra K + \Psi^{'})}{\Gamma(B \ra K + J/\Psi)}
\hs{10mm}, \hs{15mm}
S^* = \frac{\Gamma(B \ra K^* + \Psi^{'})}{\Gamma(B \ra K^* + J/\Psi)}
                                                           \label{eq:121}
\eeq
\beq
T = \frac{\Gamma(B \ra K + \eta_c)}{\Gamma(B \ra K + J/\Psi)}
\hs{10mm}, \hs{15mm}
T^* = \frac{\Gamma(B \ra K^* + \eta_c)}{\Gamma(B \ra K^* + J/\Psi)}
                                                            \label{eq:122}
\eeq
Assuming factorization and using the phase space estimates given in Part III,
we obtain :
\beq
S = 0.4438 \hs{2mm} \left( \frac{f_{\Psi^{'}}}{f_{J/\Psi}} \right)^2 \hs{2mm}
\left| \frac{F_1^{BK}(m_{\Psi^{'}}^2)}{F_1^{BK}(m_{J/\Psi}^2)} \right|^2
                                                             \label{eq:123}
\eeq
\beq
S^* = 1.0057 \hs{2mm} \left( \frac{f_{\Psi^{'}}}{f_{J/\Psi}} \right)^2 \hs{2mm}
\left| \frac{A_1^{BK^*}(m_{\Psi^{'}}^2)}{A_1^{BK^*}(m_{J/\Psi}^2)} \right|^2
\hs{2mm}
\frac{(a^{'} - b^{'}x^{'})^2 + 2 (1 + {c^{'}}^2{y^{'}}^2)}
{(a - b x)^2 + 2 ( 1 + c^2 y^2)} \
                                                             \label{eq:124}
\eeq
\beq
T = 2.5180 \hs{2mm} \left( \frac{f_{\eta_c}}{f_{J/\Psi}} \right)^2 \hs{2mm}
\left| \frac{F_0^{BK}(m_{\eta_c}^2)}{F_1^{BK}(m_{J/\Psi}^2)} \right|^2
                                                             \label{eq:125}
\eeq
\beq
T^* = 0.8706 \hs{2mm} \left( \frac{f_{\eta_c}}{f_{J/\Psi}} \right)^2 \hs{2mm}
\left| \frac{A_0^{BK^*}(m_{\Psi^{'}}^2)}{A_1^{BK^*}(m_{J/\Psi}^2)} \right|^2
\hs{2mm}
\frac{1}
{(a - b x)^2 + 2 ( 1 + c^2 y^2)} \                           \label{eq:126}
\eeq
The $\eta_c$ modes have not been experimentally observed
and the only available information refers to $J/\Psi$ and
$\Psi^{'}$ modes.
Using the PDG data \cite{RPR} collected in Table 1 of Part III,
we obtain for $S$ and $S^*$  values given in Table 4 :

\vs{3mm}
\begin{table}[thb]
\begin{center}
\begin{tabular}{|c||c||c||c||}
\hline
Ratio
& $B^{+}$
& $B^{o}$
& $B^{+}, B^o$ combined  \\
\hline
\hline
$S$  &  $0.68 \pm 0.32$
& $ < \hs{2mm} 1.07 \pm 0.30$
& $0.68 \pm 0.32$ \\
\hline
\hline
$S^*$  &  $< \hs{2mm} 1.76 \pm 0.52$
& $ 0.89 \pm 0.59$
& $0.89 \pm 0.59$ \\
\hline
\hline
\end{tabular}

\vs{3mm}
Table 4 \\
\end{center}
\end{table}
%
%

\vs{2mm}
$2^o)$
Using the values \cite{Neubert} of
$f_{J/\Psi} = (384 \pm 14) \hs{2mm} MeV$ and
$f_{\Psi^{'}} = (282 \pm 14) \hs{2mm} MeV$
as estimated from the decays $J/\Psi \ra e^{+}e^{-}$ and $\Psi^{'} \ra
e^{+}e^{-}$,
we obtain :
\beq
\left( \frac{f_{\Psi^{'}}}{f_{J/\Psi}} \right)^2 = 0.539 \pm 0.066
                                                          \label{eq:127}
\eeq
and the quantity $S$ is written :
\beq
S = [0.2392 \pm 0.0292] \hs{2mm}
\left| \frac{F_1^{BK}(m_{\Psi^{'}}^2)}{F_1^{BK}(m_{J/\Psi}^2)} \right|^2
                                                          \label{eq:128}
\eeq
In our model, the hadronic form factor $F_1^{BK}(q^2)$
has a monopole $q^2$ dependence
with a pole mass $\Lambda_F$ and we simply have :
\beq
\left| \frac{F_1^{BK}(m_{\Psi^{'}}^2)}{F_1^{BK}(m_{J/\Psi}^2)} \right|^2
= \left| \frac{\Lambda_F^2 - m_{J/\Psi}^2}
{\Lambda_F^2 - m_{\Psi^{'}}^2} \right|^2    \label{eq:130}
\eeq
This ratio of form factors is a decreasing function of $\Lambda_F^2$
and so is the ratio $S$.
At $\Lambda_F = 5 \hs{2mm} GeV$,
the prediction for $S$ is :
\beq
S(\Lambda_F = 5 \hs{2mm} GeV) = 0.4363 \pm 0.0537         \label{eq:131}
\eeq
This prediction is in agreement, within one standard deviation,
 with the experimental value
estimated in Table 2, $S_{exp} = 0.68 \pm 0.32$.
Such an agreement continues to occur for larger values of $\Lambda_F$ up to
$6.27$ $GeV$.

The range of $\Lambda_F$ depends on the three scenarios corresponding to $n_2 =
2, 1, 0$
and they are deduced from Figures 15, 16, 17 respectively.
We get :
\beqa
n_2 = 2 : \hs{10mm} & & 0.4363 \pm 0.0537 \hs{2mm} \geq \hs{2mm}
S \hs{2mm} \geq \hs{2mm} 0.3505 \pm 0.0432   \label{eq:132} \\
n_2 = 1 : \hs{10mm} & & 0.4363 \pm 0.0537 \hs{2mm} \geq \hs{2mm}
S \hs{2mm} \geq \hs{2mm} 0.3790 \pm 0.0467   \label{eq:133} \\
n_2 = 0 : \hs{10mm} & & 0.4363 \pm 0.0537 \hs{2mm} \geq \hs{2mm}
S \hs{2mm} \geq \hs{2mm} 0.4181 \pm 0.0515   \label{eq:134}
\eeqa
The errors quoted in Eq.(\ref{eq:132}), (\ref{eq:133}) and (\ref{eq:134})
are due to the uncertainty on the leptonic decay constants
$f_{\Psi^{'}}$ and $f_{J/\Psi}$.
In conclusion, the theoretical predictions of our model
for the three scenarios
agree with experimental results within one standard deviation.

\vs{2mm}
$3^o)$
The analysis of the second ratio $S^*$ is more complex because of
a large number of form factors involved.
Using Eq.(\ref{eq:127}), we get :
\beq
S^* = [0.5424 \pm 0.0664] \hs{2mm}
\left| \frac{A_1^{BK^*}(m_{\Psi^{'}}^2)}{A_1^{BK^*}(m_{J/\Psi}^2)}
 \right|^2  \hs{2mm}
\frac{(a^{'} - b^{'}x^{'})^2 + 2 (1 + {c^{'}}^2{y^{'}}^2)}
{(a - b x)^2 + 2 ( 1 + c^2 y^2)} \                        \label{eq:129}
\eeq
In our model the form factor $A_1^{BK^*}(q^2)$ is linearly decreasing
with a slope $\Lambda_1$, and we simply have
\beq
\frac{A_1^{BK^*}(m_{\Psi^{'}}^2)}{A_1^{BK^*}(m_{J\Psi}^2)}
= \frac{\Lambda_1^2 - m_{\Psi^{'}}^2}{\Lambda_1^2 - m_{J/\Psi}^2}
\label{eq:129a}
\eeq
We have computed the ratio $S^*$ for the three scenarios $n_2 = 2, 1, 0$
using the values of $\Lambda_1$, $\Lambda_2$ and $\Lambda_V$ inside
the allowed domains obtained in Part V and
represented respectively on Figures 12, 13 and 14.

The results of this scanning are :
\beqa
n_2 = 2 : \hs{15mm} & & 0.3287 \pm 0.0028 \hs{2mm} \leq \hs{2mm}
S^* \hs{2mm} \leq \hs{2mm} 0.4135 \pm 0.0038   \label{eq:135}  \\
n_2 = 1 : \hs{15mm} & & 0.3489 \pm 0.0034 \hs{2mm} \leq \hs{2mm}
S^* \hs{2mm} \leq \hs{2mm} 0.4015 \pm 0.0039   \label{eq:136}  \\
n_2 = 0 : \hs{15mm} & & 0.3763 \pm 0.0039 \hs{2mm} \leq \hs{2mm}
S^* \hs{2mm} \leq \hs{2mm} 0.3867 \pm 0.0040   \label{eq:137}
\eeqa
The errors quoted in Eqs.(\ref{eq:135}), (\ref{eq:136}) and (\ref{eq:137}) are
computed in  quadrature from those on the ratios
$f_{\Psi^{'}}/f_{J/\Psi}$, $x^D(0)$ and $y^D(0)$.
The theoretical predictions of our model for the three scenarios agree,
 within one standard deviation,
 with the experimental results estimated in Table 2 :
$S_{exp}^* = 0.89 \pm 0.59$.

\vs{2mm}
$4^o)$
Kamal and Santra \cite{Kamal} have studied the ratios $S$ and $S^*$
denoted by them respectively as $1/R$ and $1/R^{'}$.
In the case of $R$, both monopole and dipole $q^2$ dependences for $F_1^{BK}$
are considered with a pole mass $\Lambda_F = 5.43 \hs{2mm} GeV$.
 Their conclusion is that a dipole behaviour for $F_1^{BK}$ is
needed in order to obtain an agreement for $R$ between theory
and experiment in the one standard deviation limit.

The apparent contradiction between our result
( monopole for $F_1^{BK}$ ) and the one of Ref.\cite{Kamal}
is essentially due to the large experimental error of $47 \%$
for the quantity $S$ or $R$.
With $\delta = 0.47$ the relation at first order in $\delta$,
$(1 \pm \delta)^{-1} = 1 \mp \delta$ is not valid
and one standard deviation limit
for $S$ and one standard deviation limit for $R$ are different concepts.
However, since the main part of the experimental error is
due to the $K + \Psi^{'}$
mode and for that reason the consideration of one standard deviation for $S$
( where $K + \Psi^{'}$ enters in the numerator )
seems to be more relevant than for $R$.

A similar situation occurs for $S^*$ and $R^{'}$.
Here the experimental error is even larger, $66.7 \%$, and it is mainly due to
the $K^* + \Psi^{'}$ mode which enters in the numerator of $S^*$.
Again the one standard deviation limit for $S^*$ and
the one standard deviation limit for $R^{'}$ are different quantities.

Also the pole masses in Ref.\cite{Kamal} are taken only at some fixed values,
while in our approach these poles sweep inside the allowed domains of
Figs. 12 $-$ 17.

For the ratio $R^{'}$ as previously done for $\rho_L^{'}$, they propose seven
scenarios.
Furthermore, considering only in the one standard deviation limit for $R^{'}$,
they exclude four scenarios where $A_1^{BK^*}$ is either constant
or linearly decreasing with $q^2$ and conclude that
  if factorization assumption were to be held, then the only scenarios that
are consistent with experiment are those in which $A_1^{BK^*}$ rises
with $q^2$. We observe however that $R^{'}$ (or $S^*$) is not an independent
ratio
but related to the other ratios by $S^* R_{J/\Psi} = S R_{{\Psi}^{'}}$,
such that considering  $R^{'}$ (or $S^*$) alone might be inadequate.

\vs{2mm}
$5^o)$
Comparing the $K + \eta_c$ and $K + J/\Psi$ decay modes,
we now consider the ratio T
which depends on the ratio of the decay constants $f_{\eta_c}$ and
$f_{J/\Psi}$.
Unfortunately $f_{\eta_c}$ is not experimentally known
and we use theoretical estimates if we want to make predictions.

Rewriting Eq(\ref{eq:125}) in the form :
\beq
T = 2.5180 \hs{2mm} \left( \frac{f_{\eta_c}}{f_{J/\Psi}} \right)^2 \hs{2mm}
S_V( \Lambda_F )
\label{eq:139}
\eeq
where
\beq
S_V(\Lambda_F) = \left| \frac{F_0^{BK}(m_{\eta_c}^2)}
{F_1^{BK}(m_{J/\Psi}^2)}
\right|^2                    \label{eq:SV}
\eeq
We compute $S_V$ in our model
where $F_0^{BK}$ is constant and $F_1^{BK}$ has a monopole $q^2$ dependence
with the pole mass $\Lambda_F$.
As a consequence, we simply have
\beq
S_V(\Lambda_F)
= \left( 1 - \frac{m_{J/\Psi}^2}{\Lambda_F^2} \right)^2      \label{eq:140}
\eeq
The function $S_V$ is an increasing function of $\Lambda_F$.
The allowed values of $\Lambda_F$ have been discussed in
Part V (Figs. 15 $-$ 17), and we obtain :
\beqa
n_2 = 2 : & & \hs{2mm}
5 \hs{2mm} GeV \hs{1mm} \leq \hs{1mm} \Lambda_F \hs{1mm} \leq \hs{1mm}
5.71 \hs{2mm} GeV \ \hs{6mm}, \hs{6mm}
0.38 \hs{1mm} \leq \hs{1mm} S_V \hs{1mm} \leq \hs{1mm} 0.50 \label{eq:141} \\
n_2 = 1 : & & \hs{2mm}
5 \hs{2mm} GeV \hs{1mm} \leq \hs{1mm} \Lambda_F \hs{1mm} \leq \hs{1mm}
5.39 \hs{2mm} GeV \ \hs{6mm}, \hs{6mm}
0.38 \hs{1mm} \leq \hs{1mm} S_V \hs{1mm} \leq \hs{1mm} 0.45 \label{eq:142} \\
n_2 = 0 : & & \hs{2mm}
5 \hs{2mm} GeV \hs{1mm} \leq \hs{1mm} \Lambda_F \hs{1mm} \leq \hs{1mm}
5.10 \hs{2mm} GeV \ \hs{6mm}, \hs{6mm}
0.38 \hs{1mm} \leq \hs{1mm} S_V \hs{1mm} \leq \hs{1mm} 0.40 \label{eq:143}
\eeqa
As pointed out in Ref.\cite{Desh},
a measurement of the ratio $T$ will provide an opportunity to extract
the scalar decay constant $f_{\eta_c}$
from experiment.
Unfortunately the decay mode $B \ra K + \eta_c$ has not been
experimentaly observed.

There exist various theoretical ways to estimate both
$f_{\eta_c}$ and $f_{J/\Psi}$.
Using the estimate quoted in Ref.\cite{Desh},
\beq
\frac{f_{\eta_c}}{f_{J/\Psi}} = 0.993  \label{eq:144}
\eeq
we make predictions for the ratio $T$ :
\beqa
n_2 = 2 : \hs{15mm}
0.94 \hs{2mm} \leq \hs{2mm} T \hs{2mm} \leq \hs{2mm} 1.24  \label{eq:145} \\
n_2 = 1 : \hs{15mm}
0.94 \hs{2mm} \leq \hs{2mm} T \hs{2mm} \leq \hs{2mm} 1.12  \label{eq:146} \\
n_2 = 0 : \hs{15mm}
0.94 \hs{2mm} \leq \hs{2mm} T \hs{2mm} \leq \hs{2mm} 0.99  \label{eq:147}
\eeqa

\vs{2mm}
$6^o)$
We finally discuss the last ratio $T^*$ which,
under the factorization assumption, has the form given in Eq.(\ref{eq:126}).
We  compute $T^*$ for the three scenarios $n_2 = 2, 1, 0$,
using the values of $\Lambda_1, \Lambda_2$ and $\Lambda_V$ inside
the allowed domains obtained in Part V.
By writing
\beq
T^* =
\left( \frac{f_{\eta_c}}{f_{J/\Psi}}  \right)^2
\hs{2mm} \ol{T}^*                                \label{eq:148}
\eeq
we obtain
\beqa
n_2 = 2 : \hs{15mm}
0.6148 \pm 0.1108 \hs{2mm} \leq \hs{2mm} \ol{T}^* \hs{2mm} \leq \hs{2mm}
0.7740 \pm 0.1002  \label{eq:149} \\
n_2 = 1 : \hs{15mm}
0.6097 \pm 0.1259 \hs{2mm} \leq \hs{2mm} \ol{T}^* \hs{2mm} \leq \hs{2mm}
0.7375 \pm 0.1347  \label{eq:150} \\
n_2 = 0 : \hs{15mm}
0.6015 \pm 0.1415 \hs{2mm} \leq \hs{2mm} \ol{T}^* \hs{2mm} \leq \hs{2mm}
0.6217 \pm 0.1444  \label{eq:151}
\eeqa
If one accept the value (\ref{eq:144}) for the ratio of leptonic constants,
we make predictions for the ratio $T^*$ which are weakly scenario-dependent :
\beqa
n_2 = 2 :\hs{15mm}
0.50 \hs{2mm} \leq \hs{2mm} T^* \hs{2mm} \leq \hs{2mm} 0.86  \label{eq:152} \\
n_2 = 1 :\hs{15mm}
0.48 \hs{2mm} \leq \hs{2mm} T^* \hs{2mm} \leq \hs{2mm} 0.86  \label{eq:153} \\
n_2 = 0 :\hs{15mm}
0.45 \hs{2mm} \leq \hs{2mm} T^* \hs{2mm} \leq \hs{2mm} 0.76  \label{eq:154}
\eeqa

\vs{2mm}
$7^o)$
These results are now compared with those obtained in a previous paper
\cite{GKP1}.
The scenario-dependent parameter $S_V$ defined by Eq.(\ref{eq:SV})
is obviously very different from its value in the BSW model
considered in Ref.\cite{GKP1}.
The second scenario dependent parameter $S_A$ entering in $T^*$ :
\beq
\left| \frac{A_0^{BK^*}(m_{\eta_c}^2)}{A_1^{BK^*}(m_{J/\Psi}^2)} \right|^2
\equiv
\left| \frac{A_0^{BK^*}(0)}{A_1^{BK^*}(0)} \right|^2 \hs{2mm} S_A
                                                             \label{eq:155}
\eeq
is computed in our model with the result :
\beqa
n_2 = 2 :\hs{15mm}
0.767 \hs{2mm} \leq \hs{2mm} S_A \hs{2mm} \leq \hs{2mm} 0.793  \label{eq:156}
\\
n_2 = 1 :\hs{15mm}
0.832 \hs{2mm} \leq \hs{2mm} S_A \hs{2mm} \leq \hs{2mm} 0.849  \label{eq:157}
\\
n_2 = 0 :\hs{15mm}
0.966 \hs{2mm} \leq \hs{2mm} S_A \hs{2mm} \leq \hs{2mm} 0.978  \label{eq:158}
\eeqa
For the BSW model \cite{BSW}, $S_A$ is 1.069 .

The ratio $S_A/S_V$ entering in the ratio $R_{\eta_c}$ is more dependent
on the scenario than separately $S_V$ and $S_A$.
The results in our model are :
\beqa
n_2 = 2 :\hs{15mm}
1.53 \hs{2mm} \leq \hs{2mm} S_A/S_V \hs{2mm} \leq \hs{2mm} 2.09  \label{eq:159}
\\
n_2 = 1 :\hs{15mm}
1.85 \hs{2mm} \leq \hs{2mm} S_A/S_V \hs{2mm} \leq \hs{2mm} 2.19  \label{eq:160}
\\
n_2 = 0 :\hs{15mm}
2.44 \hs{2mm} \leq \hs{2mm} S_A/S_V \hs{2mm} \leq \hs{2mm} 2.54  \label{eq:161}
\eeqa
The bounds used in our previous paper \cite{GKP1},
$1 \leq S_A/S_V \leq 1.4$
are largely underestimated, essentially because of the behaviour
in $q^2$ of the hadronic form factors :
$F_0^{BK}$ constant and $A_1^{BK^*}$ linearly decreasing with $q^2$.
Therefore, the predictions of Ref.\cite{GKP1} for $R_{\eta_c}$ are
smaller than those obtained here.

\setcounter{section}{7}
\renewcommand{\theequation}{\Roman{section}.\arabic{equation}}
\setcounter{equation}{0}
\section{ \hs{3mm} Radiative Decay $B \ra K^* + \gamma$ }

\vs{2mm}
$1^o)$
The radiative decay $B \ra K^* + \gamma$ does not occur at the tree level
in the standard model. At the one loop level, we have the so-called
Penguin diagrams and, for the case consider here, the dominant contribution is
the one corresponding to the exchange of a virtual t quark.
In this approximation, the branching ratio is given by :
\beq
BR(B \ra K^* + \gamma) = \hs{2mm} \frac{\tau_B}{\hbar} \hs{2mm}
\frac{\alpha \hs{1mm} G_F^2}{256 \hs{1mm} \pi^4} \hs{2mm}
|V_{tb}|^2 \hs{2mm} |V_{ts}|^2 \hs{2mm} m_B^2 \hs{1mm}
\left( 1 - \frac{m_{K^*}^2}{m_B^2} \right)^3 \hs{1mm}
|C_7(m_b)|^2 \hs{1mm} m_b^2 \left\{ \hs{3mm} \right\}    \label{eq:8-1}
\eeq
where the quantity $ \left\{ \hs{3mm} \right\}$
depends on the hadronic form factors associated to
the tensor and pseudotensor current taken at $q^2 = 0$
and for real photon \cite{Gourdin} :
\beq
\left\{ \hs{3mm} \right\} =
\left(1 + \frac{m_s}{m_b} \right)^2 \hs{2mm} |v^{T}(0)|^2 +
\left(1 - \frac{m_s}{m_b} \right)^2 \hs{2mm} |a_2^{T}(0)|^2  \label{eq:8-2}
\eeq
The quantity $C_{7}(m_b)$ is the Wilson coefficient associated to the relevant
weak current.
It takes into account the large QCD corrections \cite{QCDc}
and it plays a determinant  role in the numerical calculation of the rate.

We use the unitarity relation of the CKM matrix
\beq
V_{tb} \hs{2mm} V_{ts}^* = - V_{cb} \hs{2mm} V_{cs}^* - V_{ub} \hs{2mm}
V_{us}^*
                                                            \label{eq:8-3}
\eeq
and we neglect the second term in the right hand side of Eq.(\ref{eq:8-3})
being $\cO(sin^2\theta_c)$ with respect to the other ones.

Numerically we take
\beq
|V_{cs}| = 0.970 \hs{5mm}, \hs{5mm}
|V_{cb}| = 0.040 \hs{5mm}, \hs{5mm}
\tau_B = 1.54 \hs{2mm} \cdot \hs{2mm} 10^{-12} \hs{2mm} s \hs{5mm}, \hs{5mm}
m_b = 4.7 \hs{2mm} GeV                        \label{eq:8-4}
\eeq
and for the Wilson coefficient, choosing $m_t = 174 \hs{2mm} GeV$ and
$\Lambda_{QCD} = 200 \hs{2mm} MeV$, we have \cite{QCDc} $C_7(m_b) = 0.325$.

The result is
\beq
BR(B \ra K^* + \gamma) = 4.45 \hs{1mm} \cdot \hs{1mm}10^{-5} \hs{2mm}
\left\{ \hs{3mm} \right\}                       \label{eq:8-5}
\eeq

The radiative decay mode $B \ra K^* + \gamma$ has been
experimentally observed by CLEO \cite{CLEO}
\beq
BR(B \ra K^* + \gamma) = (4.5 \pm 1.5 \pm 0.9)
\hs{1mm} \cdot \hs{1mm}10^{-5} = (4.50 \pm 1.75)
\hs{1mm} \cdot \hs{1mm}10^{-5} \label{eq:8-6}
\eeq
and we get
\beq
\left\{ \hs{3mm} \right\}_{exp} = 1.011 \pm 0.393   \label{eq:8-7}
\eeq

\vs{2mm}
$2^o)$
Following Isgur and Wise \cite{IW}, we assume that in the $B$ meson
at rest, the $b$ quark spinor in the weak current satisfies the free
Dirac equation for a spinor at rest : $\gamma_o b = b$.
As a consequence, the tensor (pseudotensor) current is related to the vector
(pseudovector) current
\beq
\ol{q} \hs{1mm} [\gamma_o, \gamma_j] \hs{1mm} b =
- \hs{1mm} 2 \hs{1mm} \ol{q} \hs{1mm} \gamma_j \hs{1mm} b \hs{10mm}, \hs{10mm}
\ol{q} \hs{1mm} [\gamma_o, \gamma_j] \hs{1mm} \gamma_5 \hs{1mm} b =
2 \hs{1mm} \ol{q} \hs{1mm} \gamma_j \hs{1mm} \gamma_5 \hs{1mm} b
\label{eq:8-8}
\eeq

It is then straight forward to compute the four tensor or pseudotensor
hadronic form factors in terms of the ususal BSW form factors $V, A_1, A_2$
 and $A_3$.
The two relations of interest here are :
\beq
v^{T}(0) = a_2^{T}(0) =
\left(1 + \frac{m_{K^*}}{m_B} \right) \hs{2mm} A_1^{BK^*}(0) +
\left(1 - \frac{m_{K^*}}{m_B} \right) \hs{2mm} V^{BK^*}(0)  \label{eq:8-9}
\eeq
In the limit $v^{T}(0) = a_2^{T}(0)$, the quantity $\left\{ \hs{3mm} \right\}$
of Eq.(\ref{eq:8-2}) becomes in the approximation $m_s^2 \ll m_b^2$ :
\beq
\left\{ \hs{3mm} \right\} = 2 |v^{T}(0)|^2  \label{eq:8-10}
\eeq
and using the estimate (\ref{eq:8-7}) for $\left\{ \hs{3mm} \right\}$, we
obtain
\beq
|v^{T}(0)|_{exp} = 0.71 \pm 0.14  \label{eq:8-11}
\eeq
The tensor and pseudotensor hadronic form factors have been computed
in various models, quark constituents models \cite{Desh1},
vector meson dominance models \cite{DOMING},
lattice gauge theories \cite{UKQCD}.
In most of the estimates, the equality $v^{T}(0) = a_2^{T}(0)$ is
obtained with $v^{T}(0)$
 in the range 0.5 to 1, e.g., consistent with the CLEO
result (\ref{eq:8-11}).
A recent estimate of Griffin, Masip and Mc Guigan
using the Isgur-Wise relation gives $v^{T}(0) = 0.97 \pm 0.13$ \cite{Griffin}.

It is amusing to observe that the BSW model \cite{BSW}
which fails \cite{GKP} in explaining the ratio $\rho_L$ and $R_{J/\Psi}$
in $B \ra K(K^*) + J/\Psi$  produces for $v^{T}(0)$,
using Eq.(\ref{eq:8-9}),  a value
$v^{T}(0) = 0.69$ in very nice agreement with experiment.
{\em It is clear that the radiative decay $B \ra K^* + \gamma$ is
not a very efficient filter for models.}

\vs{2mm}
$3^o)$
In our model, we compute $v^{T}(0) = a_2^{T}(0)$ using formule (\ref{eq:8-9}).
For $A_1^{BK^*}$ we use a linearly decreasing function of $q^2$
and for $V^{BK^*}$ a increasing function of $q^2$ of the dipole type,
the input values being predicted at $q^2 = t_B^{*o}$ by the Isgur-Wise
relations.
As a consequence, we obtain a relatively large $A_1^{BK^*}(0)$ and
a relatively small $V^{BK^*}(0)$.
The minimal value of $v^{T}(0)$ corresponds, in the allowed domain
of the parameter space, to $\Lambda_2 = 6 \hs{2mm} GeV$,
$\Lambda_V = 5 \hs{2mm} GeV$,
$\Lambda_1 = \Lambda_{1,\hs{1mm}MAX}(\Lambda_V, \Lambda_2)$ and
the results are in the one standard deviation limit,
including QCD corrections :
\beqa
& & n_2 = 2 :\hs{30mm} v^{T}(0) \hs{2mm} \geq \hs{2mm} 0.94 \label{eq:8-12} \\
& & n_2 = 1 :\hs{30mm} v^{T}(0) \hs{2mm} \geq \hs{2mm} 1.11 \label{eq:8-13} \\
& & n_2 = 0 :\hs{30mm} v^{T}(0) \hs{2mm} \geq \hs{2mm} 1.32 \label{eq:8-14}
\eeqa
If the assumption $v^{T}(0) = a_2^{T}(0)$ is correct,
we see that the scenario $n_2 = 2$ can accomodate the experimental result
(\ref{eq:8-11}).
A fit is clearly more difficult for the scenario $n_2 = 1$
and it seems to be impossible for the scenario $n_2= 0$.

However we must be aware of the fact that the estimate of the QCD correction,
which is scale dependent,  may have some uncertainty which
has been disregarded in the experimental error quoted in Eq.(\ref{eq:8-11}).
A theoretical error has to be added which might be
as large as 15 $\%$ \cite{Griffin}.

\setcounter{section}{8}
\renewcommand{\theequation}{\Roman{section}.\arabic{equation}}
\setcounter{equation}{0}
\section{\hs{3mm} $ D \ra \ol{K} (\ol{K}^*)$ Hadronic Form Factors}

\vs{2mm}
$1^o)$
The $B \ra K(K^*)$ and $D \ra K(K^*)$ hadronic form factors
are related by the SU(2)
heavy flavour symmetry of Isgur-Wise.
{}From the considerations of Part II, it is clear that the $q^2$-dependence
for the form factors $F_1, A_1, A_2$ and $V$, namely

\hs{5mm} (i) Same values for $n_1, n_2, n_V$ and $n_F$
in both $B$ and $D$ sectors;

\hs{4mm} (ii) The pole masses in these sectors are related
by Eqs.(\ref{eq:19}) and (\ref{eq:30}) of Part II.

For $F_0$ and $A_0$ the situation might be different in the two sectors
but the $q^2$-dependence of $F_0$ is known from that of $F_1$ and for $A_0$,
its $q^2$-dependence follows from that of $A_1$ and $A_2$ as explained
in Part II.

\vs{2mm}
$2^o)$
Let us first consider the semi-leptonic decay of $D$ mesons.
The relevant hadronic form factors are $F_1^{DK}$ for
$D \ra \ol{K} + \ell^{+} + \nu_{\ell}$
and $A_1^{DK^*}$, $A_2^{DK^*}$, $V^{DK^*}$ for
$D \ra \ol{K}^{*} + \ell^{+} + \nu_{\ell}$.
Using the dimensionless variable $t = q^2/m_D^2$,
we introduce the dimensionless function $X(t)$ :
\beq
X(t) = \frac{1}{\Gamma} \hs{2mm}
\frac{d \Gamma}{d t} \       \label{eq:9-1}
\eeq
which is independent of all parameters
entering in the semi-leptonic relevant scale factor.

We recall that the quantities $x^D(0)$, $y^D(0)$ and
$z^D(0)$
(used in this paper for normalizing the $B$ sector)
 have been extracted from
experimental data on semi-leptonic decay
in a scenario-dependent way,
 because the variation with $q^2$ of the form factors
$F_1^{DK}$, $A_1^{DK^*}$, $A_2^{DK^*}$ and $V^{DK^*}$
has not been measured.

\vs{2mm}
$3^o)$
The $q^2$ distribution in semi-leptonic decay
$D \ra \ol{K} + \ell^{+} + \nu_{\ell}$
is written in terms of the hadronic form factor $F_1^{DK}(q^2)$  as
\beq
\frac{d \Gamma(D \ra \ol{K} + \ell^{+} + \nu_{\ell})}{d q^2} \ =
\frac{G_F^2}{24 \pi^3} \hs{2mm} |V_{cs}|^2 [ K(q^2) ]^3 \hs{2mm}
| F_1^{BK}(q^2) |^2                     \label{eq:9-2}
\eeq
where the $q^2$ dependent momentum $K(q^2)$ is given by
\beq
K(q^2) = \frac{1}{2 m_D} \hs{2mm}
\left\{(m_D^2 + m_K^2 - q^2)^2 - 4 m_D^2 m_{K}^2 \right\}^{1/2} \label{eq:9-3}
\eeq
In the zero lepton mass limit, $0 \leq q^2 \leq (m_D - m_K)^2$.

In our model $F_1^{DK}(q^2)$ has a monopole $q^2$-dependence with a pole mass
$\Lambda_{DF}$ related to $\Lambda_{F}$ in the $B$ sector by Eq.(\ref{eq:19}).
Defining the dimensionless parameters :
\beq
r = \frac{m_K}{m_D} \hs{15mm}, \hs{15mm}
\alpha_F = \frac{m_D^2}{\Lambda_{DF}^2} \label{eq:9-4}
\eeq
we obtain for $X(t)$ the expression :
\beq
X(t) = \frac{1}{I(\alpha_F)} \ \hs{2mm}
\frac{[(1 + r^2 - t)^2 - 4 r^2]^{3/2}}{(1 - \alpha_F \hs{1mm} t)^2} \
\label{eq:9-5}
\eeq
where the integral $I(\alpha_F)$ is defined by the normalization condition
$X(t)$ :
\beq
I(\alpha_F) = \int^{(1 - r)^2}_{0} \hs{2mm}
\frac{[(1 + r^2 - x)^2 - 4 \hs{1mm} r^2]^{3/2}}{(1 - \alpha_F \hs{1mm} x)^2}
\hs{2mm} dx              \label{eq:9-6}
\eeq
Of course, the semi-leptonic rate is simply given by :
\beq
\Gamma(D \ra \ol{K} + \ell^{+} + \nu_{\ell}) =
\frac{G_F^2 m_D^5}{192 \pi^3} \hs{2mm} |V_{cs}|^2
\hs{2mm} |F_1^{DK}(0)|^2 \hs{2mm} I(\alpha_F)     \label{eq:9-7}
\eeq
The normalized distribution $X(t)$ for the semi-leptonic decay mode
$D \ra \ol{K} + \ell^{+} + \nu_{\ell}$ is represented on Figure 18
for values of $\alpha_F$ corresponding to the bounds on $\Lambda_F$
obtained in Part V and illustrated on Figures 15, 16 and 17.
\beqa
n_2 = 2 :\hs{10mm}
5 \hs{2mm} GeV \hs{2mm} \leq \hs{2mm} \Lambda_F \hs{2mm}
\leq \hs{2mm} 5.71 \hs{2mm} GeV \hs{10mm}
1.097 \hs{2mm} \geq \hs{2mm} \alpha_F \hs{2mm} \geq \hs{2mm} 0.630
\label{eq:9-8} \\
n_2 = 1 :\hs{10mm}
5 \hs{2mm} GeV \hs{2mm} \leq \hs{2mm} \Lambda_F \hs{2mm}
\leq \hs{2mm} 5.39 \hs{2mm} GeV \hs{10mm}
1.097 \hs{2mm} \geq \hs{2mm} \alpha_F \hs{2mm} \geq \hs{2mm} 0.787
\label{eq:9-9} \\
n_2 = 0 :\hs{10mm}
5 \hs{2mm} GeV \hs{2mm} \leq \hs{2mm} \Lambda_F \hs{2mm}
\leq \hs{2mm} 5.10 \hs{2mm} GeV \hs{10mm}
1.097 \hs{2mm} \geq \hs{2mm} \alpha_F \hs{2mm} \geq \hs{2mm} 0.999
\label{eq:9-10}
\eeqa
The distribution $X(t)$ is a monotonically decreasing function of $t$.
Its shape is not very sensitive to $\alpha_F$
except in the neighbourhood of $t = 0$.

An estimate for the slope of the $q^2$ distribution
at $q^2 = 0$ has been given by
Witherell \cite{Witherell} using two models
for the $q^2$-dependence of $F_1^{BK}(q^2)$,
an exponential form and a monopole form.
The result consistent for the two models translated in the $\alpha_F$
language is
\beq
0.76 \hs{2mm} \leq \hs{2mm} \alpha_F \hs{2mm} \leq \hs{2mm} 1.30
\label{eq:9-11}
\eeq
which means in our model
\beq
5.36 \hs{2mm} GeV \hs{2mm} \geq \hs{2mm} \Lambda_F
\hs{2mm} \geq \hs{2mm} 5.02 \hs{2mm} GeV          \label{eq:9-12}
\eeq
Therefore our bounds (\ref{eq:9-8}) $-$ (\ref{eq:9-10}) are consistent with
experiment.

\vs{2mm}
$4^o)$
The $q^2$ distribution in the semi-leptonic decay
$D \ra \ol{K}^{*} + \ell^{+} + \nu_{\ell} $
depends on the three hadronic form factors
$A_1^{DK^*}(q^2)$, $A_2^{DK^*}(q^2)$ and $V^{DK^*}(q^2)$.
We have three possible polarizations for the final $K^*$,
$\lambda = 0, \pm 1$.

It is convenient to define dimensionless parameters :
\beq
r^* = \frac{m_{K^*}}{m_D} \hs{4mm}, \hs{4mm} t = \frac{q^2}{m_D^2}
\hs{4mm}, \hs{4mm}
\alpha_j = \frac{m_D^2}{\Lambda_{Dj}^2} \hs{20mm};
\hs{5mm} j = 1, 2, V \label{eq:9-13}
\eeq
where the pole masses in the $B$ and $D$ sectors are related by
Eq.(\ref{eq:30}).

The fixed $q^2$ distribution is given by :
\beq
\frac{d \Gamma(D \ra \ol{K}^{*} + \ell^{+} + \nu_{\ell})_{\ld}}{d t} =
\frac{G_F^2 m_D^5}{192 \pi^3} \hs{2mm} |V_{cs}|^2 \hs{2mm}
 (1 + r^*)^2 \hs{2mm} |A_1^{DK^*}(0)|^2 \hs{2mm}
M_{\lambda}(t ; \a_1, \a_2, \a_V)           \label{eq:9-14}
\eeq
where
\beqa
M_L(t; \a_1, \a_2) & = & k(t) \hs{2mm}
\left| \ta(t) \hs{2mm} \frac{1}{(1 - \a_1 t)^{n_1}}
- \tb(t) \hs{2mm} \frac{x^D(0)}{(1 - \a_2 t)^{n_2}} \right|^2
\label{eq:9-15} \\
M_{\pm}(t; \a_1, \a_V) & = & t \hs{2mm} k(t) \hs{2mm}
\left| \frac{1}{(1 - \a_1 t)^{n_1}}
\mp \tc(t) \hs{2mm} \frac{y^D(0)}{(1 - \a_V t)^{n_V}} \right|^2
\label{eq:9-16}
\eeqa
with k(t) given by :
\beq
k(t) = \left\{(1 + {r^*}^2 - t)^2 - 4 {r^*}^2 \right\}^{1/2}   \label{eq:9-17}
\eeq
The coefficients $\ta(t), \hs{1mm} \tb(t) $ and $\tc(t)$ are :
\beqa
\ta(t) & = & \frac{1 - {r^*}^2 - t}{2 r^*}     \label{eq:9-18} \\
\tb(t) & = & \frac{k^2(t)}{2 r^* (1 + r^*)^2}  \label{eq:9-19} \\
\tc(t) & = & \frac{k(t)}{(1 + r^*)^2}          \label{eq:9-20}
\eeqa
We define the integrals :
\beq
I_{\ld}(\a_1, \a_2, \a_V) =
\int_{0}^{(1 - r^*)^2} \hs{2mm} M_{\ld}(t; \a_1, \a_2, \a_V) \hs{2mm} dt
\label{eq:921a}
\eeq
and the integrated rate is given by :
\beq
\Gamma(D \ra \ol{K}^{*} + \ell^{+} + \nu_{\ell}) =
\frac{G_F^2 m_D^5}{192 \pi^3} \hs{2mm} |V_{cs}|^2 \hs{2mm}
 (1 + r^*)^2 \hs{2mm} |A_1^{DK^*}(0)|^2 \hs{2mm}
\Sigma_{\ld} \hs{2mm}
I_{\ld}(\a_1, \a_2, \a_V)           \label{eq:9-21}
\eeq
The normalized $q^2$ distribution is  :
\beq
X(t) = \frac{ \Sigma_{\ld} \hs{2mm} M_{\ld}(t ; \a_1, \a_2, \a_V) }
{ \Sigma_{\ld} \hs{2mm} I_{\ld}(t ; \a_1, \a_2, \a_V) }  \label{eq:9-22}
\eeq
Of course for the scenarios selected in Part V, we have $n_1 = -1, n_V = 2 $
and three possible values for $n_2$; $n_2 = 2, 1, 0$.
We have computed $X(t)$ in these three cases by using the PDG values \cite{RPR}
for $x^D(0)$ and $y^D(0)$.
The parameters $\a_1, \a_2, \a_V$ $-$ or equivalently
$\Lambda_1, \Lambda_2, \Lambda_V$ $-$
are constrainted to stay inside the allowed domains represented
on Figures 12, 13 and 14.
The results are shown on Figures 19, 20 and 21.
As in the previous case, the largest sensitivity of $X(t)$
to the parameters $\a_i$
is in the neighbourhood of $t = 0$.

In an analogous way, it is possible to study the $q^2$ distributions
for the polarization parameters $\rho_L^{\it sl}(t)$
and $\cA_{LR}^{\it sl}(t)$ :
\beqa
\rho_L^{\it sl}(t) & = &
\frac{M_L(t ; \a_1, \a_2, \a_V)}
{\Sigma_{\ld} \hs{2mm} M_{\ld}(t ; \a_1, \a_2, \a_V) } \label{eq;923} \\
\cA_{LR}^{\it sl}(t) & = &
\frac{M_{-}(t ; \a_1, \a_2, \a_V) - M_{+}(t ; \a_1, \a_2, \a_V)}
{M_{-}(t ; \a_1, \a_2, \a_V) + M_{+}(t ; \a_1, \a_2, \a_V)}  \label{eq:924}
\eeqa
We only give here the integrated ratios $\rho_L^{\it sl}$
and $\cA_{LR}^{\it sl}$
where the functions $M_{\ld}(t ; \a_1, \a_2, \a_V)$ in Eq.(\ref{eq:9-15})
and (\ref{eq:9-16}) are replaced by their integrals over $t$,
$I_{\ld}(\a_1, \a_2, \a_V)$.
The results for the three cases $n_2 = 2, 1, 0$ are the following :
\beqa
 n_2 = 2 : \hs{10mm}
& &
0.516 \hs{1mm} \leq \hs{1mm} \rho_L^{\it sl} \hs{1mm} \leq \hs{1mm} 0.541
                                          \label{eq:9-24} \\
& &
0.885 \hs{1mm} \geq \hs{1mm} \cA_{LR}^{\it sl} \hs{1mm} \geq \hs{1mm} 0.829 \no
\\
\cr
 n_2 = 1 : \hs{10mm}
& &
0.526 \hs{1mm} \leq \hs{1mm} \rho_L^{\it sl} \hs{1mm} \leq \hs{1mm} 0.541
                                         \label{eq:9-25} \\
& &
0.904 \hs{1mm} \geq \hs{1mm} \cA_{LR}^{\it sl} \hs{1mm} \geq \hs{1mm} 0.857 \no
\\
\cr
 n_2 = 0 : \hs{10mm}
& &
0.536 \hs{1mm} \leq \hs{1mm} \rho_L^{\it sl} \hs{1mm} \leq \hs{1mm} 0.538
                                       \label{eq:9-26} \\
& &
0.904 \hs{1mm} \geq \hs{1mm} \cA_{LR}^{\it sl} \hs{1mm} \geq \hs{1mm} 0.892 \no
\eeqa
In Eqs.(\ref{eq:9-24}) - (\ref{eq:9-26}) the results are presented in such a
way to
exhibit a correlation between the largest (smallest) $\rho_L^{\it sl}$
and the smallest (largest) $\cA_{LR}^{\it sl}$.

\vs{2mm}
$5^o)$
Let us now consider the hadronic two body decays
$D^o \ra \ol{K} (K^{-*}) + \pi^{+} (\rho^{+})$.
Assuming factorization to be justified in the $D$ sector,
the various decay rates can be determined by using the hadronic
$D \ra \ol{K} (\ol{K}^*)$ form factors studied here.

At the quark level, the tree level diagram is of the spectator type
and the various rates are written in the following way :
\beq
BR = BR_0 \hs{2mm} \cdot \hs{2mm} \left( \frac{f_{u\ol{d}}}{m_D} \right)^2
\hs{2mm} \cdot \hs{2mm} PS \hs{2mm} \cdot \hs{2mm} FF       \label{eq:9-27}
\eeq
The scale $BR_0$ is given by :
\beq
BR_0 = \left( \frac{G_F m_D^2}{\sqrt{2}} \right)^2 \hs{2mm}
|V_{cs}|^2 \hs{2mm} |V_{ud}|^2 \hs{2mm} \frac{m_D}{8 \pi} \hs{2mm}
a_1^2 \hs{2mm} \frac{ \tau_{D_o} }{\hbar}              \label{eq:9-28}
\eeq
where $a_1$ is the BSW \cite{BSW} parameter
for color favoured processes.
As in the Part III,
$PS$ is a phase space factor,
$f_{u\ol{d}}$ is a leptonic decay constant $f_{\pi^{+}}$ or $f_{\rho^{+}}$
which are experimentally known and
$FF$ depends on the hadronic form factors.

The results are :
\beqa
& & a) \hs{2mm} D^o \ra K^{-} + \pi^{+} \hs{6mm}, \hs{6mm} PS = 0.3993
\hs{5mm}, \hs{5mm} FF  =  |F_0^{DK}(m_{\pi}^2)|^2          \label{eq:929} \\
& & b) \hs{2mm} D^o \ra K^{-} + \rho^{+} \hs{6mm}, \hs{6mm} PS = 0.1936
\hs{5mm}, \hs{5mm} FF  =  |F_1^{DK}(m_{\rho}^2)|^2         \label{eq:9-30} \\
& & c) \hs{2mm} D^o \ra {K^*}^{-} + \pi^{+} \hs{5mm}, \hs{5mm} PS = 0.2216
\hs{5mm}, \hs{5mm} FF  =  |A_0^{DK^*}(m_{\pi}^2)|^2        \label{eq:931} \\
& & d) \hs{2mm} D^o \ra {K^*}^{-} + \rho^{+} \hs{5mm}, \hs{5mm} PS = 0.0843
\no \\
& & \hs{40mm} FF  =  |A_1^{DK^*}(m_{\rho}^2)|^2
\left\{(a_D - b_D x_D)^2 + 2 [ 1  + {c_D}^2 {y_D}^2 ] \right\}  \label{eq:9-32}
\eeqa
where
\beq
a_D = 1.5310 \hs{7mm},\hs{7mm} b_D = 0.2419 \hs{7mm},\hs{7mm} c_D = 0.2087
\label{eq:9-33}
\eeq
and
\beq
x_D = \frac{A_2^{DK^*}(m_{\rho}^2)}{A_1^{DK^*}(m_{\rho}^2)} \hs{10mm},
\hs{10mm}
y_D = \frac{V^{DK^*}(m_{\rho}^2)}{A_1^{DK^*}(m_{\rho}^2)}      \label{eq:9-34}
\eeq
In order to test the assumption of factorization in the $D$ sector, we shall
study two ratios of rates :
\beq
R_{\pi} = \frac{\Gamma(D^o \ra {K^*}^{-} + \pi^{+})}
{\Gamma(D^o \ra K^{-} + \pi^{+})}      \hs{10mm}, \hs{10mm}
T_D = \frac{\Gamma(D^o \ra K^{-} + \rho^{+})}
{\Gamma(D^o \ra K^{-} + \pi^{+})}                  \label{eq:9-35}
\eeq
Using Eq.(\ref{eq:929}) and (\ref{eq:931}), we obtain
\beq
R_{\pi} = 0.5550 \hs{2mm} \left| \frac{A_0^{DK^*}(m_{\pi}^2)}
{F_0^{DK}(m_{\pi}^2)} \right|^2                     \label{eq:9-36}
\eeq
It is legitimate to neglect the variation of the form factors $F_0$ and $A_0$
between
$q^2 = 0$ and $q^2 = m_{\pi}^2$.
Using the normalization conditions \cite{BSW},
\beqa
F_0^{DK}(0) & = & F_1^{DK}(0)    \label{eq:9-37a} \\
A_0^{DK^*}(0) & = & \frac{m_D + m_{K^*}}{2 m_{K^*}} \hs{2mm} A_1^{DK^*}(0)
- \frac{m_D - m_{K^*}}{2 m_{K^*}} \hs{2mm} A_2^{DK^*}(0)   \label{eq:9-37}
\eeqa
and the values at $q^2 = 0$ of the form factors as given by the PDG \cite{RPR}
:
\beq
F_1^{DK}(0) = 0.75 \pm 0.03 \hs{5mm}, \hs{5mm}
A_1^{DK^*}(0) = 0.56 \pm 0.04  \hs{5mm}, \hs{5mm}
A_2^{DK^*}(0) = 0.40 \pm 0.08              \label{eq:9-38}
\eeq
we obtain :
\beq
A_0^{DK^*}(0) = 0.6473 \pm 0.0757 \label{eq:9-39a}
\eeq
and the theoretical prediction for $R_{\pi}$ is
\beq
( R_{\pi} )_{th} = 0.4134 \pm 0.1022        \label{eq:9-39}
\eeq
The experimental result is \cite{RPR} :
\beq
( R_{\pi} )_{exp} = 1.22 \pm 0.16        \label{eq:9-40}
\eeq
The discrepancy between theory and experiment is very large.
The theoretical prediction is clearly scenario-independent
and it is unlikely that final state interaction would be able to fill the gap
between theory and experiment.
The most probable explanation is a failure of factorization in the $D$ sector.

Let us now consider the quantity $T_D$.
Using Eq.(\ref{eq:929}) and (\ref{eq:9-30}), we get
\beq
T_D = 0.4850 \hs{2mm} \left( \frac{f_{\rho}}{f_{\pi}} \right)^2 \hs{2mm}
\left| \frac{F_1^{DK}(m_{\rho}^2)}{F_0^{DK}(m_{\pi}^2)} \right|^2
\label{eq:9-41}
\eeq
For the leptonic decay constants $f_{\pi}$ and $f_{\rho}$,
we use the experimental values
\beq
f_{\pi} = 131.7 \hs{2mm} MeV \hs{20mm} f_{\rho} = 212 \hs{2mm} MeV
\label{eq:9-42}
\eeq
and $T_D$ is written as :
\beq
T_D = 1.2566 \hs{2mm}
\left| \frac{F_1^{DK}(m_{\rho}^2)}{F_0^{DK}(m_{\pi}^2)} \right|^2
\label{eq:9-43}
\eeq
Neglecting as previously the variation of $F_0$
between $q^2 = 0$ and $q^2 = m_{\pi}^2$ and
using the normalization condition (\ref{eq:9-37a}), we get
\beq
\frac{F_1^{DK}(m_{\rho}^2)}{F_0^{DK}(m_{\pi}^2)} \simeq
\frac{F_1^{DK}(m_{\rho}^2)}{F_1^{DK}(0)}       \label{eq:9-44}
\eeq
In this paper we have used a monopole $q^2$-dependence for $F_1$, $n_F = 1$
and we have
\beq
\frac{F_1^{DK}(m_{\rho}^2)}{F_1^{DK}(0)} =
\frac{1}{1 - \frac{m_{\rho}^2}{\Lambda_{DF}^2}} \label{eq:9-45}
\eeq
where the pole masses $\Lambda_F$ and $\Lambda_{DF}$
in the $B$ and $D$ sectors are related by Eq.(\ref{eq:19}).
With the constraint $\Lambda_F \geq 5 \hs{2mm} GeV$,
the maximal value of $T_D$ is obtained with $\Lambda_{DF} = 1.78 \hs{2mm} GeV$
$-$ which corresponds to $\Lambda_F$ = 5 $GeV$ $-$
with the result.
\beq
{(T_D)}_{th} < 1.8946    \label{eq:9-46}
\eeq
The experimental value is significantly larger \cite{RPR} :
\beq
{(T_D)}_{exp} = 2.59 \pm 0.34    \label{eq:9-47}
\eeq
Again the factorization assumption seems to be in a bad shape
in the $D$ sector.

\setcounter{section}{9}
\renewcommand{\theequation}{\Roman{section}.\arabic{equation}}
\setcounter{equation}{0}
\section{ \hs{3mm} Critical Discussions and Conclusions }

\vs{2mm}
$1^o)$
We have shown that the assumption of factorization is not ruled out by
experimental
data for the colour-suppressed decay modes of the $B$ meson,
$B \ra K (K^*) + J/\Psi (\Psi^{'})$.
The failure pointed out in Ref.\cite{GKP}
might be due to inadequate choices of hadronic form factors
and the aim of this paper is essentially to exhibit
possible $q^2$-dependences that
are able to explain experimental data and  particularly the ratios
$\rho_L$ and $R_{J/\Psi}$.
Of course the possibility to understand experiment is not necessary
a proof of factorization.

Let us first summarize the assumptions and constraints contained in our model.

\vs{2mm}
(A) Assumptions :

\ben
\item
Factorization holds for color supressed $B$ decays and final state strong
interaction effects can be neglected.

\item
The SU(2) heavy flavour symmetry between the $b$ and $c$ quarks is realized
by the Isgur-Wise relations \cite{IW}.

\item
The input experimental data in the $D$ sector are taken from the analysis
of semi-leptonic decays
$D \ra \ol{K} + \ell^{+} + \nu_{\ell}$ and
$D \ra \ol{K}^{*} + \ell^{+} + \nu_{\ell}$ in the form of the values
at $q^2 = 0$ of the $D \ra K (K^*)$ hadronic form factors\cite{RPR}.

\een

\vs{2mm}
(B) The experimental constraints are :

\ben
\item
The experimental rates for $B \ra K + J/\Psi$, $B \ra K^* + J/\Psi$,
$B \ra K + \Psi^{'}$ and $B \ra K^{*} + \Psi^{'}$ used in the form
of ratios of rates $R_{J/\Psi}$, $R_{\Psi^{'}}$, $S$ and $S^*$
[defined respectively by Eq.(\ref{eq:70}) and Eq.(\ref{eq:121})].

\item
The observed longitudinal polarization fraction
$\rho_L$ in $B \ra K^* + J/\Psi$
[defined by Eq.(\ref{eq:66})].

\een

\vs{2mm}
(C) The theoretical constraints are :

\ben
\item
The explicit form of the $q^2$ dependence of the hadronic form factors $F_1$,
$A_1$, $A_2$, $V$ choosen as $\left[ 1 - q^2/\Lambda^2 \right]^{-n}$ with
$ n = -1, 0, 1, 2$.

\item
The pole masses $\Lambda$ of the various form factors in the $B$ sector are
limited to
the $(5 - 6) \hs{2mm} GeV$ range in order to relate them in a likely way to
$b\ol{s}$
bound state masses.

\item
The ratios of form factors $\mu^B(q^2)$ and $\lambda^B(q^2)$
defined in Eqs.(\ref{eq:11})
and (\ref{eq:41}) are assumed to be independent of $q^2$ and related in a
natural way
to the pole masses $\Lambda_F$ and $\Lambda_2$
by Eqs.(\ref{eq:21}) and (\ref{eq:50}).
\een

\vs{2mm}
$2^o)$
Among our three theoretical assumptions,
the weakest one seems to us the third, i.e., the use of the experimental values
at $q^2 =0$ of the $D \ra K(K^*)$ hadronic form factors as deduced
in a scenario-dependent way from the experimental triple angular distribution
of
semi-leptonic decay.
We must emphasize that  measurements of the $q^2$-dependence of
these form factors are not available
and their values at $q^2 = 0$ are obtained by
extrapolation at $q^2 = 0$ of experimental data, assuming monopole
$q^2$-dependence
for all form factors.

Consider first the simplest case of the form factor $F_1^{DK}(q^2)$.
The two decay modes $D^{+} \ra \ol{K}^o \hs{2mm} \ell^{+} \hs{2mm} \nu_{\ell}$
and $D^{o} \ra K^{-} \hs{2mm} \ell^{+} \hs{2mm} \nu_{\ell}$
are expected to have equal
rates, the associated weak current $c\ol{s}$ being isoscalar.
Experimentally these two rates differ by few standard deviations \cite{RPR}
and some average dominated by the most accurately measured $D^o$ mode is used
for extracting the quantity $F_1^{DK}(0)$ and the slope of $F_1^{DK}(q^2)$
at $q^2 = 0 $ from the measured $q^2$ distribution.
In order to have an estimate of the uncertainties of the analysis,
let us notice that the rate value used by Witherell \cite{Witherell} is
10 $\%$ higher than the one of the PDG \cite{RPR}
both values being given with 5 $\%$ errors.
In spite of this difference, the values obtained by the PDG \cite{RPR},
$F_1^{DK}(0) = 0.75 \pm 0.03$ and by Witherell \cite{Witherell},
$F_1^{DK}(0) = 0.77 \pm 0.04$ are very similar.

We have considered the problem of determining $F_1^{BK}(0)$ from the
semi-leptonic rate by using Eq.(\ref{eq:9-7}).
The result depends on the parameter $\a_F$ simply related to the pole masses
$\Lambda_{DF}$ and $\Lambda_{F}$ in the $D$ and $B$ sectors.

The results are :
\beqa
& & n_2 = 2 :\hs{10mm}
5 \hs{2mm} GeV \hs{1mm} \leq \hs{1mm} \Lambda_F
\hs{1mm} \leq \hs{1mm} 5.71 \hs{2mm} GeV \hs{5mm}, \hs{5mm}
0.68 \hs{1mm} \leq \hs{1mm} F_1^{DK}(0)
\hs{1mm} \leq \hs{1mm} 0.79               \label{eq:CON-1} \\
& & n_2 = 1 :\hs{10mm}
5 \hs{2mm} GeV \hs{1mm} \leq \hs{1mm} \Lambda_F
\hs{1mm} \leq \hs{1mm} 5.39 \hs{1mm} GeV  \hs{5mm}, \hs{5mm}
0.68 \hs{1mm} \leq \hs{1mm} F_1^{DK}(0)
\hs{1mm} \leq \hs{1mm} 0.77               \label{eq:CON-2} \\
& & n_2 = 0 :\hs{10mm}
5 \hs{2mm} GeV \hs{1mm} \leq \hs{1mm} \Lambda_F
\hs{1mm} \leq \hs{1mm} 5.10 \hs{2mm} GeV  \hs{5mm}, \hs{5mm}
0.68 \hs{1mm} \leq \hs{1mm} F_1^{DK}(0)
\hs{1mm} \leq \hs{1mm} 0.73               \label{eq:CON-3}
\eeqa
The numerical value of the rate used in this analysis is the PDG value of
$(8.2 \pm 0.4) \hs{1mm} 10^{10} \hs{1mm} s^{-1}$.
We notice that the slope results presented by Witherell \cite{Witherell}
correspond in our language to
$5.02 \hs{2mm} GeV  \leq \Lambda_F \leq 5.36 \hs{2mm} GeV$.

We now comment  on
$D \ra \ol{K}^* \hs{2mm} \ell^{+} \hs{2mm} \nu_{\ell}$
data, for which the $q^2$-dependences
for the relevant hadronic form factors,
$A_1^{K^*}, A_2^{DK^*}$ and $V^{DK^*}$, are not yet available.
By fitting the $q^2$ distribution and the angular distribution due to the
$K^* \ra K + \pi$ decay, the ratios $x^D(0)$ and $y^D(0)$ are obtained
assuming for the form factors a monopole $q^2$-dependence
with the pole masses given by the lowest lying
$c\ol{s}$ bound states, $D_s^*(2110)$ and $D_s^{**}(2560)$.
The results of the PDG \cite{RPR} and of Witherell \cite{Witherell}
are nearly identical.
The quantity $A_1^{DK^*}(0)$ is then obtained from the transition rate.
In this case the semi-leptonic decay rates
$\Gamma(D^o \ra {K^*}^{-} + \ell^{+} + \nu_{\ell})$
and
$\Gamma(D^+ \ra \ol{K^*}^{o} + \ell^{+} + \nu_{\ell})$
are equal within errors.
However the average value of the rate used by the PDG
and by Witherell differs by $11 \%$
and as a consequence the values of $A_1^{DK^*}(0)$ obtained in the two cases
are different :
\beq
PDG \hs{1mm} \cite{RPR}
\hs{5mm} A_1^{DK^*}(0) = 0.56 \pm 0.04 \hs{8mm}, \hs{8mm}
 W \hs{1mm} \cite{Witherell}
\hs{5mm} A_1^{DK^*}(0) = 0.61 \pm 0.05      \label{eq:CON-4}
\eeq
We study the problem of determining $A_1^{DK^*}(0)$ from the semi-leptonic
decay
rate by using Eq.(\ref{eq:9-21}).
We vary the parameters $\Lambda_1, \Lambda_2, \Lambda_V$ $(\a_1, \a_2, \a_V)$
in the allowed domains determined in Part V and the results are :
\beqa
& & n_2 = 2 :\hs{20mm} 0.601 \hs{2mm} \leq \hs{2mm}
A_1^{DK^*}(0) \hs{2mm} \leq \hs{2mm} 0.645    \label{eq:CON-5} \\
& & n_2 = 1 :\hs{20mm} 0.618 \hs{2mm} \leq \hs{2mm}
A_1^{DK^*}(0) \hs{2mm} \leq \hs{2mm} 0.645    \label{eq:CON-6} \\
& & n_2 = 0 :\hs{20mm} 0.638 \hs{2mm} \leq \hs{2mm}
A_1^{DK^*}(0) \hs{2mm} \leq \hs{2mm} 0.647    \label{eq:CON-7}
\eeqa
Our $A_1^{DK^*}(0)$ is to be compared to the PDG one $0.56 \pm 0.04$,
since the numerical value of the rate used in our analysis is the PDG value
$(4.6 \pm 0.4) \hs{1mm} 10^{10} \hs{1mm} s^{-1}$.

The errors quoted for $F_1^{DK}(0)$ and $A_1^{DK^*}(0)$ in references
\cite{RPR} and \cite{Witherell} do not take into account the theoretical
uncertainties due to the $q^2$-dependence used
for analysing the experimental data.
The formal results previously obtained with our model illustrate clearly
the point and it seems to us that the errors have been underestimated.

Finally let us point out that the various ratios studied in this paper
have different types of dependence
with respect to the form factor values at $q^2 = 0$ in the $D$ sector.

\hs{5mm} (i). \hs{3mm}
$\cA_{LR}$ and $\cA_{LR}^{'}$ depend only on $y^D(0)$.

\hs{4mm} (ii). \hs{3mm}
$\rho_{L}$ and $\rho_{L}^{'}$ depend on $x^D(0)$ and $y^D(0)$.

\hs{3mm} (iii). \hs{3mm}
$R_{J/\Psi}$ and $R_{\Psi^{'}}$ depend on $x^D(0)$, $y^D(0)$ and $z^D(0)$.

\hs{2.5mm} (iv). \hs{3mm}
$R_{\eta_c}$ depends on $x^D(0)$ and $z^D(0)$.

\hs{4mm} (v). \hs{3mm}
$S$ and $T$ are independent of these three ratios.

\hs{2.5mm} (vi). \hs{3mm}
$S^*$ and $T^*$ depend on $x^D(0)$ and $y^D(0)$.

For the semi-leptonic normalized distribution $X(t)$, it is independent
on these ratios in the
$D \ra \ol{K} + \ell^{+} + \nu_{\ell}$ case and it depends on $x^D(0)$ and
$y^D(0)$
in the $D \ra \ol{K}^{*} + \ell^{+} + \nu_{\ell}$ mode.

\vs{2mm}
$3^o)$
In addition to the necessity of an improvement
in accuracy for the observed rates
$-$ in particular, those involving the $\Psi^{'}$
are still badly known with
errors at the $45 \%$ or $64 \%$ level $-$
we need a detection of the $K + \eta_c$ and $K^* + \eta_c$ modes.
On the polarization side, only the ratio $\rho_L$ has been measured
and it was an important quantity to be taken into account for the fits.
It is clear that the knowledge of $\cA_{LR}$, $\rho_L^{'}$ and $\cA_{LR}^{'}$
predicted by our model would considerably help
in reducing the size of the allowed
domains in the parameter space.

The best situation would be to select only one scenario
and a small domain in the parameter space.
The worse situation for our model would be that the new measurements exclude
the three presently remaining scenarios.
Our approach being purely phenomenological is not connected with
any theoretical calculation like, for instance, QCD sum rule
or lattice gauge theories and our model is certainly not the unique way
to compute $B \ra K (K^*)$ hadronic form factors.

However if we are in the best situation previously mentioned, it will be
necessary to provide a theoretical support to
the so determined hadronic form factors
and for that, results of Ref.\cite{QCDSUM} seem to be
in a good shape because of the unusual $q^2$ behaviour prediction for
$A_1(q^2)$.

If we are in the worse situation, it will be reasonable to think seriously
of the role played by non-factorizable contributions.

\section*{ Acknowledgements}

Y. Y. K would like to thank the Commissariat \`a l'Energie Atomique
of France for the award of a fellowship
and especially  G. Cohen-Tannoudji and J. Ha\"{\i}ssinski
for their encouragements.

\appendix

\renewcommand{\theequation}{\Alph{section}.\arabic{equation}}
\subsection*{Appendix A}
\setcounter{equation}{0}
\addtocounter{section}{1}

\ben
  \item
Error on $\rho_L$ ( Eq:(\ref{eq:68}) ).
\beq
\sam \rho_L = 2 \rho_L (1 - \rho_L)
\left\{ \left[ \frac{b x}{a - b x} \right]^2
        \left( \frac{\sam x^D}{x^D} \right)^2 +
        \left[ \frac{2 c^2 y^2}{1 + c^2 y^2} \right]^2
        \left( \frac{\sam y^D}{y^D} \right)^2 \right\}^{1/2} \label{eq:A1}
\eeq

  \item
Error on $N = (a - bx)^2 + 2 (1 + c^2 y^2)$
\beq
\sam N = 2 \left\{ [ b x (a - b x) ]^2
           \left( \frac{\sam x^D}{x^D} \right)^2 +
           \left[ 2 c^2 y^2 \right]^2
        \left( \frac{\sam y^D}{y^D} \right)^2 \right\}^{1/2} \label{eq:A2}
\eeq

   \item
Error on $R_{J/\Psi}$ ( Eq:(\ref{eq:70}) ).
\beq
\sam R_{J/\Psi} = R_{J/\Psi} \left\{
            \left( \frac{ \sam N}{N} \right)^2 +
4 \left( \frac{ \sam z^D}{z^D} \right)^2 \right\}^{1/2}   \label{eq:A3}
\eeq

    \item
{}From the PDG values in the $D$ sector at $q^2 = 0$ \cite{RPR}.
\beq
\frac{\sam x^D}{x^D} = \frac{15}{73} \hs{5mm}, \hs{5mm}
\frac{\sam y^D}{y^D} = \frac{25}{189} \hs{5mm}, \hs{5mm}
\left( \frac{\sam z^D}{z^D} \right)^2 =
\left( \frac{1}{14} \right)^2 + \left( \frac{1}{25} \right)^2  \label{eq:A4}
\eeq

    \item
In the $\Psi^{'}$ case, we make the situations
$a \ra a^{'}, b \ra b^{'}, c \ra c^{'}, x \ra x^{'}, y \ra y^{'}$
and $z \ra z^{'}$.

    \item
Error on $S$ ( Eq:(\ref{eq:123}) ).
\beq
\sam S = 2 \hs{2mm} S
\left\{ \left( \frac{ \sam f_{J/\Psi}}{f_{J/\Psi}} \right)^2 +
        \left( \frac{ \sam f_{\Psi}^{'}}{f_{\Psi}^{'}} \right)^2
\right\}^{1/2}                                                \label{eq:A5}
\eeq

    \item
Error on $S^*$ ( Eq:(\ref{eq:124}) ).
\beqa
\sam S^* &=& 2 \hs{2mm} S^*
\left\{ \left[ \frac{ \sam S}{2 S} \right]^2 +
        \left[ \frac{b x (a - b x)}{N} -
\frac{b^{'} x^{'} (a^{'} - b^{'}x^{'})}{N^{'}} \right]^2
\left( \frac{\sam x^D}{x^D} \right)^2 \right. \no \\
& & \hs{40mm} \left.
+ \hs{2mm} 4
\left[ \frac{c^2 y^2}{N} - \frac{{c^{'}}^2{y^{'}}^2}{N^{'}} \right]^2
\left( \frac{\sam y^D}{y^D} \right)^2
\right\}^{1/2}                                        \label{eq:A6}
\eeqa

\een

\subsection*{Appendix B}
\setcounter{equation}{0}
\addtocounter{section}{1}

\hs{5mm}
{\bf Semi-leptonic decay $ D \ra \ol{K} \hs{2mm} e^{+} \hs{2mm} \nu_{e}$} :

Using for the hadronic form factor $F_1^{DK}(q^2)$ a monopole $q^2$-dependence,
the decay rate is given by Eq.(\ref{eq:9-7}) where the integral
$I(\a_F)$ is defined in Eq.(\ref{eq:9-6}).
{}From the experimental rate, it is possible to deduce the value of
$F_1^{DK}(0)$ if the slope parameter $\a_F$ is known.

The computation of $I(\a_F)$ has been done both numerically and analytically.
In the case of interest here :
\beq
\frac{1}{(1 + r)^2} \hs{2mm} < \hs{2mm} \a_F
\hs{2mm} < \hs{2mm} \frac{1}{(1 - r)^2}
\hs{5mm}, \hs{5mm}
r = \frac{m_K}{m_D}    \label{eq:B1}
\eeq
the result is :
\beqa
I(\a_F) & = & \hs{2mm} \frac{1 - r^2}{2 \a_F^3} \
\left\{ -6 + 9 (1 + r^2) \a_F - 2 (1 - r^2)^2 \a_F^2 \right\} \no \\
& - & \hs{2mm} \frac{3}{\a_F^4} \
\left\{ 1 - 2 (1 + r^2) \a_F + (1 + r^4) \a_F^2 \right\} \ln r   \no \\
& + & \hs{2mm} \frac{3}{\a_F^4} \ \hs{2mm}
[1 - (1 + r^2) \a_F]
\sqrt{ [(1 + r)^2 \a_F - 1][1 - (1 - r)^2 \a_F ] } \no \\
& \cdot & \hs{2mm} \left\{ \frac{\pi}{2} \
- Arctg
\frac{(1 + r^2) - (1 - r^2)^2 \a_F }
{(1 - r^2) \sqrt{ [(1 + r)^2 \a_F - 1][1 - (1 - r)^2 \a_F ]  } } \
\right\} \label{eq:B2}
\eeqa

Of course such a formula can be used for any semi-leptonic decay of the type
${\bf 0}^{-}$ $\ra$ ${\bf 0}^{-}  +  e^{+} + \nu_{e}$
provided the corresponding form factor has a monopole $q^2$-dependence.

\newpage
%
%

\def\pr{{\sl Phys. Rev.}~}
\def\prl{{\sl Phys. Rev. Lett.}~}
\def\pl{{\sl Phys. Lett.}~}
\def\np{{\sl Nucl. Phys.}~}
\def\zp{{\sl Z. Phys.}~}



\renewcommand{\textfraction}{0}
\clearpage

\section*{\hs{3mm} Figure Captions and Table Captions}

\section*{A. \hs{2mm} Figure captions}
\vspace{0.5cm}

\ben

\item
{\bf Figure 1} :
Parameters for the $B \ra K$ transitions.
On the horizontal axis is $\Lambda_F^2$,
on the left vertical axis is the parameter $\mu$,
on the right vertical axis is the pole mass $\Lambda_{DF}^2$.
The relevant relations are
$$
\mu^B = \frac{m_B^2 - m_K^2}{\Lambda_F^2} \hs{3mm}, \hs{3mm}
\mu^D =\frac{\mu^B - \sigma}{1 + \sigma \mu^B} \hs{3mm}, \hs{3mm}
\ol{\mu}^D = \frac{m_D^2 - m_K^2}{\Lambda_{DF}^2}
$$
$$
\Lambda_{DF}^2 = \frac{m_c}{m_b} \hs{1mm} \Lambda_F^2 -
\left( 1 - \frac{m_c}{m_b} \right) (m_b m_c - m_K^2)
$$
The values of $\Lambda_F$ and $\Lambda_{DF}$ used by Bauer-Stech-Wirbel
are indicated by the point BSW and they are $\Lambda_F$ = 5.43 $GeV$ and
$\Lambda_{DF}$ = 2.11 $GeV$.

\item
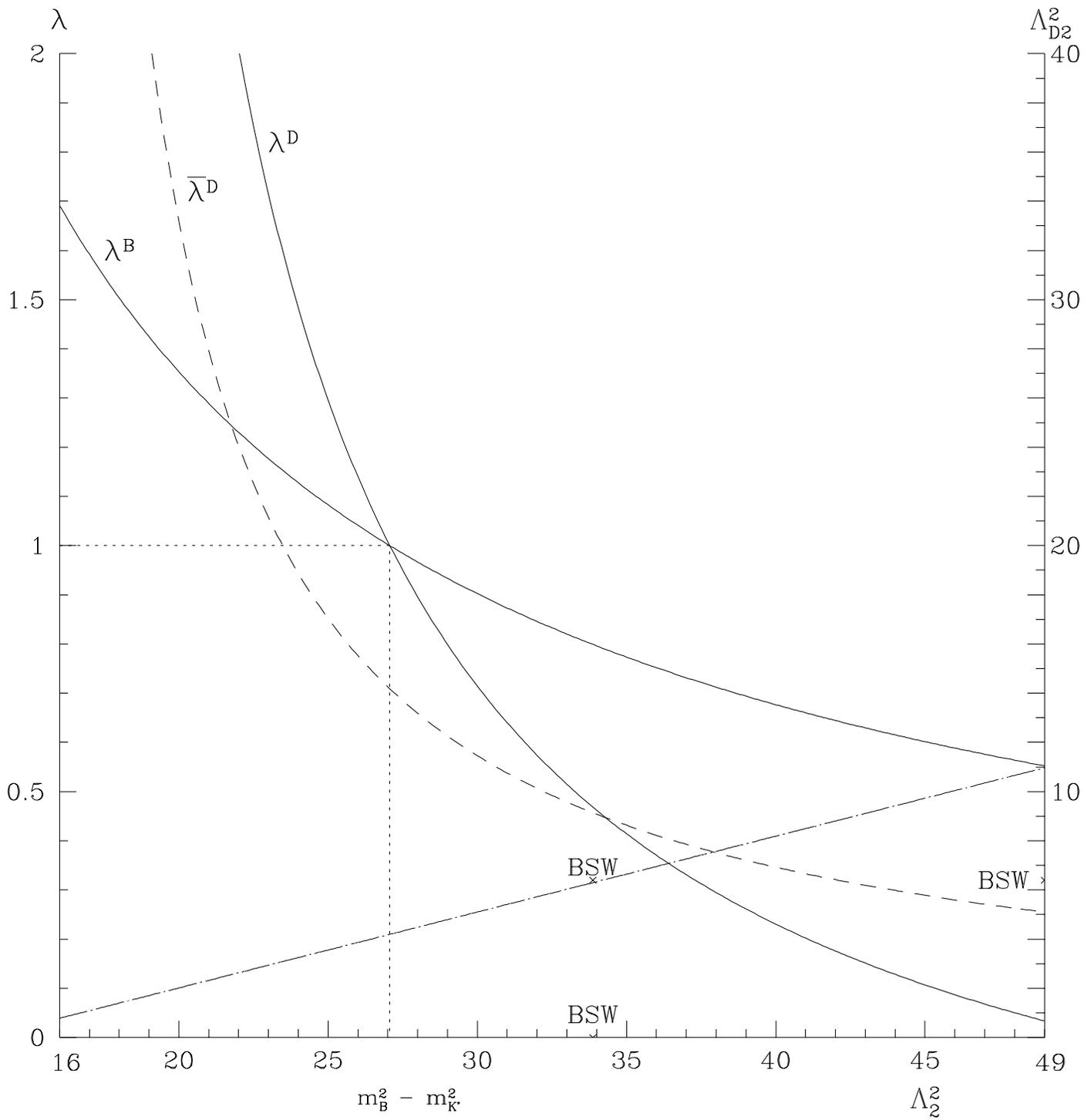
{\bf Figure 2} :
Parameters for the $B \ra K^*$ transitions.
On the horizontal axis is $\Lambda_2^2$,
on the left vertical axis is the parameter $\ld$,
on the right vertical axis is the pole mass $\Lambda_{D2}^2$.
The relevant relations are
$$
\ld^B = \frac{m_B^2 - m_{K^*}^2}{\Lambda_2^2} \hs{3mm}, \hs{3mm}
\ld^D =\frac{\ld^B - \sigma}{1 + \sigma \ld^B} \hs{3mm}, \hs{3mm}
\ol{\ld}^D = \frac{m_D^2 - m_{K^*}^2}{\Lambda_{D2}^2}
$$
$$
\Lambda_{D2}^2 = \frac{m_c}{m_b} \hs{1mm} \Lambda_2^2 -
\left( 1 - \frac{m_c}{m_b} \right) (m_b m_c - m_{K^*}^2)
$$
The values of $\Lambda_2$ and $\Lambda_{D2}$ used by Bauer-Stech-Wirbel
are indicated by the point BSW and they are $\Lambda_2$ = 5.82 $GeV$ and
$\Lambda_{D2}$ = 2.53 $GeV$.

\item
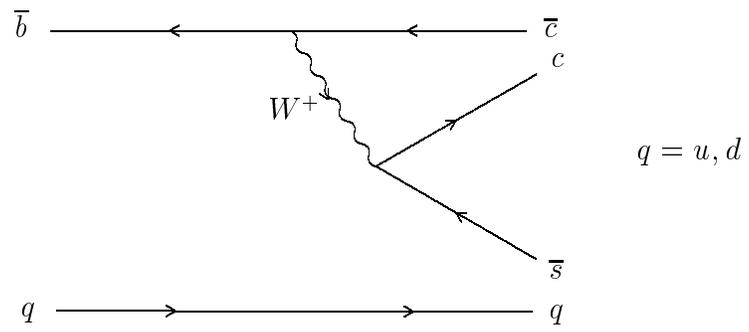
{\bf Figure 3} :
The quark level colour-suppressed diagram
$\ol{b}q \ra (\ol{c}c) + (\ol{s}q) $.

\item
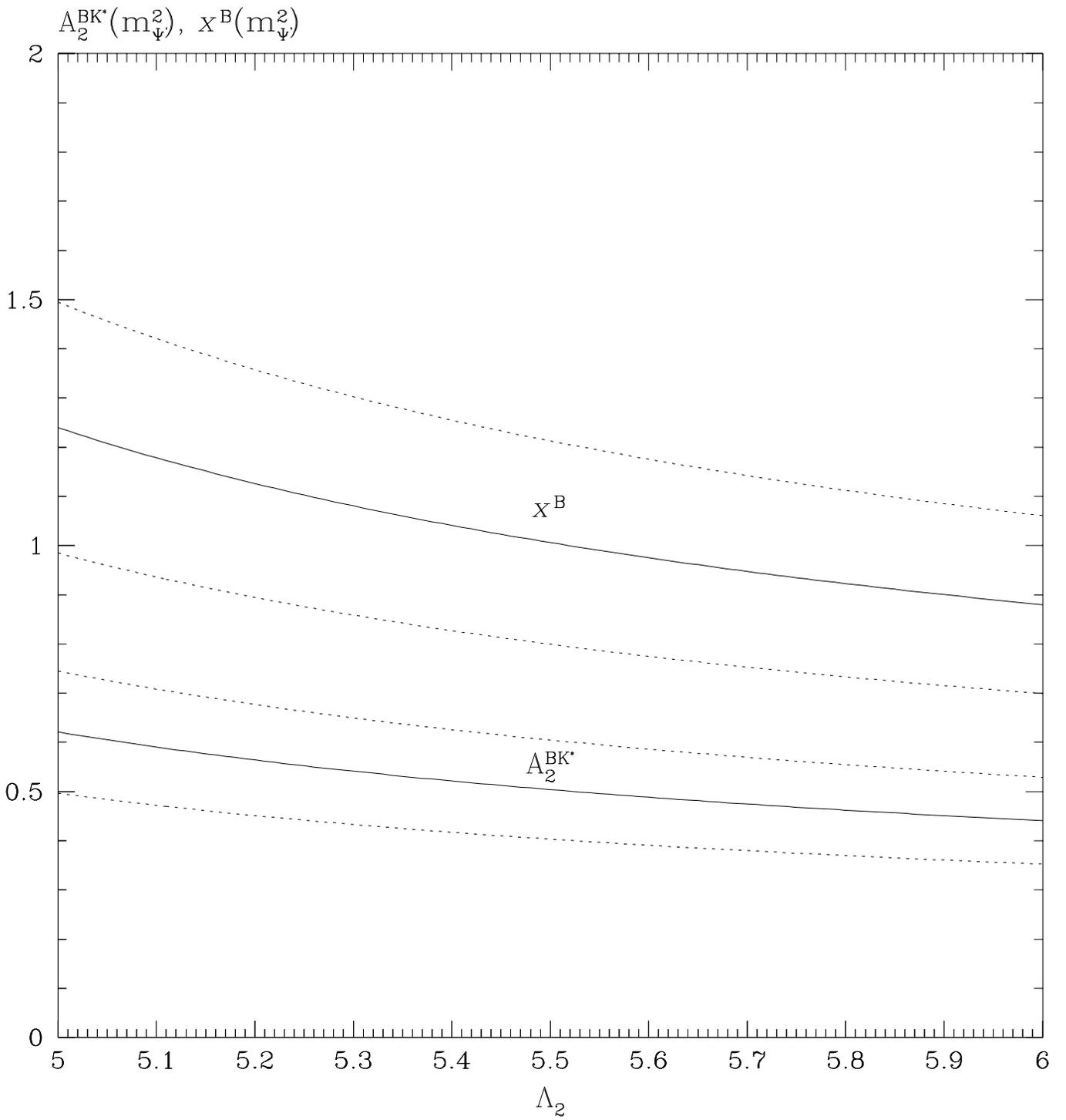
{\bf Figure 4} :
$A_2^{BK*}(m_{\Psi^{'}}^2)$ and $x^B(m_{\Psi^{'}}^2)$ as functions of
$\Lambda_2$ for $ 5 \hs{2mm}GeV \leq \Lambda_2 \leq 6 \hs{2mm}GeV$.
One standard deviations are indicated with dotted points.

\item
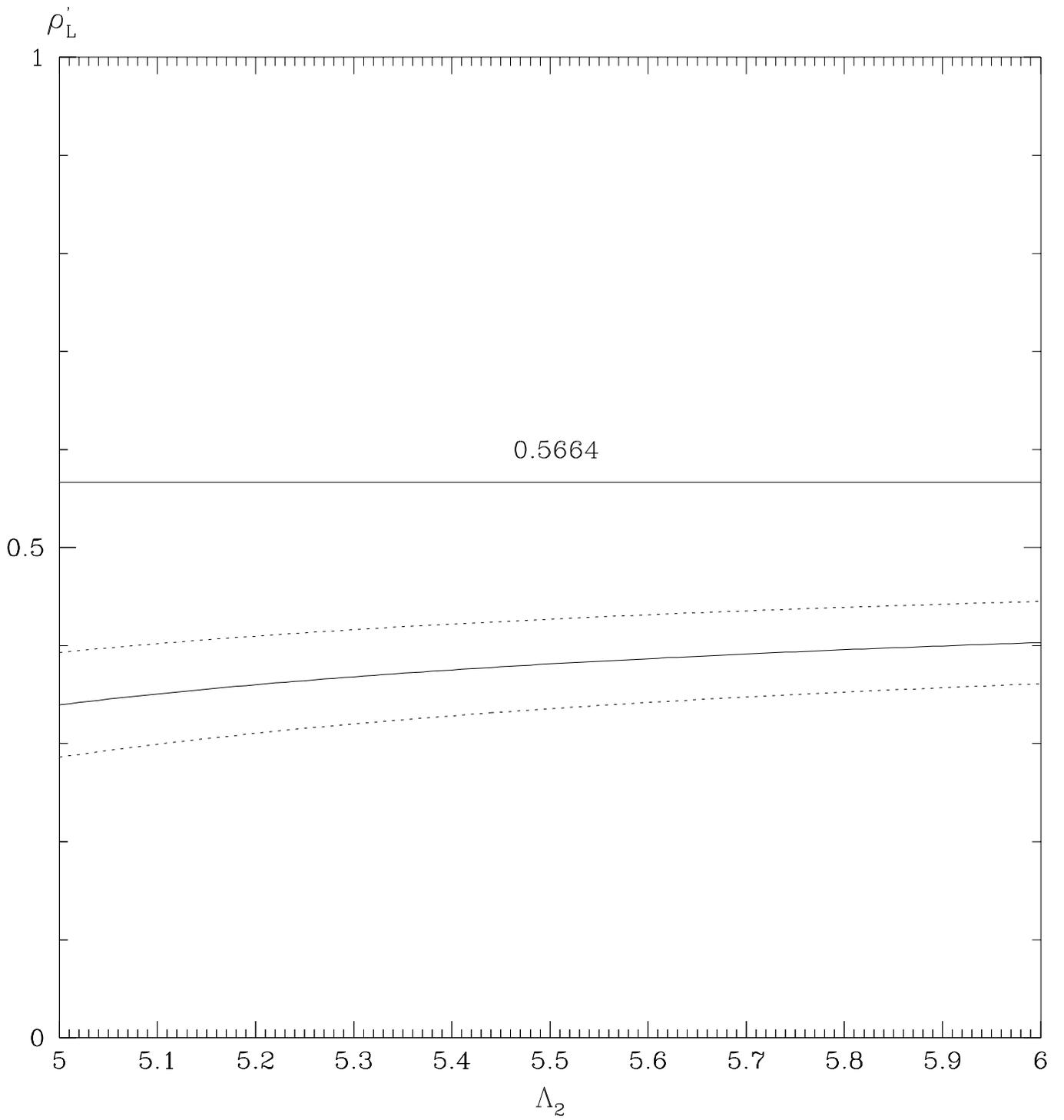
{\bf Figure 5} :
The fractional longitudinal polarization $\rho_L^{'}$ as a function
of $\Lambda_2$ for $5 \hs{2mm}GeV \leq \Lambda_2 \leq 6 \hs{2mm}GeV$.
One standard deviations are indicated with dotted points.
Our scenario-independent upper limit $\rho_L^{'} \leq 0.5664$ is indicated.

\item
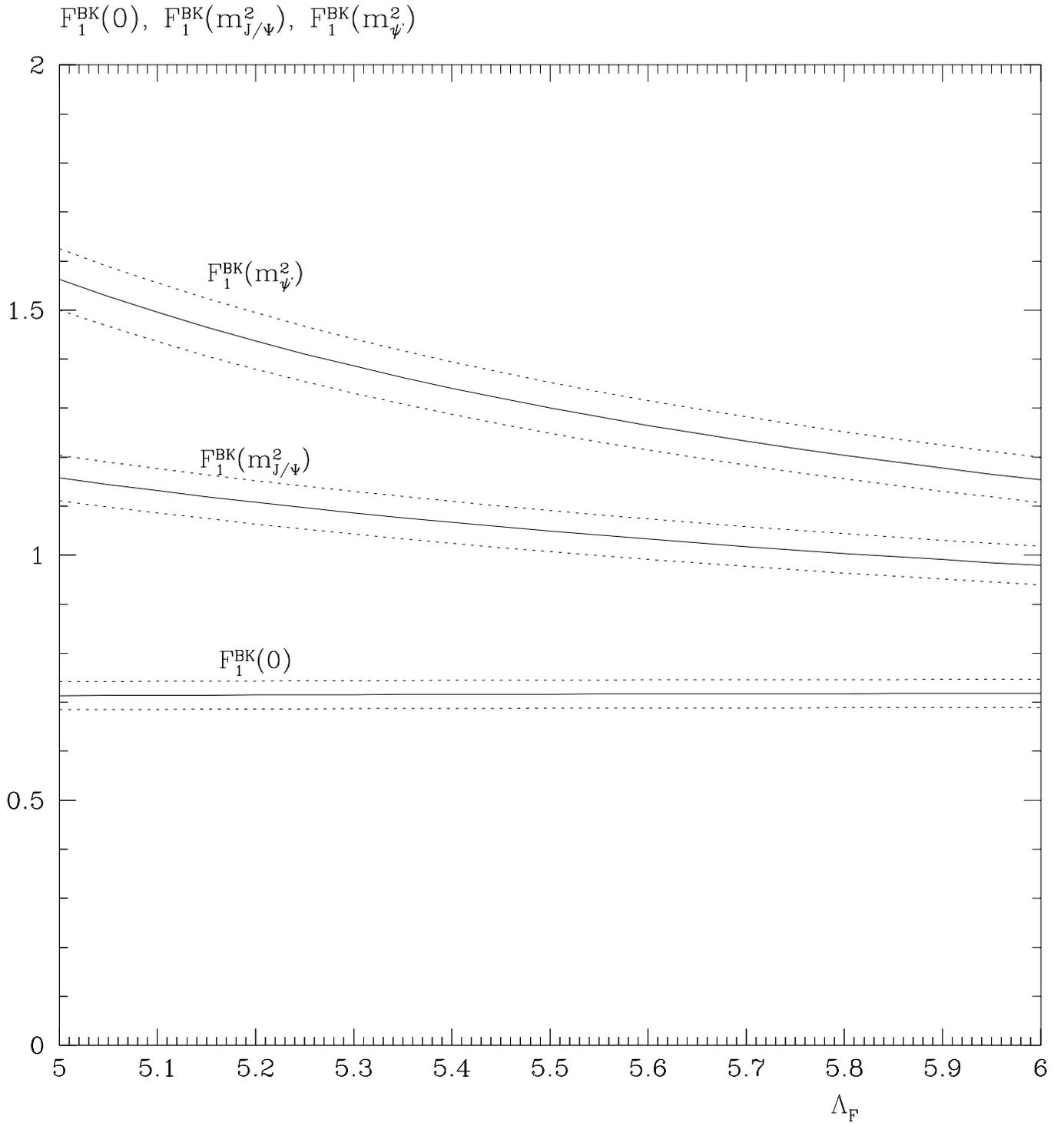
{\bf Figure 6} :
The hadronic form factor $F_1^{BK}(q^2)$ at $q^2 =0$, $q^2 = m^2_{J/\Psi}$
and $q^2 = m_{\Psi^{'}}^2$ as functions of $\Lambda_F$ for
$5 \hs{2mm}GeV \leq \Lambda_F \leq 6 \hs{2mm}GeV$.
One standard deviations are indicated with dotted points.

\item
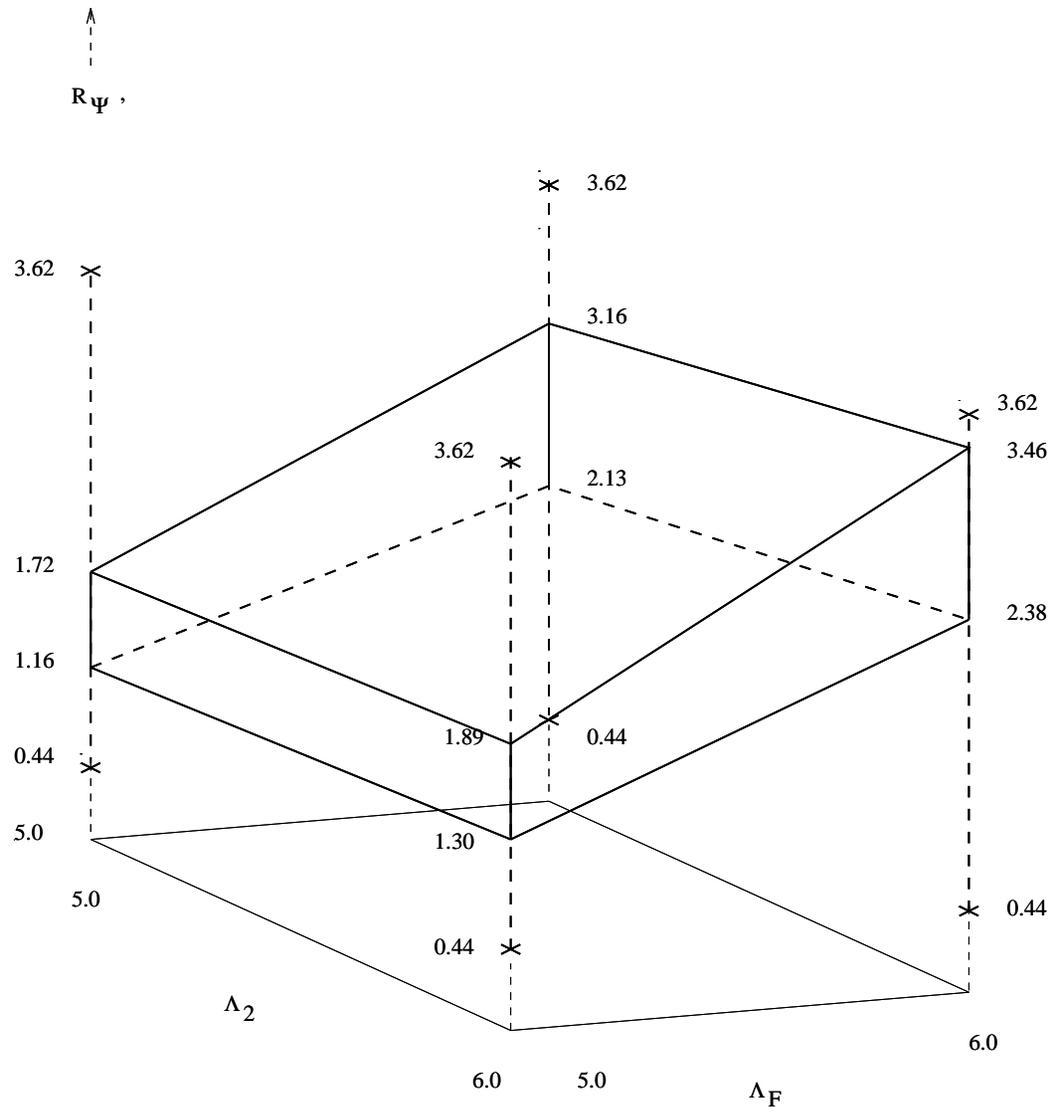
{\bf Figure 7} :
The ratio of rates $R_{\Psi^{'}}$ as a function of $\Lambda_2$
and $\Lambda_F$ for
$5 \hs{2mm}GeV \leq \Lambda_2, \Lambda_F \leq 6 \hs{2mm}GeV$.
The cross points indicate the one standard deviation experimental limits
$0.44 \leq R_{\Psi^{'}} \leq 3.62$.

\item
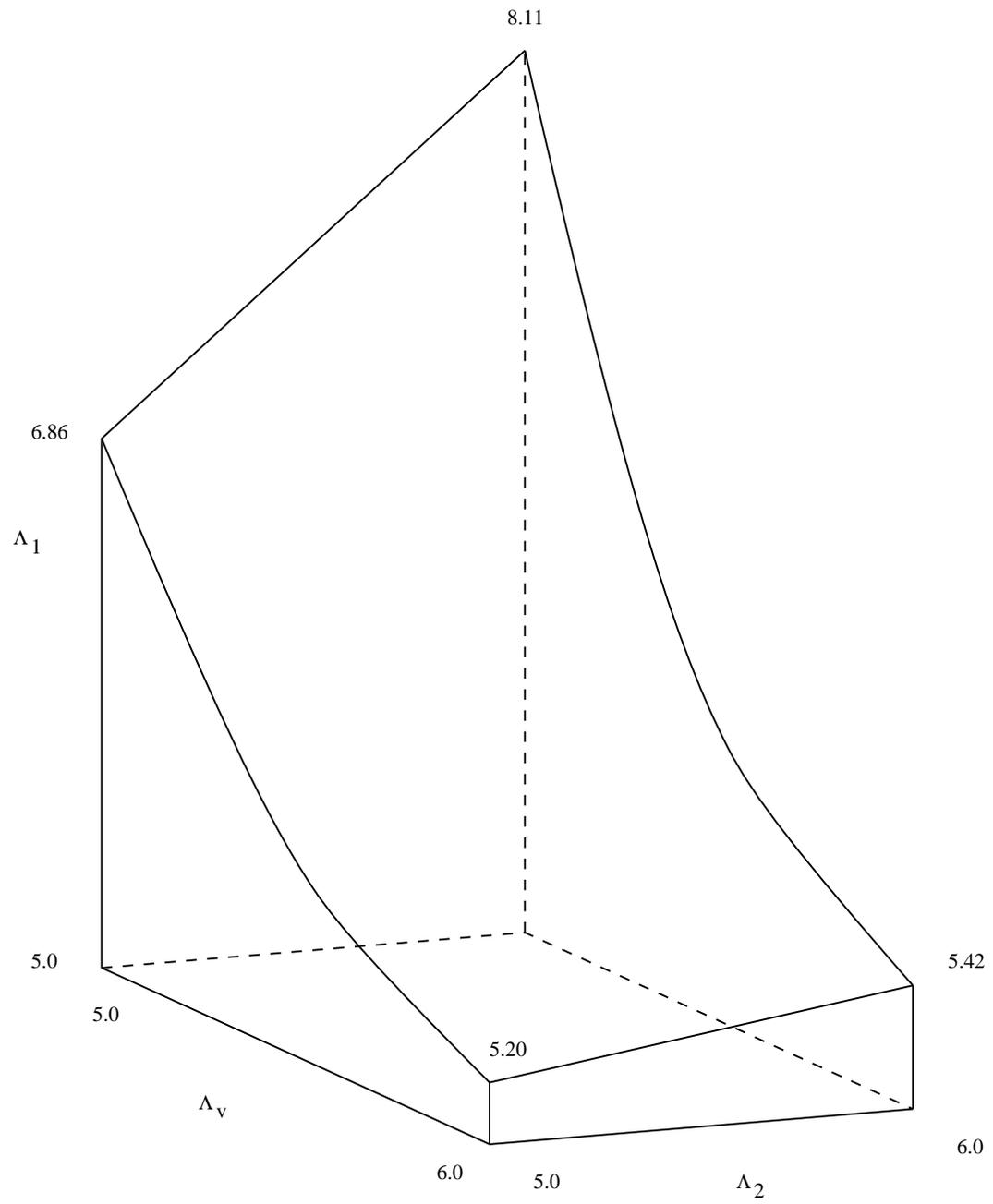
{\bf Figure 8} :
The allowed domain in the $\Lambda_1$, $\Lambda_2$, $\Lambda_V$ space
due to the constraints $\rho_L + \sam \rho_L \geq 0.70$
 for $\Lambda_2, \Lambda_V$ $\in$
(5 - 6) $GeV$, $\Lambda_1 \geq 5 \hs{2mm} GeV$.
The scenario is $[n_1, n_2, n_V] = [-1, 2, 2]$.

\item
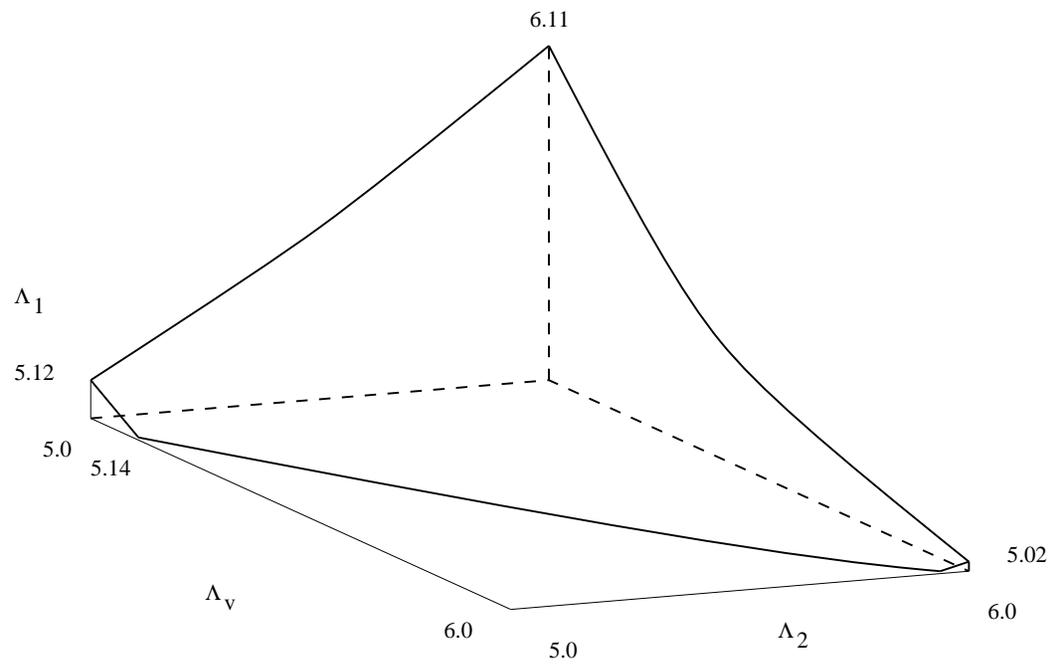
{\bf Figure 9} :
Same as Figure 8 for the scenario [-1, 1, 2].

\item
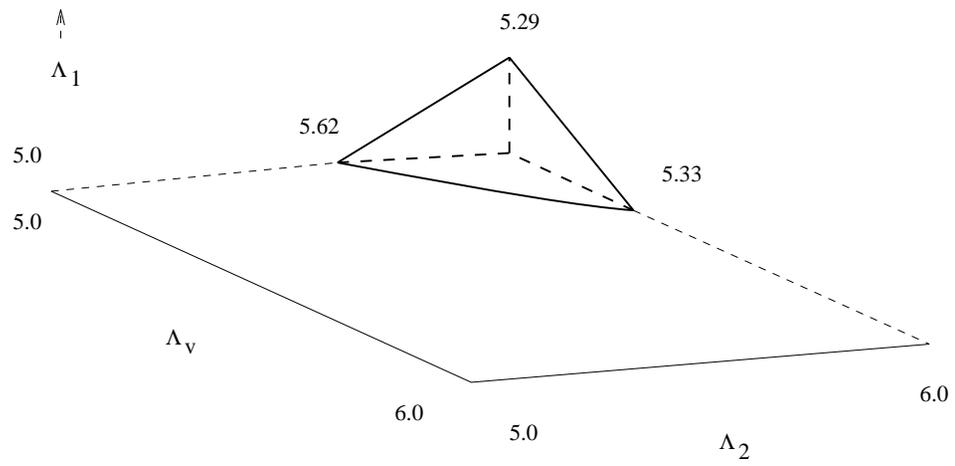
{\bf Figure 10} :
Same as Figure 8 for the scenario [-1, 0, 2].

\item
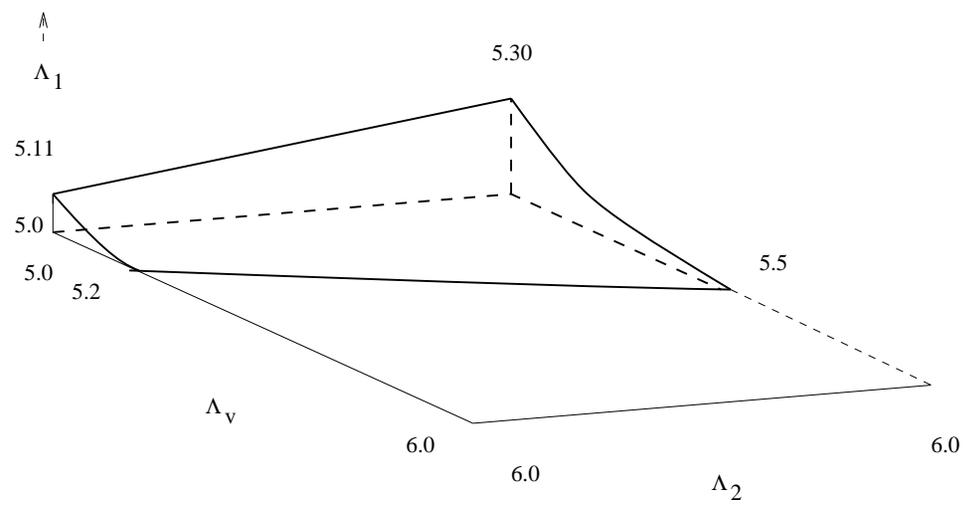
{\bf Figure 11} :
Same as Figure 8 for the scenario [-1, 2, 1].

\item
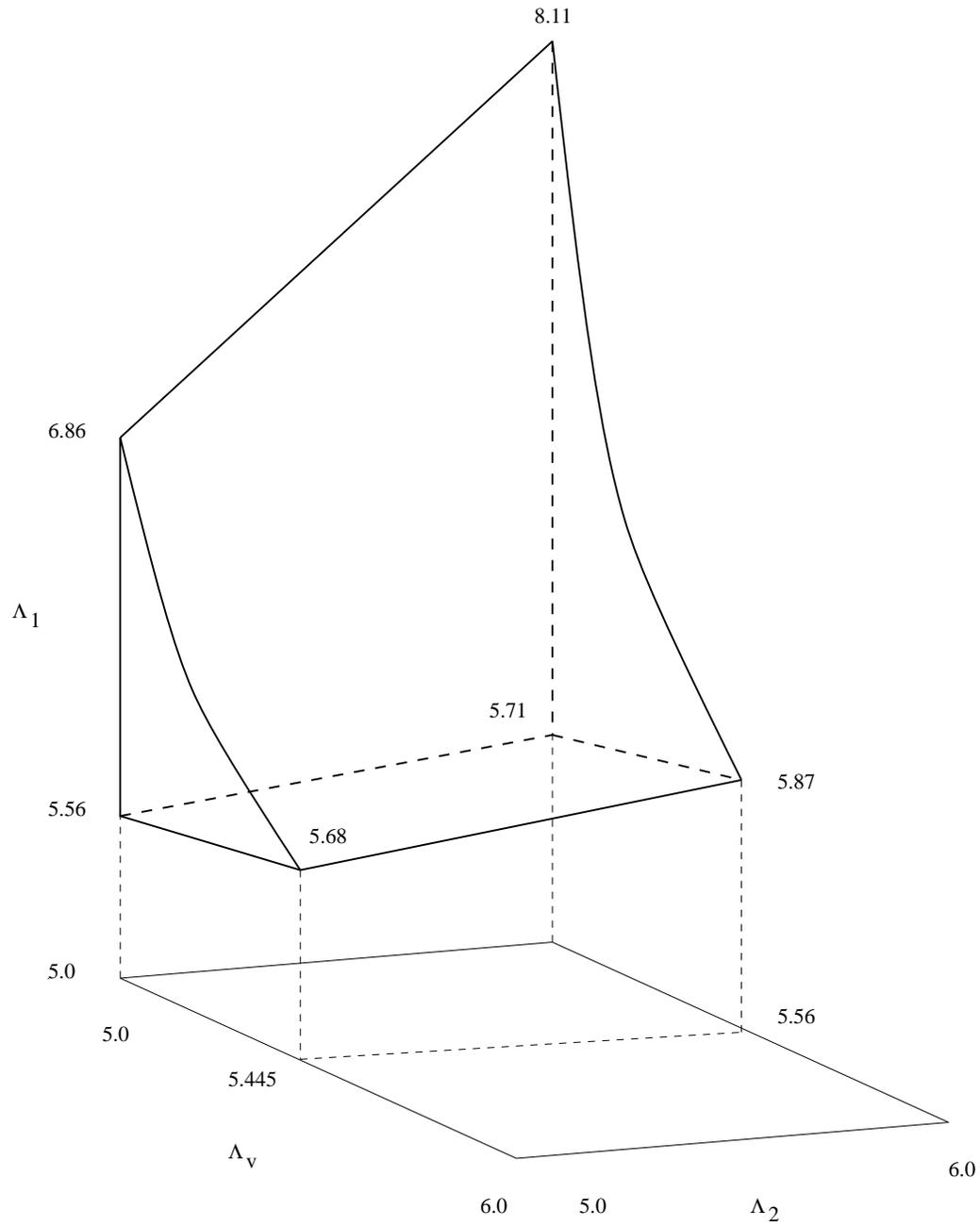
{\bf Figure 12} :
The allowed domain in the $\Lambda_1$, $\Lambda_2$, $\Lambda_V$ space
due to the constraints $\rho_L + \sam \rho_L \geq 0.70$
and $R_{J/\Psi} - \sam R_{J/\Psi} \leq 2.20$ for $\Lambda_2, \Lambda_V$ $\in$
(5 - 6) $GeV$, $\Lambda_1 \geq 5 \hs{2mm} GeV$.
The scenario is $[n_1, n_2, n_V] = [-1, 2, 2]$.
$\sam \rho_L$ and $\sam R_{J/\Psi}$ are theoretical errors
induced by experimental errors of $x^D(0), y^D(0)$ and $z^D(0)$.

\item
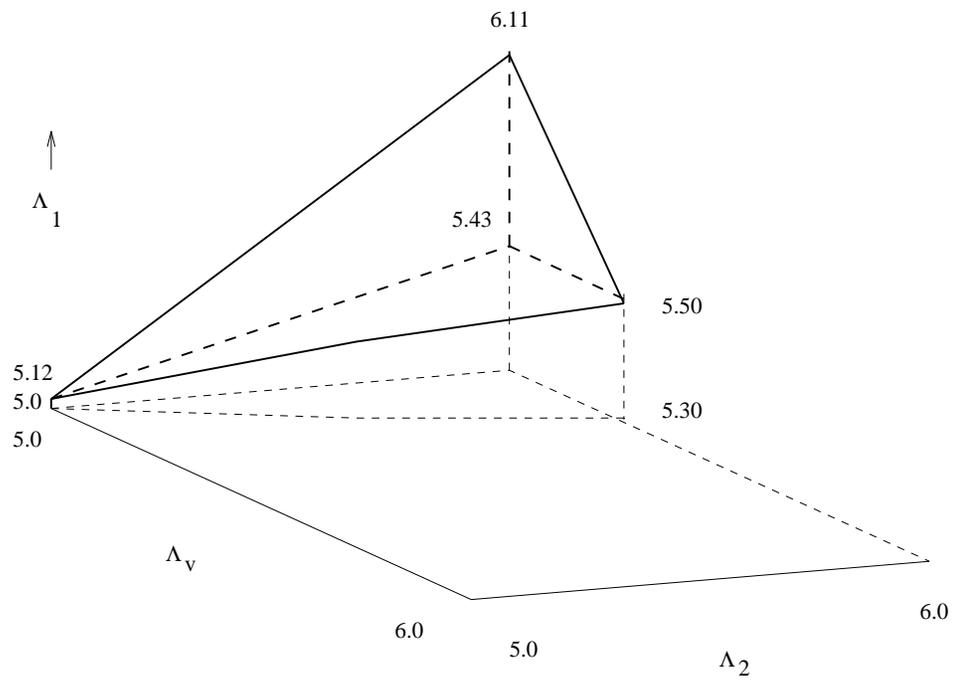
{\bf Figure 13} :
Same as Figure 12 for the scenario [-1, 1, 2].

\item
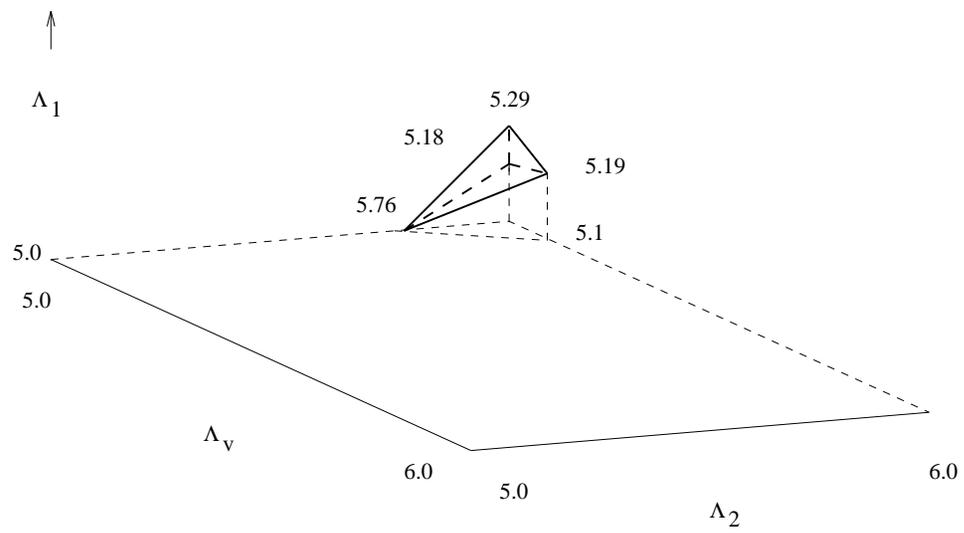
{\bf Figure 14} :
Same as Figure 12 for the scenario [-1, 0, 2].

\item
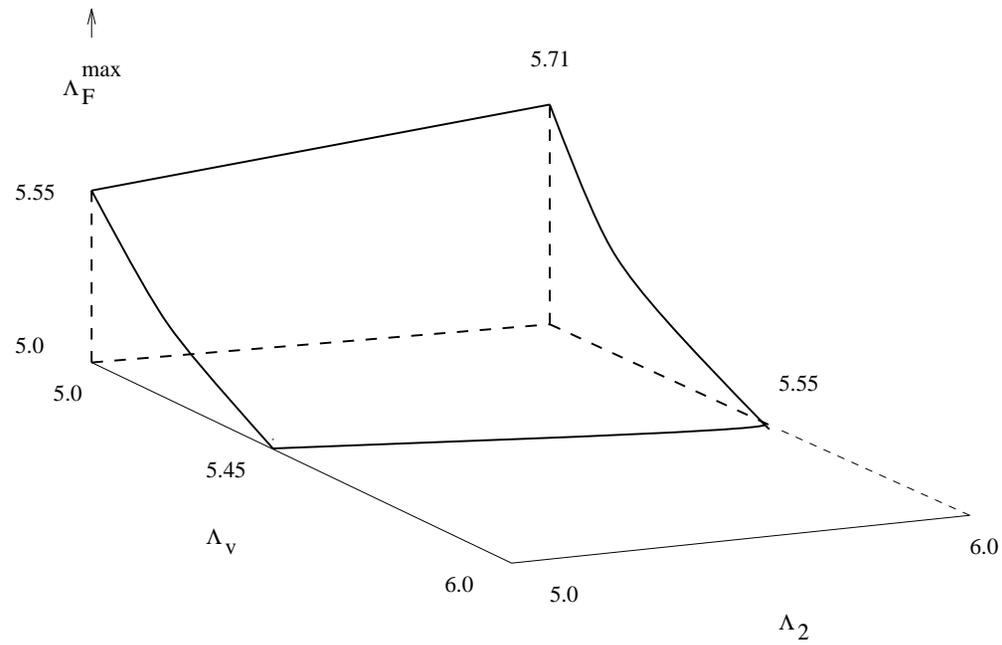
{\bf Figure 15} :
The allowed domain in the $\Lambda_F$, $\Lambda_2$, $\Lambda_V$ space
due to the constraint

$\Lambda_{1, \hs{1mm}MIN}(\Lambda_2, \Lambda_V, \Lambda_F = 5 \hs{2mm} GeV)
\leq \Lambda_1 \leq \Lambda_{1, \hs{1mm}MAX}(\Lambda_2, \Lambda_V)$
with $\Lambda_F \geq 5 \hs{1mm} GeV$.
The scenario is $[n_1, n_2, n_V] = [-1, 2, 2]$.

\item
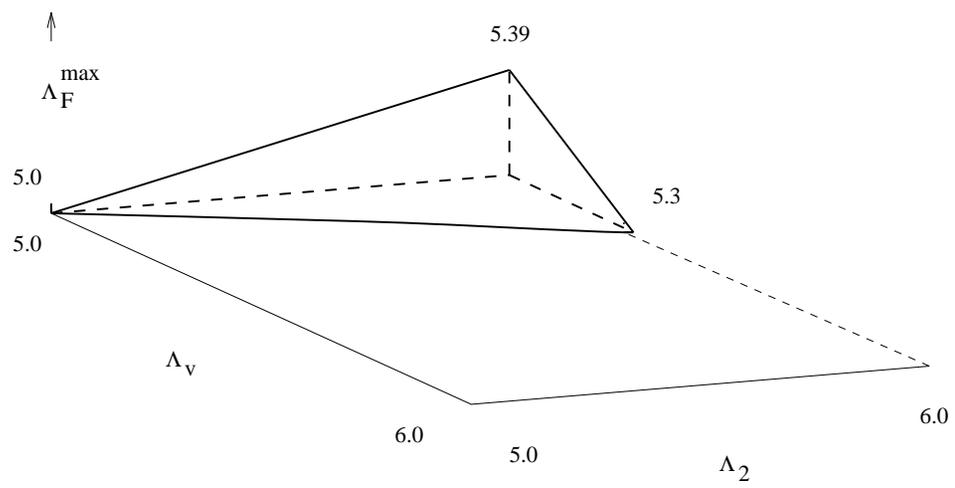
{\bf Figure 16} :
Same as Figure 15 for the scenario [-1, 1, 2].

\item
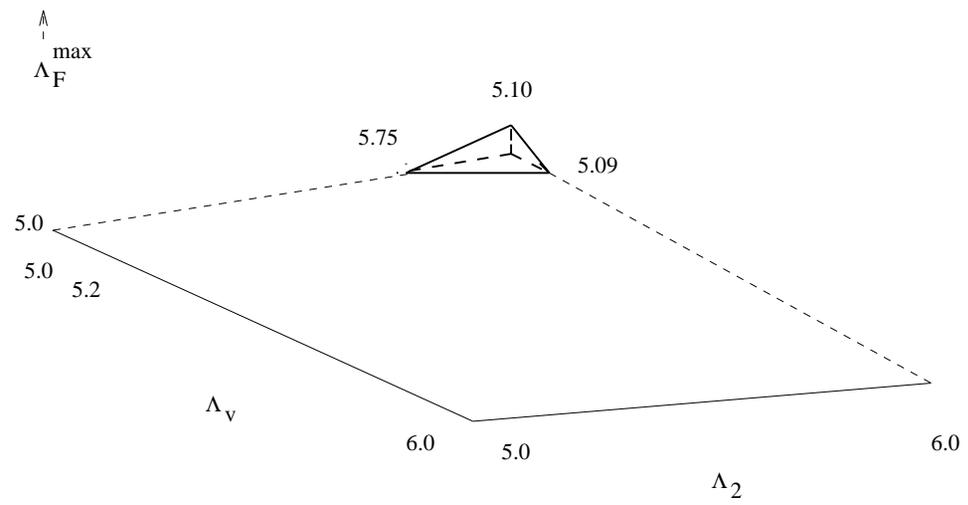
{\bf Figure 17} :
Same as Figure 15 for the scenario [-1, 0, 2].

\item
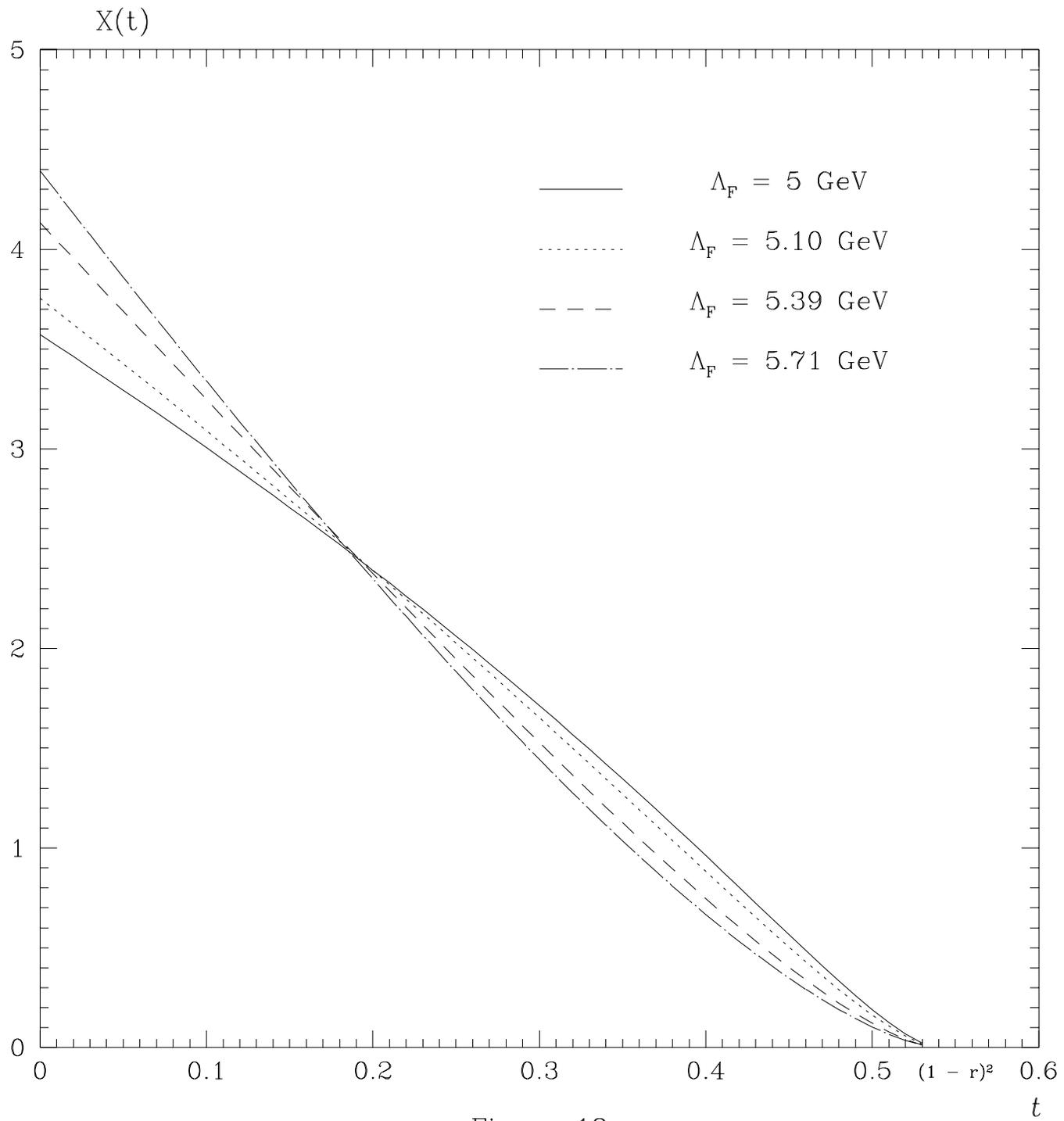
{\bf Figure 18} :
The normalized dimensionless distribution $X(t)$ for the semi-leptonic
decay $D \ra \ol{K} + \ell^{+} + \nu_{\ell}$.
The scenario $n_2 = 2$ corresponds to
$5 \hs{2mm} GeV \hs{1mm} \leq \hs{1mm} \Lambda_F \hs{1mm} \leq \hs{1mm}
5.71 \hs{2mm} GeV$, the scenarios $n_2 =1 $ corresponds
to
$5 \hs{2mm} GeV \hs{1mm} \leq \hs{1mm} \Lambda_F \hs{1mm} \leq \hs{1mm}
5.39 \hs{2mm} GeV$ and the scenario $n_2 = 0$ corresponds to
$5 \hs{2mm} GeV \hs{1mm} \leq \hs{1mm} \Lambda_F \hs{1mm} \leq \hs{1mm}
5.10 \hs{2mm} GeV$.
By Eq.(\ref{eq:19}) the pole masses $\Lambda_{DF}$ in the $D$ sector
can be obtained from $\Lambda_F$ given here.

\item
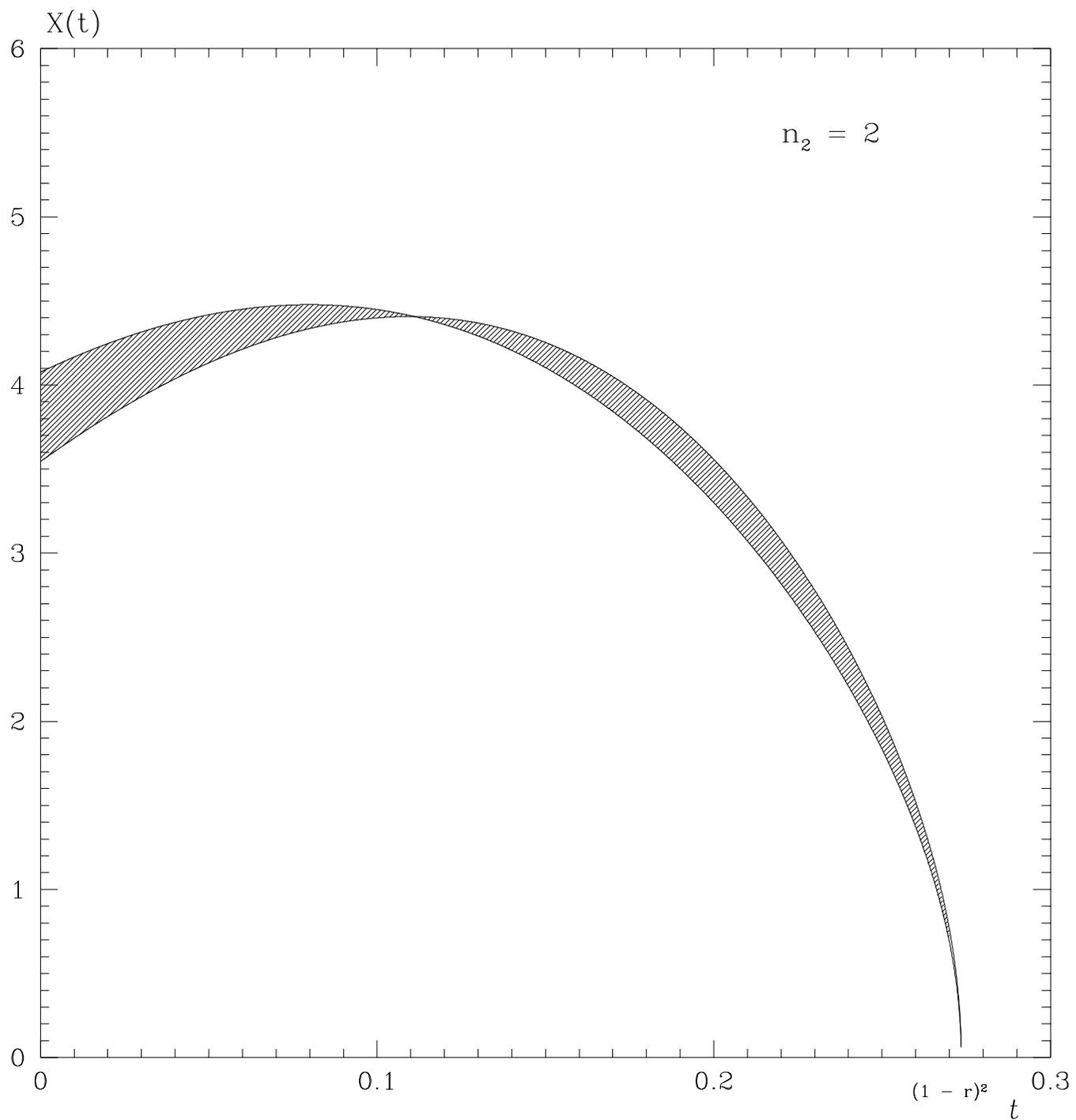
{\bf Figure 19} :
The normalized dimensionless distribution
$X(t)$ for the semi-leptonic decay
$D \ra \ol{K}^{*} + \ell^{+} + \nu_{\ell}$.
The scenario is $[n_1, n_2, n_V] = [-1, 2, 2]$.
The thickness of the curve is due to the $\Lambda_1,
\Lambda_2, \Lambda_V $ ranges.

\item
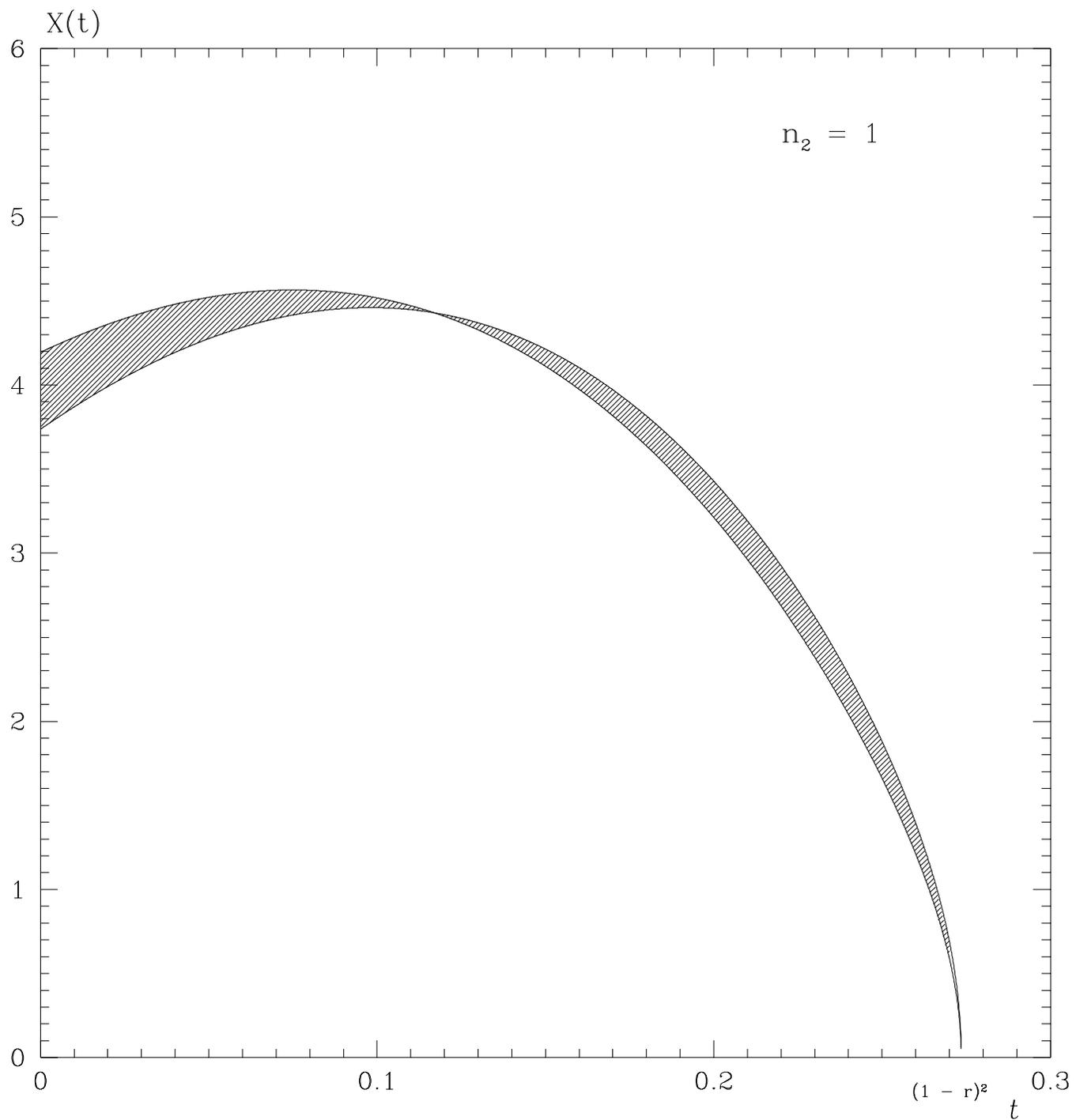
{\bf Figure 20} :
Same as Figure 19 for the scenario [-1, 1, 2].

\item
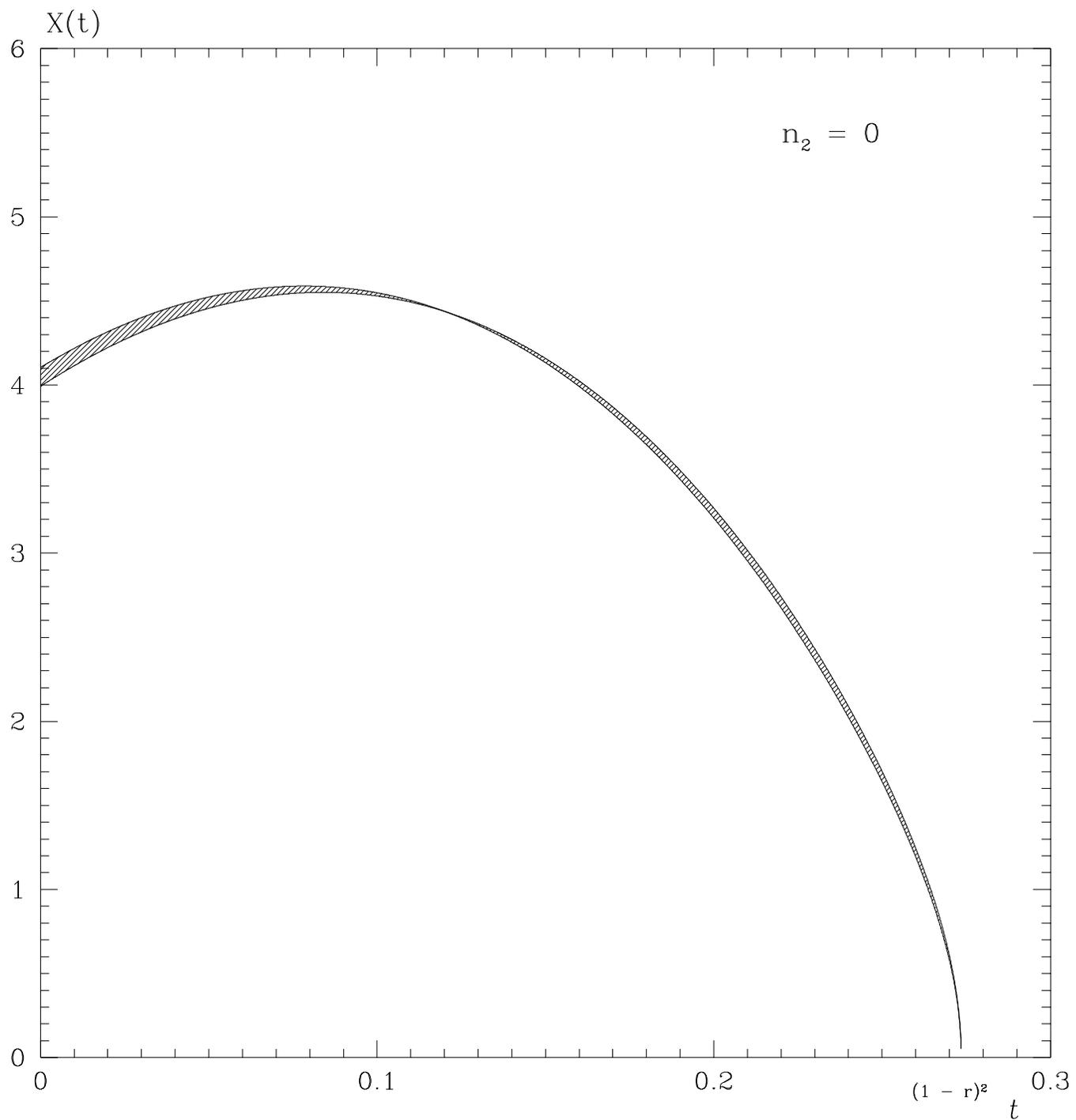
{\bf Figure 21} :
Same as Figure 19 for the scenario [-1, 0, 2].

\een

\clearpage

\section*{B. \hs{2mm} Table Captions}

\begin{enumerate}

\item
{\bf Table 1} : \\
Experimental data of the decay rates
for $K^{(*)} + \Psi^{'}$ and $K^{(*)} + J/\Psi$
as averaged by PDG \cite{RPR}.

\item
{\bf Table 2} : \\
Experimental data and the averaged values
for the ratios $R_{\Psi^{'}}$, $R_{J/\Psi}$.

\item
{\bf Table 3} : \\
Numerical results for $\rho_L$, $\rho_L + \sam \rho_L$,
$R_{J/\Psi}$ and
$R_{J/\Psi} - \sam R_{J/\Psi}$
at three values of $\Lambda_1$ ;
$\Lambda_{1,\hs{1mm}MAX}$, $\Lambda_{1,\hs{1mm}MIN}$
and an intermediate value between two extreme values
in three scenarios ($n_2 = 2, 1,0$).

\item
{\bf Table 4} : \\
Experimental data and the averaged values
for the ratios $S$, $S^*$.

\end{enumerate}

\end{document}